\documentclass{article}

\usepackage[utf8]{inputenc}

\usepackage[T1]{fontenc}
\usepackage{parskip}
\usepackage{ragged2e}
\usepackage{enumitem}
\usepackage{float}
\usepackage{lscape}
\usepackage{rotating}

\usepackage{fancyhdr}
\usepackage{etoolbox}
\usepackage{geometry}
 \newcommand{\tmpx}{}

\fancypagestyle{sec}{\lhead{\tmpx}}

\usepackage{amsmath}
\usepackage{amsfonts}
\usepackage{amssymb}
\usepackage{mathtools}
\usepackage{amsthm}
\usepackage{bbm}
\usepackage{nccmath} 
\usepackage{scalerel}
\usepackage{actuarialsymbol}
\usepackage{mleftright}

\usepackage{multirow}
\usepackage{multicol}
\usepackage{tabularx}

\usepackage{graphicx}
\usepackage[table]{xcolor}
\usepackage{array}
\usepackage{subcaption}

\usepackage{xurl}
\usepackage{thmtools, thm-restate}
\usepackage{enumitem}
\usepackage{titlesec}
\usepackage[absolute,overlay]{textpos}

\usepackage{scrlayer}
\DeclareNewLayer[
  background,
  rightmargin,
  contents={%
    \parbox[b][\layerheight][c]{\dimexpr\footskip+\footheight\relax}{%
      \hfill\rotatebox{90}{\pagemark}}}
]{lscape.foot}
\DeclareNewLayer[
  background,
  textarea,
  addhoffset=\dimexpr-\headsep-\headheight\relax,
  width=\dimexpr\headsep+\headheight\relax,
  contents={\hfill\rotatebox{90}{\headmark}\hspace*{\headsep}}
]{lscape.head}
\DeclareNewPageStyleByLayers{lscape}{lscape.foot,lscape.head}

\usepackage[sectionbib, round]{natbib}
\usepackage[colorlinks, linkcolor = black,  linkcolor = blue,
urlcolor  = blue,
citecolor = blue,
anchorcolor = blue]{hyperref}
\usepackage{etoolbox}
\usepackage{bookmark}

\titleformat{\subsubsection}
{\normalfont\normalsize\bfseries}{\thesubsubsection}{1em}{}
\titlespacing*{\subsubsection}
{0pt}{3.25ex plus 1ex minus .2ex}{1.5ex plus .2ex}
 
 \titleformat{\paragraph}[runin]
{\itshape}{\theparagraph .}{1em}{}

\makeatletter
\newcounter{author}
\renewcommand*\author[1]{%
  \stepcounter{author}%
  \ifnum\c@author=1
    \gdef\@author{#1}%
  \else
    \xdef\@author{\unexpanded\expandafter{\@author\and#1}}%
  \fi
  \csgdef{author@\the\c@author}{#1}}
\newcommand*\email[1]{%
  \csgdef{email@\the\c@author}{#1}}
\newcommand*\orcid[1]{%
  \csgdef{orcid@\the\c@author}{#1}}
\newcommand*\address[1]{%
  \csgdef{address@\the\c@author}{#1}}
\AtEndDocument{%
  \xdef\author@count{\the\c@author}%
  \c@author=1
  \print@authors}
\newcommand*\print@authors{%
  \ifnum\c@author>\author@count
  \else
    \print@author{\the\c@author}%
    \advance\c@author by 1
    \expandafter\print@authors
  \fi}
\newcommand*\print@author[1]{%
  \par\medskip
  \begin{tabular}{@{}l@{}}%
    \textsc{\csuse{author@#1}}\\
    \csuse{address@#1}\\
    \textit{E-Mail}:
    \href{mailto:\csuse{email@#1}}{\csuse{email@#1}}\\
    \textit{ORCiD}:
    \href{\csuse{orcid@#1}}{\csuse{orcid@#1}}
  \end{tabular}}
\patchcmd{\NAT@test}{\else \NAT@nm}{\else \NAT@hyper@{\NAT@nm}}{}{}
\makeatother

\numberwithin{equation}{section}

\newtagform{normalsize}[\normalsize]{\normalsize(}{\normalsize)}

\usetagform{normalsize}

\numberwithin{equation}{section}
\theoremstyle{definition}

\numberwithin{definition}{section}
\theoremstyle{thoerem}
\newtheorem{theorem}{Theorem}
\numberwithin{theorem}{section}
\theoremstyle{plain}
\newtheorem{proposition}{Proposition}
\numberwithin{proposition}{section}
\theoremstyle{plain}

\numberwithin{corollary}{section}
\theoremstyle{remark}
\newtheorem{remark}{Remark}
\numberwithin{remark}{section}
\mathtoolsset{showonlyrefs}

\providecommand{\keywords}[1]
{
  \small	
  \textbf{\textit{Keywords---}} #1
}

\author{Jos\'e Miguel Flores-Contró}
\address{\textit{Department of Actuarial Science} \\ \textit{Faculty of Business and Economics} \\
\textit{University of Lausanne} \\ \textit{Lausanne, Switzerland}}
\email{josemiguel.florescontro@unil.ch}
\orcid{https://orcid.org/0000-0001-9332-8811}
\author{S\'everine Arnold}
\address{\textit{Department of Actuarial Science} \\ \textit{Faculty of Business and Economics} \\
\textit{University of Lausanne} \\ \textit{Lausanne, Switzerland}}
\email{severine.arnold@unil.ch}
\orcid{https://orcid.org/0000-0001-9851-6328}

\title{The Role of Direct Capital Cash Transfers Towards Poverty and Extreme Poverty Alleviation - An Omega Risk Process}

\begin{document}

\date{\vspace{-2ex}}

\maketitle

\begin{abstract}
Trapping refers to the event when a household falls into the area of poverty. Households that live or fall into the area of poverty are said to be in a poverty trap, where a poverty trap is a state of poverty from which it is difficult to escape without external help. Similarly, extreme poverty is considered as the most severe type of poverty, in which households experience severe deprivation of basic human needs. In this article, we consider an Omega risk process with deterministic growth and a multiplicative jump (collapse) structure to model the capital of a household. It is assumed that, when a household's capital level is above a certain capital barrier level that determines a household's eligibility for a capital cash transfer programme, its capital grows exponentially. As soon as its capital falls below the capital barrier level, the capital dynamics incorporate external support in the form of direct transfers (capital cash transfers) provided by donors or governments. Otherwise, once trapped, the capital grows only due to the capital cash transfers. Under this model, we first derive closed-form expressions for the trapping probability and then do the same for the probability of extreme poverty, which only depends on the current value of the capital given by some extreme poverty rate function. Numerical examples illustrate the role of capital cash transfers on poverty and extreme poverty dynamics.
\vspace{0.5cm}


\end{abstract}

\section{Introduction} \label{Introduction-Section1}

In development economics, households that live or fall below the poverty line are said to be in a poverty trap,  where a poverty trap is a state of poverty from which it is difficult to escape without external help. Similarly, extreme poverty refers to the most severe type of poverty, characterised by severe deprivation of basic human needs, including food, safe drinking water, sanitation facilities, health, shelter, education and information \citep{Book:UnitedNations1996}.

According to the \cite{Book:WorldBank2018}, the number of people living in extreme poverty declined from $36\%$ in 1990 to $10\%$ of the world\rq s population in 2015. However, this downward trend has been decelerating throughout the years. Indeed, recent research published by the United Nations University World Institute for Development Economics Research (UNU-WIDER) shows that,  due to the COVID-19 crisis, global poverty could increase for the first time since 1990 \citep{Article:Hoy2020}, therefore threatening one of the global public\rq s priority: ending poverty. In 2015, owing to the importance of the topic, world leaders agreed on seventeen Sustainable Development Goals (SDGs) which engage not only public and private sectors but also society in attaining a better and more sustainable future for all. Among these goals, eradicating extreme poverty by 2030 is at the top of the list of priorities, followed by other targets among which, the reduction of at least by half of the proportion of people living in poverty and the implementation of appropriate social protection programmes, stand out (SDG 1: End poverty in all its forms everywhere) \citep{Book:UnitedNations2015}.

Poverty is not an individualised condition, as it does not affect only those who are poor. That is, poverty causes enormous economic, social and psychological costs to both the poor and the non-poor. Crime, access to and affordability of health care and economic productivity are just a few examples of common global concerns that are exacerbated by poverty \citep{Book:Rank2021}. Child poverty is a clear example of how poverty affects us all. For instance, children who grow up in poverty are much more likely to commit crime as adults \citep{Article:Bjerk2007}. More crime means higher correction costs and a rise in private spending on crime prevention (e.g. in buying alarms and locks). Similarly, growing up in poverty can have harmful effects on a person\rq s health \citep{Article:Brooks1997,Article:Case2002,Book:Ravallion2016}. This causes hospitals and health insurers to spend more on the treatment of preventable diseases \citep{Book:ChildrensDefenseFund1994}, jeopardising access to and affordability of health care. Lastly, poor children are often less exposed to education \citep{Book:Rank2014} and they may therefore have fewer qualifications, which in turn translates into lower paid and more unstable jobs. This results in lower economic productivity in adulthood for poor children. Specifically, for the United States of America, \cite{Article:McLaughlin2018} indicate the aggregate annual cost of child poverty amounts to USD 1.0298 trillion, representing 5.4\% of the country\rq s gross domestic product (GDP). Moreover, \cite{Article:McLaughlin2018} also estimate that, for every dollar spent on reducing childhood poverty, the country would save at least seven dollars with respect the economic costs of poverty. 

Cash transfer programmes are one of the main social protection strategies to reduce poverty and are therefore considered important mechanisms to help achieve SDG 1. In their simplest form, these programmes transfer cash, whether in small, regular amounts, or as lump sums, to people living below the poverty line and are generally funded by governments, international organisations, donors or nongovernmental organisations (NGOs) \citep{Book:WorldBank2012}. Moreover, cash transfers are usually classified as unconditional (UCTs) or conditional (CCTs), with the former not requiring beneficiaries to undertake any specific actions nor meet any conditions whereas the latter needs them to have some specific behavioural conditions in exchange of the cash transfer \citep{Article:Baird2014}, such as enrolling children in school or taking them to regular health check-ups \citep{Article:Handa2006}.

Adopting a ruin-theoretic approach, this article studies the impact of regular UCTs on poverty and extreme poverty dynamics and, particularly, their effectiveness in reducing the likelihood of a household living in poverty and extreme poverty. Previous research has addressed the role of UCTs as a pathway out of extreme poverty for households. \cite{Book:Handa2016} study two programmes, the Child Grant Programme (CGP) and the Multiple Category Targeted Programme (MCP), which were implemented in 2010 by the Ministry of Community Development, Mother and Child Health (MCDMCH) of the Government of Zambia. The authors find that both of these UCTs go far beyond their primary objective of protecting food security and consumption, as they also have an enormous impact on households\rq \ productive capacity. Although a flat transfer of USD 12 per month may not permanently lift households out of the poverty trap, their results suggest these programmes can help raise the standards of living of the country\rq s population. In the same way,  \cite{Book:Ambler2017} show that the Benazir Income Support Program (BISP), an UCT initiative introduced in 2008 by the Government of Pakistan, has increased women empowerment in the country, frequently associated with economic growth \citep{Article:Duflo2012}, which at the same time has been linked with poverty reduction \citep{Book:Adams2003}. As a matter of fact, UCTs have recently gained popularity as a cost-effective social protection strategy to attain some public policy objectives, including poverty alleviation \citep{Article:Aker2013,Article:Baird2014,Article:Blattman2014,Article:Haushofer2016,Article:Jensen2017,Article:Pega2022}.

Despite the growing interest in studying the impact of UCTs on poverty dynamics over the years, most studies have adopted an empirical approach. This article is an attempt to attach a mathematically based theoretical framework to the vast empirical literature. In this paper, we extend the model proposed by \cite{Article:Kovacevic2011}. Here, a household\rq s capital process $X = \{X_t\}_{t\geq 0}$ grows exponentially at a rate $r > 0$, which incorporates household rates of consumption, income generation and investment or savings, above a critical capital level (or poverty line) $x^{*} > 0$, whereas below a capital barrier level $B > x^{*}$, the capital also integrates external support in the form of direct transfers (capital cash transfers) provided by donors or governments at a rate $c_{{\scaleto{T}{4pt}}}> 0$. At time $T_{i}$, the $i\text{th}$ capital loss event time, the capital process jumps (downwards) to $Z_{i} \cdot X_{T_{i}}$, where $\{Z_i\}_{i=1}^\infty$ is a sequence of i.i.d. random variables with distribution function $G_{Z}$ supported in $(0,1]$, representing the proportions of remaining capital after each loss event (in the present paper, it will be regularly assumed the random variables are $Beta(\alpha,1)-$distributed). A more comprehensive picture of this model is introduced in Section \ref{TheCapitalModel-Section2}. 

The probability of falling (trapping probability) and the moment at which a household falls (trapping time) into the poverty trap have recently attracted the interest of some researchers (see, for example, \cite{Article:Kovacevic2011}, \cite{Article:Azais2015}, \cite{Article:Flores-Contro2021}, \cite{Article:Henshaw2023} and \cite{Article:Flores-Contro2024}). These studies focus on analysing the behaviour of a household\rq s capital above the critical capital, hence overlooking its evolution below this threshold. That is, under this set up, a household\rq s capital process is killed at the trapping time $\tau^{\scaleto{\text{ {\fontfamily{qcr}\selectfont P}}}{2.5pt}}_{x}:=\inf \left\{t \geq 0: X_{t}<x^{*} \mid X_{0}=x\right\}$. In this article, we assume households may escape from the poverty trap only due to external support received in the form of capital cash transfers. Therefore, we define the random variable $\tau^{\scaleto{\text{ {\fontfamily{qcr}\selectfont EP}}}{2.5pt}}_{x}$ for $x \in (0,\infty)$ as the time of extreme poverty i.e. the moment at which a household becomes extremely poor and $\psi^{\scaleto{\text{ {\fontfamily{qcr}\selectfont EP}}}{2.5pt}}\left(x\right)= \mathbbm{P}\left(\tau^{\scaleto{\text{ {\fontfamily{qcr}\selectfont EP}}}{2.5pt}}_{x}<\infty\right)$ as the probability of extreme poverty. Hence, under this new set up, a household\rq s capital process is killed at the time of extreme poverty. The approach taken here differs from the aforementioned studies, where the area of poverty $\Lambda = [0, x^{*}]$ was considered as an absorbing state from which it was not possible to escape. To explore these ideas, we consider an Omega risk process, which in classical risk theory, distinguishes between ruin (negative surplus) and bankruptcy (going out of business). Thus, it is assumed that, even with negative surplus levels, an insurance company can do business as usual and continue until bankruptcy occurs. 

The Omega model was first introduced in \cite{Article:Albrecher2011}, where closed-form formulas for the expected discounted dividends until bankruptcy under a dividend barrier strategy are obtained for the case in which the surplus of an insurance company is modeled as a Brownian motion. Similarly, \cite{Article:Gerber2012},  \cite{Article:Albrecher2013b} and \cite{Article:Wang2016} derive explicit expressions for the expected discounted penalty function at bankruptcy and the probability of bankruptcy when the surplus of an insurance company is modeled as a Brownian motion, a compound Poisson risk model with exponential claim sizes and an Ornstein-Uhlenbeck process, respectively. Certainly, the Omega model has been extensively studied during the last decade in the actuarial science literature, with researchers incorporating the bankruptcy concept into traditional ruin models. A particular clear example of this is in \cite{Article:Cui2016}, where an Omega model with surplus-dependent tax payments and capital injections in a time-homogeneous diffusion setting is studied. This work not only incorporates features from the Omega model \citep{Article:Albrecher2011} but also from traditionally well-studied ruin models such as the risk model with tax \citep{Article:Albrecher2007} and the risk model with capital injections \citep{Article:Albrecher2014}. More recently, \cite{Article:Gao2019} and \cite{Article:He2019} obtain analytical results for the expected discounted penalty function and the probability of bankruptcy for surplus processes under three- and two-step premium rate settings, respectively. In like manner, \cite{Article:Gao2022} also derive results for the expected discounted dividends until bankruptcy for a jump-diffusion surplus process with a two-step premium rate under a dividend barrier strategy.  Besides, alternative versions of the Omega model have also been considered. For instance,  \cite{Article:Kaszubowski2019} allows the surplus process to evolve below zero but assumes it is killed once it falls below some fixed level $-d<0$.
 
Under the classical risk theory set up, the probability of bankruptcy is quantified by a bankruptcy rate function $\omega\left(x\right)$, where $x$ represents the value of the negative surplus. The bankruptcy rate function is defined in such a way in which the probability of bankruptcy increases when the deficit grows. Consequently, for the household capital process, the bankruptcy event is swaped for the extreme poverty one and an extreme poverty rate function $\omega\left(X_{s}\right)$, which is assumed to be locally bounded and dependent on the capital level below the critical capital $x^{*}$, is defined on $\left(0, x^{*}\right]$. Namely, for some capital $X_{s}<x^{*}$ and no prior extreme poverty event, the probability of extreme poverty on the time interval $\left[s,s+dt\right)$ is  given by $\omega\left(X_{s}\right)dt$. Moreover, we assume that $\omega\left(\cdot\right)\geq0$ and $\omega\left(x\right)\geq\omega\left(y\right)$ for $0<x\leq y$ to reflect the likelihood of extreme poverty does not decrease as the capital approaches zero. Clearly, when $\omega\left(y\right)\equiv \infty$ for all $y<x^{*}$, the probability of extreme poverty is equal to the trapping probability $\psi^{\scaleto{\text{ {\fontfamily{qcr}\selectfont P}}}{2.5pt}}\left(x\right)=\mathbbm{P}\left(\tau^{\scaleto{\text{ {\fontfamily{qcr}\selectfont P}}}{2.5pt}}_{x}<\infty\right)$.

In general, UCTs target the poor. However, in recent years, cash transfer programmes have reached unprecedented levels of coverage. For example, in 2020, in response to the COVID-19 pandemic, one out of six people in the world received at least one cash transfer payment \citep{Book:Gentilini2022}. As a consequence of this expansion, it is now more common to encounter UCTs targeting other population groups, such as the vulnerable non-poor (those living just above the poverty line). One example is Ingreso Solidario, an UCT programme in Colombia that was implemented in April 2020 as a response to the COVID-19 pandemic. Ingreso Solidario provided monthly transfers of approximately USD 40 to eligible households: poor households not covered by pre-existing social programmes and non-poor households deemed vulnerable based on an assessment of their living conditions \citep{Book:Vera-Cossio2023}. The capital model considered in this article allows for the assessment of targeted UCTs, either to the poor only (letting $B \rightarrow x^{*+}$) or to both poor and vulnerable non-poor households (when $B > x^{*}$), on poverty dynamics. Moreover, the capital model is in line with the idea that spending on poverty reduction and prevention can help save on the economic costs of poverty. As such, when capital cash transfers target only the poor, the essential aim of the UCT programme is to lift households out of poverty. On the other hand, when the UCT programme targets both the poor and the vulnerable non-poor, the programme hopes to prevent the vulnerable non-poor from falling into poverty, apart from lifting the poor out of poverty. Nevertheless, both settings pursue one same objective: poverty reduction.

Particular attention should be paid to the fact that the targeted UCTs considered in this article, either to the poor only or to both poor and vulnerable non-poor households, prevent households from becoming extremely poor, as extreme poverty implies poverty (recall that a household is at risk of becoming extremely poor only when its capital lies below the critical capital $x^{*}$ or, in other words, a household can become extremely poor only when it is already poor). This is consistent with how extreme poverty is currently measured. For instance, the World Bank uses the International Poverty Line (IPL), set at USD 2.15 per person per day, to measure extreme poverty \citep{Book:Jolliffe2022}. The IPL is also the most relevant poverty line to measure poverty in low-income countries, whereas in other countries, other poverty lines are used to measure poverty. For example, the poverty line is set at USD 3.65 and USD 6.85 per person per day, in lower and upper middle-income countries, respectively \citep{Book:Jolliffe2022}. According to the World Bank's definition of extreme poverty, it is clear how extreme poverty implies poverty. In general, extreme poverty differs from conventional poverty in that it has greater depth (degree of deprivation), larger length (duration over time) and greater breadth (the number of dimensions such as illiteracy and malnutrition, among others) \citep{Article:Emran2014}. Because of these characteristics, the economic costs of extreme poverty are also expected to be higher than those of conventional poverty. Hence, extreme poverty should be avoided by all means, and should be considered and studied separately.

The remainder of the paper is structured as follows. In Section \ref{TheCapitalModel-Section2}, we introduce the capital model, with special emphasis on its behavior inside and outside the poverty area. Explicit equations, their solutions and numerical illustrations for the trapping probability are given in Section \ref{WhenandHowHouseholdsBecomePoor?-Section3} for the particular case in which the remaining proportions of capital are $Beta(\alpha,1)-$distributed. In particular, a comparison between the trapping probability of the original capital model introduced by \cite{Article:Kovacevic2011} and the one proposed in this article is presented in Appendix \ref{Appendix B: Effects of Underlying Factors on the Trapping Probability}. The event of extreme poverty and the time when it occurs are discussed in Section \ref{WhenandHowHouseholdsBecomeExtremelyPoor?-Section4}. In addition, closed-form solutions and numerical illustrations for the probability of extreme poverty are derived in Section \ref{WhenandHowHouseholdsBecomeExtremelyPoor?-Section4}, assuming constant and exponential extreme poverty rate functions for the particular case in which the remaining proportions of capital are $Beta(\alpha,1)-$distributed. Following \cite{Article:Albrecher2013b}, Section \ref{MonteCarloSimulation-Section5} illustrates how to approximate the probability of extreme poverty for more general cases by making use of an efficient Monte Carlo simulation method. Finally, concluding remarks are discussed in Section \ref{Conclusion-Section6}.

\section{The Capital Model} \label{TheCapitalModel-Section2}

This paper extends the capital process originally proposed in \cite{Article:Kovacevic2011}, where an individual household\rq s income $I_{t}$ at time $t$ comprises consumption $C_{t}$ and savings (investments) $S_{t}$. Hence, as in the original capital process, income dynamics are given by

\vspace{0.3cm}

\begin{align}
    I_{t} = C_{t} + S_{t}. \label{TheCapitalModel-Section2-Equation1}
\end{align}

\vspace{0.3cm}

Moreover, consumption is an increasing function of income and its dynamics are given by

\vspace{0.3cm}

\begin{align}
    C_{t}= \begin{cases} I_{t}  \hspace{2.7cm} \textit{ if } I_{t}\leq I^{*},\\ I^{*}+a\left(I_{t}-I^{*}\right) \hspace{0.5cm} \textit{ if } I_{t}> I^{*}, \end{cases}\label{TheCapitalModel-Section2-Equation2}
\end{align}

\begin{figure}[H]
	\begin{subfigure}[b]{0.5\linewidth}
  		\includegraphics[width=7.5cm, height=7.5cm]{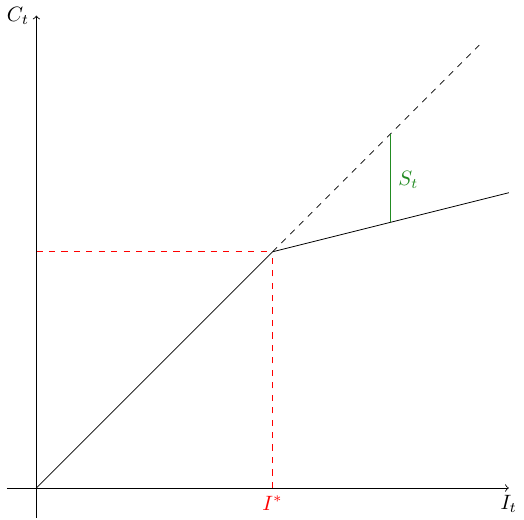}
		\caption{}
  		\label{TheCapitalModel-Section2-Figure1-a}
	\end{subfigure}
	\begin{subfigure}[b]{0.5\linewidth}
  		\includegraphics[width=7.5cm, height=7.5cm]{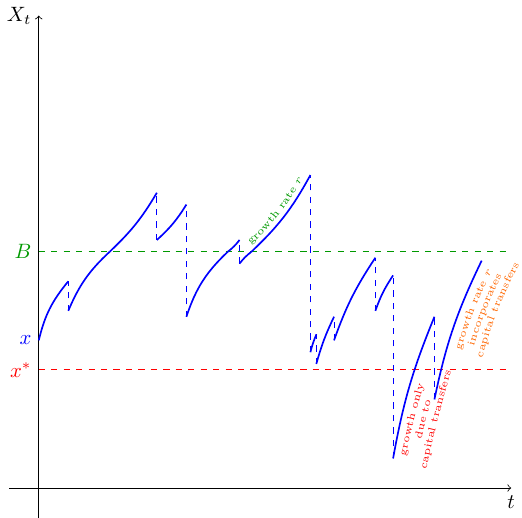}
		\caption{}
  		\label{TheCapitalModel-Section2-Figure1-b}
	\end{subfigure}
	\caption{(a) Consumption and savings (b) Trajectory of the stochastic process $X_{t}$.}
	\label{TheCapitalModel-Section2-Figure1}
\end{figure}

\vspace{0.3cm}

where $0<a<1$. It is assumed that permanent consumption below $I^{*}$ might result in severe adverse effects on health \citep{Article:Kovacevic2011}. Figure \ref{TheCapitalModel-Section2-Figure1-a} shows the dynamics of consumption and savings. Consider the accumulated capital $X_{t}$ up to time $t$ follows the dynamics

\vspace{0.3cm}

\begin{align}
    \frac{dX_{t}}{dt}=c_{{\scaleto{S}{4pt}}}S_{t}, \label{TheCapitalModel-Section2-Equation3} 
\end{align}

\vspace{0.3cm}

with $0<c_{{\scaleto{S}{4pt}}}<1$, and income is generated through capital 

\vspace{0.3cm}

\begin{align}
    I_{t}=b X_{t}, \label{TheCapitalModel-Section2-Equation4}
\end{align}

\vspace{0.3cm}

where $0<b$ holds.

Putting all these pieces together and defining $x^{*}\cdot b = I^{*}$, one gets the dynamical system

\vspace{0.3cm}

\begin{align}
    \frac{dX_{t}}{dt}= r \cdot \left[X_{t}-x^{*}\right]^{+}, \label{TheCapitalModel-Section2-Equation5} 
\end{align}

\vspace{0.3cm}

where $r = (1-a) \cdot b \cdot c_{{\scaleto{S}{4pt}}} > 0$ and $x^{*}>0$ represents the threshold below which a household lives in poverty, also interpreted as the amount of capital needed to acquire the critical income $I^{*}$ as a perpetuity \citep{Article:Kovacevic2011}. 

We now also consider direct transfers (capital cash transfers) provided by donors or governments only to those deemed eligible. Assume a household qualifies to be a beneficiary of the unconditional capital cash transfer programme when its accumulated capital $X_{t}$ up to time $t$ is below some capital barrier level $B > x^{*}$ and that the external support will be provided at a rate $c_{{\scaleto{T}{4pt}}} > 0$. The main objective of the proposed UCTs is to reduce the gap between the capital barrier level and the accumulated capital $X_{t}$ up to time $t$ for those households with capital levels below the capital barrier level $B > x^{*}$. Under this framework, one gets the dynamical system

\vspace{0.3cm}

\begin{align}
    \frac{dX_{t}}{dt}= r \cdot \left[X_{t}-x^{*}\right]^{+} +  c_{{\scaleto{T}{4pt}}} \cdot [B - X_{t}]^{+}.\label{TheCapitalModel-Section2-Equation6}
\end{align}

\vspace{0.3cm}

In line with the ideology that households are susceptible to the occurrence of capital losses, including severe illness, the death of a household member or breadwinner and catastrophic events such as floods and earthquakes, we model the occurrence of these events with a Poisson process with intensity $\lambda$ and consider the capital process follows the dynamics of \eqref{TheCapitalModel-Section2-Equation6} in between events. On the occurrence of a loss, the household's capital at the event time is reduced by a random proportion $0\leq 1-Z_{i}\leq1$. Hence, the fraction of the capital not destroyed at the event time is given by $Z_{i}.$ The sequence $\{Z_i\}_{i=1}^\infty$ is independent of the Poisson process and i.i.d. with common distribution function $G_{Z}$. A trajectory of the capital process $X_{t}$ is shown in Figure \ref{TheCapitalModel-Section2-Figure1-b}.

Here, the trajectories of the piecewise-deterministic process \citep{Article:Davis1984} behave as follows: if the capital lies above the capital barrier level $B > x^{*}$, then the capital grows exponentially at a rate $r$, whereas if the capital lies above the critical capital $x^{*}$ but below the capital barrier level $B > x^{*}$, then the capital growth is composed by both the individual household rate $r$ and the external support rate $c_{{\scaleto{T}{4pt}}}$; otherwise, the capital growth only incorporates the external support rate $c_{{\scaleto{T}{4pt}}}$. Note that, both the critical capital $x^{*}$ and the capital barrier level $B > x^{*}$ act as equilibrium levels for the process. That is, the further above the current value of the process is from the critical capital $x^{*}$, the faster the capital will depart from the critical capital $x^{*}$ at the individual household rate $r$. Similarly, the further below the current value of the process is from the capital barrier level $B > x^{*}$, the faster the capital will grow to the capital barrier level $B > x^{*}$  at the external support rate $c_{{\scaleto{T}{4pt}}}$ (there is a \lq \lq $B-$reverting\rq \rq \ effect where $c_{{\scaleto{T}{4pt}}}$ is the rate of reversion).

\vspace{0.3cm}

\section{When and How Households Become Poor?} \label{WhenandHowHouseholdsBecomePoor?-Section3}

In this Section \ref{WhenandHowHouseholdsBecomePoor?-Section3}, we will study the trapping time $\tau^{{\scaleto{\text{ {\fontfamily{qcr}\selectfont P}}}{2.5pt}}}_{x}$, which is defined as the time at which a household with initial capital $x \ge x^{*}$ falls into the area of poverty. That is,

\vspace{0.3cm}

\begin{align}
    \tau^{{\scaleto{\text{ {\fontfamily{qcr}\selectfont P}}}{2.5pt}}}_{x} :=\inf \left\{t \geq 0: X_{t}<x^{*} \mid X_{0}=x\right\}.
    \label{WhenandHowHouseholdsBecomePoor?-Section3-Equation1}
\end{align}

\vspace{0.3cm}

Note that, we use the superscript ${\lq\lq\scaleto{\text{{\fontfamily{qcr}\selectfont P}}}{4pt}}$\rq \rq \ to distinguish trapping-related variables and functions. Our analysis will involve the expected discounted penalty function, a concept commonly used in actuarial science \citep{Article:Gerber1998}. The expected discounted penalty function contains information on the trapping time $\tau^{{\scaleto{\text{ {\fontfamily{qcr}\selectfont P}}}{2.5pt}}}_{x}$ itself and two related random variables, the capital surplus prior to the trapping time $X_{\tau^{{\scaleto{\text{ {\fontfamily{qcr}\selectfont P}}}{2.5pt}}-}_{x}}- x^{*}$ and the capital deficit at the trapping time $\mid X_{\tau^{\scaleto{\text{ {\fontfamily{qcr}\selectfont P}}}{2.5pt}}_{x}}-x^{*}\mid$. 

For a force of interest $\delta \ge 0$ and initial capital $x \ge x^{*}$, the expected discounted penalty function is defined as

\vspace{0.3cm}

\begin{align}
    m^{\scaleto{\text{ {\fontfamily{qcr}\selectfont P}}}{2.5pt}}_{\delta}(x)= \mathbb{E}\left[w^{\scaleto{\text{ {\fontfamily{qcr}\selectfont P}}}{2.5pt}}(X_{\tau^{{\scaleto{\text{ {\fontfamily{qcr}\selectfont P}}}{2.5pt}}-}_{x}}- x^{*},\mid X_{\tau^{\scaleto{\text{ {\fontfamily{qcr}\selectfont P}}}{2.5pt}}_{x}}-x^{*}\mid)e^{-\delta \tau^{\scaleto{\text{ {\fontfamily{qcr}\selectfont P}}}{2.5pt}}_{x}} \mathbbm{1}_{\{\tau^{\scaleto{\text{ {\fontfamily{qcr}\selectfont P}}}{2.5pt}}_{x} < \infty\}}\right],
    \label{WhenandHowHouseholdsBecomePoor?-Section3-Equation2}
\end{align}

\vspace{0.3cm}

where $\mathbbm{1}_{\{A\}}$ is the indicator function of a set $A$, and $w^{\scaleto{\text{ {\fontfamily{qcr}\selectfont P}}}{2.5pt}}(x_{1}, x_{2})$, for $0 \leq x_{1},  x_{2} < x^{*} $, is a non-negative penalty function of $x_{1}$, the capital surplus prior to the trapping time, and $x_{2}$, the capital deficit at the trapping time. For more details on the so-called Gerber-Shiu risk theory, interested readers may wish to consult \cite{Book:Kyprianou2013}. The function $m^{\scaleto{\text{ {\fontfamily{qcr}\selectfont P}}}{2.5pt}}_{\delta}(x)$ is useful for deriving results in connection with joint and marginal distributions of $\tau^{{\scaleto{\text{ {\fontfamily{qcr}\selectfont P}}}{2.5pt}}}_{x}$, $X_{\tau^{{\scaleto{\text{ {\fontfamily{qcr}\selectfont P}}}{2.5pt}}-}_{x}}- x^{*}$ and $\mid X_{\tau^{\scaleto{\text{ {\fontfamily{qcr}\selectfont P}}}{2.5pt}}_{x}}-x^{*}\mid$. For instance, \eqref{WhenandHowHouseholdsBecomePoor?-Section3-Equation2} could also be viewed in terms of a Laplace transform when $\delta$ is serving as the argument. Indeed, if we let $w^{\scaleto{\text{ {\fontfamily{qcr}\selectfont P}}}{2.5pt}}(x_{1}, x_{2})=1$, \eqref{WhenandHowHouseholdsBecomePoor?-Section3-Equation2} is the Laplace transform of the trapping time $\tau^{{\scaleto{\text{ {\fontfamily{qcr}\selectfont P}}}{2.5pt}}}_{x}$ \footnote{We know from probability theory that, for a continuous random variable $Y$, with probability density function $f_{Y}$, the Laplace transform of $f_{Y}$ is given by the expected value $\mathcal{L}\{f_{Y}\}\left(s\right)=\mathbb{E}\left[e^{-sY}\right]$.}. Another choice is $w^{\scaleto{\text{ {\fontfamily{qcr}\selectfont P}}}{2.5pt}}(x_{1}, x_{2}) = \mathbbm{1}_{\{ x_{1} \leq x, x_{2} \leq y\}}$ for $\delta =0$, for which \eqref{WhenandHowHouseholdsBecomePoor?-Section3-Equation2} leads to the joint distribution function of the capital surplus prior to the trapping time and the capital deficit at the trapping time. It is not difficult to realise that, by appropriately choosing a penalty function $w^{\scaleto{\text{ {\fontfamily{qcr}\selectfont P}}}{2.5pt}}(x_{1}, x_{2})$ and force of interest $\delta$, various risk quantities can be modeled. A non-exhaustive list of such risk quantities is given in \cite{Article:He2023}. In this article, we are mainly interested in studying the impact of capital cash transfers on the probability of falling into the poverty trap. Thus, we will focus our analysis on the risk quantity $\psi^{\scaleto{\text{ {\fontfamily{qcr}\selectfont P}}}{2.5pt}}(x) = \mathbb{P}(\tau^{\scaleto{\text{ {\fontfamily{qcr}\selectfont P}}}{2.5pt}}_{x} < \infty)$, which can be derived by choosing $w^{\scaleto{\text{ {\fontfamily{qcr}\selectfont P}}}{2.5pt}}(x_{1}, x_{2}) =1$ and $\delta=0$ in \eqref{WhenandHowHouseholdsBecomePoor?-Section3-Equation2}. 

Following \cite{Article:Gerber1998}, our goal is to derive a functional equation for $m^{\scaleto{\text{ {\fontfamily{qcr}\selectfont P}}}{2.5pt}}_{\delta}(x)$ by applying the law of iterated expectations to the right-hand side of \eqref{WhenandHowHouseholdsBecomePoor?-Section3-Equation2}.

We point out that $m_{\delta}^{\scaleto{\text{ {\fontfamily{qcr}\selectfont P}}}{2.5pt}}(x)$ has different sample paths for $x \geq B$ and $x^{*} \leq x < B$. Hence, we distinguish the two situations by writing $m^{\scaleto{\text{ {\fontfamily{qcr}\selectfont P}}}{2.5pt}}_{\delta}(x) = m^{\scaleto{\text{ {\fontfamily{qcr}\selectfont P}}}{2.5pt}}_{\delta,u}(x)$ for  $x \geq B$ and $m^{\scaleto{\text{ {\fontfamily{qcr}\selectfont P}}}{2.5pt}}_{\delta}(x) = m^{\scaleto{\text{ {\fontfamily{qcr}\selectfont P}}}{2.5pt}}_{\delta, l}(x)$  for $x^{*} \leq x < B$. Similarly, we write $\psi^{\scaleto{\text{ {\fontfamily{qcr}\selectfont P}}}{2.5pt}}_{u}(x) = \mathbb{P}(\tau^{\scaleto{\text{ {\fontfamily{qcr}\selectfont P}}}{2.5pt}}_{x} < \infty)$ for $x \geq B$ and $\psi^{\scaleto{\text{ {\fontfamily{qcr}\selectfont P}}}{2.5pt}}_{l}(x) = \mathbb{P}(\tau^{\scaleto{\text{ {\fontfamily{qcr}\selectfont P}}}{2.5pt}}_{x} < \infty)$ for $x^{*} \leq x < B$. \vspace{0.3cm}

\begin{remark}
	Note that, when $B = x^{*}$, above the critical capital $x^{*}$, the dynamics of the capital process follow those of the original process \citep{Article:Kovacevic2011}. Thus, the trapping probability $\psi^{\scaleto{\text{ {\fontfamily{qcr}\selectfont P}}}{2.5pt}}(x)$ and the expected discounted penalty function at the trapping time $m^{\scaleto{\text{ {\fontfamily{qcr}\selectfont P}}}{2.5pt}}_{\delta}(x)$, are equivalent to those studied in \cite{Article:Henshaw2023} and \cite{Article:Flores-Contro2024}, respectively. Clearly, this also holds true when $c_{{\scaleto{T}{4pt}}}=0$. Appendix \ref{Appendix B: Effects of Underlying Factors on the Trapping Probability} evidences this behaviour for a set of selected parameters.
	
\end{remark}

\vspace{0.3cm}

\begin{theorem}\label{WhenandHowHouseholdsBecomePoor?-Section3-Theorem1}
When $x \geq B$, we have

\vspace{0.3cm}

{\allowdisplaybreaks
\begin{align}
       m^{\scaleto{\text{ {\fontfamily{qcr}\selectfont P}}}{2.5pt}}_{\delta,u}(x)&=\frac{\lambda}{r}(x-x^{*})^{\frac{\lambda+\delta}{r}}\int^{\infty}_{x}\frac{1}{(u-x^{*})^{\frac{\lambda+\delta}{r}+1}}\left[\int^{1}_{B/u}m^{\scaleto{\text{ {\fontfamily{qcr}\selectfont P}}}{2.5pt}}_{\delta,u}(u \cdot z)dG_{Z}(z)\right. \\ \\ 
       & \left.+ \int^{B/u}_{x^{*}/u}m^{\scaleto{\text{ {\fontfamily{qcr}\selectfont P}}}{2.5pt}}_{\delta,l}(u \cdot z)dG_{Z}(z)+A^{\scaleto{\text{ {\fontfamily{qcr}\selectfont P}}}{2.5pt}}(u)\right]du,
        \label{WhenandHowHouseholdsBecomePoor?-Section3-Equation3}
\end{align}
}

\vspace{0.3cm}

and when $x^{*} \leq x < B$, we have

\vspace{0.3cm}

{\allowdisplaybreaks
\begin{align}
       m^{\scaleto{\text{ {\fontfamily{qcr}\selectfont P}}}{2.5pt}}_{\delta,l}(x)&=\frac{\lambda}{r-c_{{\scaleto{T}{4pt}}}}(x+x^{**})^{\frac{\lambda+\delta}{r-c_{{\scaleto{T}{2pt}}}}}\int^{B}_{x}\frac{1}{(u+x^{**})^{\frac{\lambda+\delta}{r-c_{{\scaleto{T}{2pt}}}}+1}}\left[\int^{1}_{x^{*}/u}m^{\scaleto{\text{ {\fontfamily{qcr}\selectfont P}}}{2.5pt}}_{\delta,l}(u \cdot z)dG_{Z}(z) +A^{\scaleto{\text{ {\fontfamily{qcr}\selectfont P}}}{2.5pt}}(u)\right]du \\ \\
       &+ \frac{\lambda}{r}(B-x^{*})^{\frac{\lambda+\delta}{r}}\left(\frac{x+x^{**}}{B+x^{**}}\right)^{\frac{\lambda + \delta}{r-c_{{\scaleto{T}{2pt}}}}}\int^{\infty}_{B}\frac{1}{(v-x^{*})^{\frac{\lambda+\delta}{r}+1}}\left[\int^{1}_{B/v}m^{\scaleto{\text{ {\fontfamily{qcr}\selectfont P}}}{2.5pt}}_{\delta,u}(v \cdot z)dG_{Z}(z) \right. \\ \\
       &\left. + \int^{B/v}_{x^{*}/v}m^{\scaleto{\text{ {\fontfamily{qcr}\selectfont P}}}{2.5pt}}_{\delta,l}(v \cdot z)dG_{Z}(z) +A^{\scaleto{\text{ {\fontfamily{qcr}\selectfont P}}}{2.5pt}}(v)\right]dv,
        \label{WhenandHowHouseholdsBecomePoor?-Section3-Equation4}
\end{align}
}

\vspace{0.3cm}

where the function $A^{\scaleto{\text{ {\fontfamily{qcr}\selectfont P}}}{2.5pt}}(u)$ is given by

\vspace{0.3cm}

{\allowdisplaybreaks
 \begin{align}
	A^{\scaleto{\text{ {\fontfamily{qcr}\selectfont P}}}{2.5pt}}(u):=\int_{0}^{x^{*}/u}w^{\scaleto{\text{ {\fontfamily{qcr}\selectfont P}}}{2.5pt}}(u-x^{*}, x^{*}-u\cdot z)dG_{Z}(z).
	\label{WhenandHowHouseholdsBecomePoor?-Section3-Equation5}
\end{align}
}

\end{theorem}

\vspace{0.3cm}

See Appendix \ref{ProofofTheorem3.1} for proof of Theorem \ref{WhenandHowHouseholdsBecomePoor?-Section3-Theorem1}.

\vspace{0.3cm}

\begin{remark}
	We point out that the Integral Equations (IEs) \eqref{WhenandHowHouseholdsBecomePoor?-Section3-Equation3} and \eqref{WhenandHowHouseholdsBecomePoor?-Section3-Equation4} allow us to consider the differentiability of the functions $m^{\scaleto{\text{ {\fontfamily{qcr}\selectfont P}}}{2.5pt}}_{\delta,u}(x)$ and $m^{\scaleto{\text{ {\fontfamily{qcr}\selectfont P}}}{2.5pt}}_{\delta,l}(x)$. For instance, it is easy to see from \eqref{WhenandHowHouseholdsBecomePoor?-Section3-Equation3} and \eqref{WhenandHowHouseholdsBecomePoor?-Section3-Equation4} that $m^{\scaleto{\text{ {\fontfamily{qcr}\selectfont P}}}{2.5pt}}_{\delta,u}(x)$ and $m^{\scaleto{\text{ {\fontfamily{qcr}\selectfont P}}}{2.5pt}}_{\delta,l}(x)$ are differentiable in $(B,\infty)$ and $(x^{*},B)$, respectively. Furthermore, they satisfy the following condition

\vspace{0.3cm}

\begin{align}
	m^{\scaleto{\text{ {\fontfamily{qcr}\selectfont P}}}{2.5pt}}_{\delta,u}(B) = m^{\scaleto{\text{ {\fontfamily{qcr}\selectfont P}}}{2.5pt}}_{\delta,l}(B^{-}).	\label{WhenandHowHouseholdsBecomePoor?-Section3-Equation6}
\end{align}

\end{remark}

\vspace{0.3cm}

Now, by differentiating the IEs \eqref{WhenandHowHouseholdsBecomePoor?-Section3-Equation3} and \eqref{WhenandHowHouseholdsBecomePoor?-Section3-Equation4}, we obtain the Integro-Differential Equations (IDEs) for $m^{\scaleto{\text{ {\fontfamily{qcr}\selectfont P}}}{2.5pt}}_{\delta,u}(x)$ and $m^{\scaleto{\text{ {\fontfamily{qcr}\selectfont P}}}{2.5pt}}_{\delta,l}(x)$ in the following theorem

\vspace{0.3cm}

\begin{theorem}\label{WhenandHowHouseholdsBecomePoor?-Section3-Theorem2}
When $x \geq B$, we have

\vspace{0.3cm}

{\allowdisplaybreaks
\begin{align}
       &r(x-x^{*}) m'^{\scaleto{\text{ {\fontfamily{qcr}\selectfont P}}}{2.5pt}}_{\delta,u}(x) - (\lambda + \delta) m^{\scaleto{\text{ {\fontfamily{qcr}\selectfont P}}}{2.5pt}}_{\delta,u}(x) + \lambda \left[\int^{1}_{B/x}m^{\scaleto{\text{ {\fontfamily{qcr}\selectfont P}}}{2.5pt}}_{\delta,u}(x\cdot z) dG_{Z}(z)\right. \\ \\
       & \left. + \int^{B/x}_{x^{*}/x} m^{\scaleto{\text{ {\fontfamily{qcr}\selectfont P}}}{2.5pt}}_{\delta,l}(x\cdot z) dG_{Z}(z) + A^{\scaleto{\text{ {\fontfamily{qcr}\selectfont P}}}{2.5pt}}(x) \right] = 0,
        \label{WhenandHowHouseholdsBecomePoor?-Section3-Equation7}
\end{align}
}

\vspace{0.3cm}

and when $x^{*} \leq x < B$, we have

\vspace{0.3cm}

{\allowdisplaybreaks
\begin{align}
        (r - c_{{\scaleto{T}{4pt}}})(x+x^{**}) m'^{\scaleto{\text{ {\fontfamily{qcr}\selectfont P}}}{2.5pt}}_{\delta,l}(x) - (\lambda + \delta) m^{\scaleto{\text{ {\fontfamily{qcr}\selectfont P}}}{2.5pt}}_{\delta,l}(x) + \lambda \left[\int^{1}_{x^{*}/x}m^{\scaleto{\text{ {\fontfamily{qcr}\selectfont P}}}{2.5pt}}_{\delta,l}(x\cdot z) dG_{Z}(z) + A^{\scaleto{\text{ {\fontfamily{qcr}\selectfont P}}}{2.5pt}}(x) \right] = 0.
        \label{WhenandHowHouseholdsBecomePoor?-Section3-Equation8}
\end{align}
}

\vspace{0.3cm}

In addition, the boundary conditions for $m^{\scaleto{\text{ {\fontfamily{qcr}\selectfont P}}}{2.5pt}}_{\delta,u}(x)$ and $m^{\scaleto{\text{ {\fontfamily{qcr}\selectfont P}}}{2.5pt}}_{\delta,l}(x)$ are given by \eqref{WhenandHowHouseholdsBecomePoor?-Section3-Equation6}, $\lim\limits_{x\to\infty} m^{\scaleto{\text{ {\fontfamily{qcr}\selectfont P}}}{2.5pt}}_{\delta,u}(x) = 0$ and $m^{\scaleto{\text{ {\fontfamily{qcr}\selectfont P}}}{2.5pt}}_{\delta,l}(x^{*}) = \frac{1}{\lambda + \delta} \left[c_{{\scaleto{T}{4pt}}}(B-x^{*}) m'^{\scaleto{\text{ {\fontfamily{qcr}\selectfont P}}}{2.5pt}}_{\delta,l}(x^{*})+\lambda A^{\scaleto{\text{ {\fontfamily{qcr}\selectfont P}}}{2.5pt}}(x^{*})\right]$.

\end{theorem}

\vspace{0.3cm}

\begin{remark}
Equation \eqref{WhenandHowHouseholdsBecomePoor?-Section3-Equation8} for $m^{\scaleto{\text{ {\fontfamily{qcr}\selectfont P}}}{2.5pt}}_{\delta,l}(x)$ is independent of $m^{\scaleto{\text{ {\fontfamily{qcr}\selectfont P}}}{2.5pt}}_{\delta,u}(x)$. However, $m^{\scaleto{\text{ {\fontfamily{qcr}\selectfont P}}}{2.5pt}}_{\delta,l}(x)$ is subject to the boundary condition \eqref{WhenandHowHouseholdsBecomePoor?-Section3-Equation6} which is involved with $m^{\scaleto{\text{ {\fontfamily{qcr}\selectfont P}}}{2.5pt}}_{\delta,u}(x)$. Furthermore, it is easy to see from \eqref{WhenandHowHouseholdsBecomePoor?-Section3-Equation6}, \eqref{WhenandHowHouseholdsBecomePoor?-Section3-Equation7} and \eqref{WhenandHowHouseholdsBecomePoor?-Section3-Equation8} that $m^{\scaleto{\text{ {\fontfamily{qcr}\selectfont P}}}{2.5pt}}_{\delta,u}(x)$ and $m^{\scaleto{\text{ {\fontfamily{qcr}\selectfont P}}}{2.5pt}}_{\delta,l}(x)$ satisfy 

\vspace{0.3cm}

\begin{align}
        m'^{\scaleto{\text{ {\fontfamily{qcr}\selectfont P}}}{2.5pt}}_{\delta,u}(B)=m'^{\scaleto{\text{ {\fontfamily{qcr}\selectfont P}}}{2.5pt}}_{\delta,l}(B^{-}).
        \label{WhenandHowHouseholdsBecomePoor?-Section3-Equation9}
\end{align}

\vspace{0.3cm}

\end{remark}

\subsection{The Trapping Time} \label{TheTrappingTime-Subsection31}

Sometimes it is easier to work with a transformation rather than with the original distribution function of a random variable. In this article, we focus on studying the Laplace transform of the random variables of interest. The Laplace transform of a random variable characterises the probability distribution uniquely and is known to be a powerful tool in probability theory and, in particular, quite useful when studying nonnegative random variables. Recall that, by specifying the penalty function such that $w^{\scaleto{\text{ {\fontfamily{qcr}\selectfont P}}}{2.5pt}}(x_{1}, x_{2}) = 1$, $m^{\scaleto{\text{ {\fontfamily{qcr}\selectfont P}}}{2.5pt}}_{\delta}(x)$ becomes the Laplace transform of the trapping time, also interpreted as the expected present value of a unit payment due at the trapping time. 

Thus, with $w^{\scaleto{\text{ {\fontfamily{qcr}\selectfont P}}}{2.5pt}}\left(x_{1},x_{2}\right)=1$, equations \eqref{WhenandHowHouseholdsBecomePoor?-Section3-Equation7} and \eqref{WhenandHowHouseholdsBecomePoor?-Section3-Equation8} can then be written such that when $x \geq B$,

\vspace{0.3cm}

\begin{align}
	0&=r(x-x^{*}) m'^{\scaleto{\text{ {\fontfamily{qcr}\selectfont P}}}{2.5pt}}_{\delta,u}(x) - (\lambda + \delta) m^{\scaleto{\text{ {\fontfamily{qcr}\selectfont P}}}{2.5pt}}_{\delta,u}(x) + \lambda \left[\int^{1}_{B/x}m^{\scaleto{\text{ {\fontfamily{qcr}\selectfont P}}}{2.5pt}}_{\delta,u}(x\cdot z) dG_{Z}(z) \right. \\ \\
	&\left. + \int^{B/x}_{x^{*}/x} m^{\scaleto{\text{ {\fontfamily{qcr}\selectfont P}}}{2.5pt}}_{\delta,l}(x\cdot z) dG_{Z}(z) + G_{Z}\left(\frac{x^{*}}{x}\right) \right],
\label{TheTrappingTime-Subsection31-Equation1}
\end{align}

\vspace{0.3cm}

and when $x^{*} \leq x <  B$,

\vspace{0.3cm}

\begin{align}
	0&=(r - c_{{\scaleto{T}{4pt}}})(x+x^{**}) m'^{\scaleto{\text{ {\fontfamily{qcr}\selectfont P}}}{2.5pt}}_{\delta,l}(x) - (\lambda + \delta) m^{\scaleto{\text{ {\fontfamily{qcr}\selectfont P}}}{2.5pt}}_{\delta,l}(x) + \lambda \left[\int^{1}_{x^{*}/x}m^{\scaleto{\text{ {\fontfamily{qcr}\selectfont P}}}{2.5pt}}_{\delta,l}(x\cdot z) dG_{Z}(z)
	+ G_{Z}\left(\frac{x^{*}}{x}\right) \right]. \\ \\
	\label{TheTrappingTime-Subsection31-Equation2}
\end{align}

\vspace{0.3cm}

\begin{proposition}\label{TheTrappingTime-Subsection21-Proposition1}

Consider a household capital process with initial capital $x\ge x^{*}$, capital growth rate $r$, capital barrier level $B$, capital transfer rate $c_{{\scaleto{T}{4pt}}}$, intensity $\lambda > 0$ and remaining proportions of capital with distribution $Beta(\alpha, 1)$ where $\alpha >0$; that is, $Z_{i}\sim Beta(\alpha, 1)$. The Laplace transform of the trapping time is given by

\vspace{0.3cm}

\begin{align}
	m^{\scaleto{\text{ {\fontfamily{qcr}\selectfont P}}}{2.5pt}}_{\delta}\left(x\right)&=
		\begin{cases}
		\begin{aligned}
		 A^{\scaleto{\text{ {\fontfamily{qcr}\selectfont P}}}{2.5pt}}_{2,u}{y^{\scaleto{\text{ {\fontfamily{qcr}\selectfont P}}}{2.5pt}}_{u}(x)}^{-b^{\scaleto{\text{ {\fontfamily{qcr}\selectfont P}}}{2.5pt}}_{u}} { }_{2} F_{1}\left(b^{\scaleto{\text{ {\fontfamily{qcr}\selectfont P}}}{2.5pt}}_{u}, b^{\scaleto{\text{ {\fontfamily{qcr}\selectfont P}}}{2.5pt}}_{u}-c^{\scaleto{\text{ {\fontfamily{qcr}\selectfont P}}}{2.5pt}}_{u}+1 ; b^{\scaleto{\text{ {\fontfamily{qcr}\selectfont P}}}{2.5pt}}_{u}-a^{\scaleto{\text{ {\fontfamily{qcr}\selectfont P}}}{2.5pt}}_{u}+1 ; {y^{\scaleto{\text{ {\fontfamily{qcr}\selectfont P}}}{2.5pt}}_{u}(x)}^{-1}\right) \hspace{1.75cm} \textit{for} \hspace{0.3cm} x \geq B,\end{aligned} \\ \\
		\begin{aligned}
		A^{\scaleto{\text{ {\fontfamily{qcr}\selectfont P}}}{2.5pt}}_{1,l}&{y^{\scaleto{\text{ {\fontfamily{qcr}\selectfont P}}}{2.5pt}}_{l}(x)}^{-a^{\scaleto{\text{ {\fontfamily{qcr}\selectfont P}}}{2.5pt}}_{l}} { }_{2} F_{1}\left(a^{\scaleto{\text{ {\fontfamily{qcr}\selectfont P}}}{2.5pt}}_{l}, a^{\scaleto{\text{ {\fontfamily{qcr}\selectfont P}}}{2.5pt}}_{l}-c^{\scaleto{\text{ {\fontfamily{qcr}\selectfont P}}}{2.5pt}}_{l}+1 ; a^{\scaleto{\text{ {\fontfamily{qcr}\selectfont P}}}{2.5pt}}_{l}-b^{\scaleto{\text{ {\fontfamily{qcr}\selectfont P}}}{2.5pt}}_{l}+1 ; {y^{\scaleto{\text{ {\fontfamily{qcr}\selectfont P}}}{2.5pt}}_{l}(x)}^{-1}\right)\\ \\ & +A^{\scaleto{\text{ {\fontfamily{qcr}\selectfont P}}}{2.5pt}}_{2,l}{y^{\scaleto{\text{ {\fontfamily{qcr}\selectfont P}}}{2.5pt}}_{l}(x)}^{-b^{\scaleto{\text{ {\fontfamily{qcr}\selectfont P}}}{2.5pt}}_{l}} { }_{2} F_{1}\left(b^{\scaleto{\text{ {\fontfamily{qcr}\selectfont P}}}{2.5pt}}_{l}, b^{\scaleto{\text{ {\fontfamily{qcr}\selectfont P}}}{2.5pt}}_{l}-c^{\scaleto{\text{ {\fontfamily{qcr}\selectfont P}}}{2.5pt}}_{l}+1 ; b^{\scaleto{\text{ {\fontfamily{qcr}\selectfont P}}}{2.5pt}}_{l}-a^{\scaleto{\text{ {\fontfamily{qcr}\selectfont P}}}{2.5pt}}_{l}+1 ; {y^{\scaleto{\text{ {\fontfamily{qcr}\selectfont P}}}{2.5pt}}_{l}(x)}^{-1}\right)  \hspace{0.8cm} \textit{for}  \hspace{0.3cm} x^{*} \leq x < B,\end{aligned}
	\end{cases}
\label{TheTrappingTime-Subsection31-Equation3}
\end{align}

\vspace{0.3cm}

where $\delta \ge 0$ is the force of interest for valuation, ${ }_{2} F_{1}\left(\cdot \right)$ is Gauss\rq s Hypergeometric Function as defined in \eqref{Appendix A: Mathematical Proofs-Equation10}, $y^{\scaleto{\text{ {\fontfamily{qcr}\selectfont P}}}{2.5pt}}_{u}(x)=\frac{x}{x^{*}}$, $a^{\scaleto{\text{ {\fontfamily{qcr}\selectfont P}}}{2.5pt}}_{u}=\frac{-(\delta + \lambda - \alpha r) - \sqrt{(\delta + \lambda -\alpha r)^{2}+4 r \alpha \delta}}{2r}$, $b^{\scaleto{\text{ {\fontfamily{qcr}\selectfont P}}}{2.5pt}}_{u}=\frac{-(\delta + \lambda - \alpha r) + \sqrt{(\delta + \lambda -\alpha r)^{2}+4 r \alpha \delta}}{2r}$, $c^{\scaleto{\text{ {\fontfamily{qcr}\selectfont P}}}{2.5pt}}_{u}=c^{\scaleto{\text{ {\fontfamily{qcr}\selectfont P}}}{2.5pt}}_{l}=\alpha$, $y^{\scaleto{\text{ {\fontfamily{qcr}\selectfont P}}}{2.5pt}}_{l}(x)=-\frac{x}{x^{**}}$ with $x^{**}=\frac{c_{{\scaleto{T}{2.5pt}}}B-rx^{*}}{r-c_{{\scaleto{T}{2.5pt}}}}$, $a^{\scaleto{\text{ {\fontfamily{qcr}\selectfont P}}}{2.5pt}}_{l}=\frac{-(\delta + \lambda - \alpha \left(r-c_{{\scaleto{T}{2.5pt}}}\right)) - \sqrt{(\delta + \lambda -\alpha \left(r-c_{{\scaleto{T}{2.5pt}}}\right))^{2}+4 \left(r - c_{{\scaleto{T}{2.5pt}}}\right) \alpha \delta}}{2\left(r - c_{{\scaleto{T}{2.5pt}}}\right)}$, $b^{\scaleto{\text{ {\fontfamily{qcr}\selectfont P}}}{2.5pt}}_{l}=\frac{-(\delta + \lambda - \alpha \left(r- c_{{\scaleto{T}{2.5pt}}}\right)) + \sqrt{(\delta + \lambda -\alpha \left(r - c_{{\scaleto{T}{2.5pt}}}\right))^{2}+4 \left(r - c_{{\scaleto{T}{2.5pt}}}\right) \alpha \delta}}{2\left(r - c_{{\scaleto{T}{2.5pt}}}\right)}$ and the constants $A^{\scaleto{\text{ {\fontfamily{qcr}\selectfont P}}}{2.5pt}}_{2,u}$, $A^{\scaleto{\text{ {\fontfamily{qcr}\selectfont P}}}{2.5pt}}_{1,l}$ and $A^{\scaleto{\text{ {\fontfamily{qcr}\selectfont P}}}{2.5pt}}_{2,l}$ are given by \eqref{Appendix A: Mathematical Proofs-Equation15}, \eqref{Appendix A: Mathematical Proofs-Equation16}, and \eqref{Appendix A: Mathematical Proofs-Equation17}, respectively.

\end{proposition}

\vspace{0.3cm}

The mathematical proof of Proposition \ref{TheTrappingTime-Subsection21-Proposition1} is presented in Appendix \ref{ProofofProposition3.1.1}.

\vspace{0.3cm}

\begin{remark}
As mentioned previously, the Laplace transform of the trapping time approaches the trapping probability as $\delta$ tends to zero, i.e. 
    
\vspace{0.3cm}
    
\begin{align}
        \lim _{\delta \downarrow 0} m^{\scaleto{\text{ {\fontfamily{qcr}\selectfont P}}}{2.5pt}}_{\delta}(x) =\mathbb{P}(\tau^{\scaleto{\text{ {\fontfamily{qcr}	\selectfont P}}}{2.5pt}}_{x}<\infty)\equiv\psi^{\scaleto{\text{ {\fontfamily{qcr}\selectfont P}}}{2.5pt}}(x),
        \label{TheTrappingTime-Subsection21-Equation9}
\end{align}
    
\vspace{0.3cm}
    
for $\alpha > \frac{\lambda}{r}$. If the net profit condition $\alpha > \frac{\lambda}{r}$ does not hold, then trapping would be certain \citep{Article:Henshaw2023}.
     
\end{remark}   

\vspace{0.3cm}

\begin{figure}[H]
	\begin{subfigure}[b]{0.5\linewidth}
  		\includegraphics[width=8cm, height=8cm]{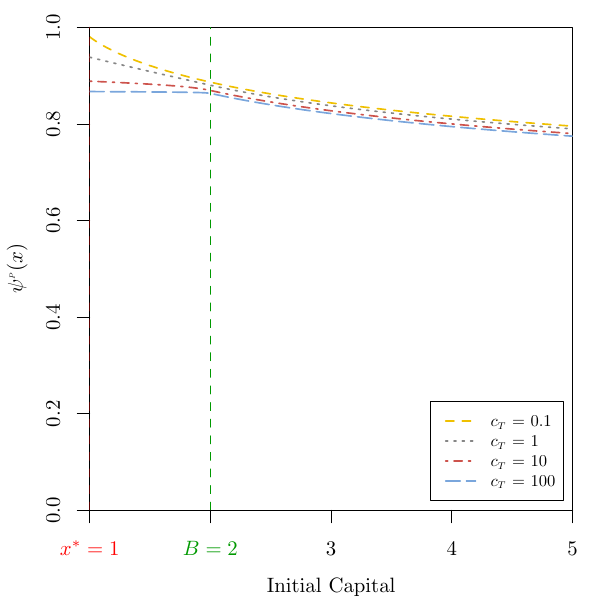}
		\caption{}
  		\label{TheTrappingTime-Subsection31-Figure1-a}
	\end{subfigure}
	\begin{subfigure}[b]{0.5\linewidth}
  		\includegraphics[width=8cm, height=8cm]{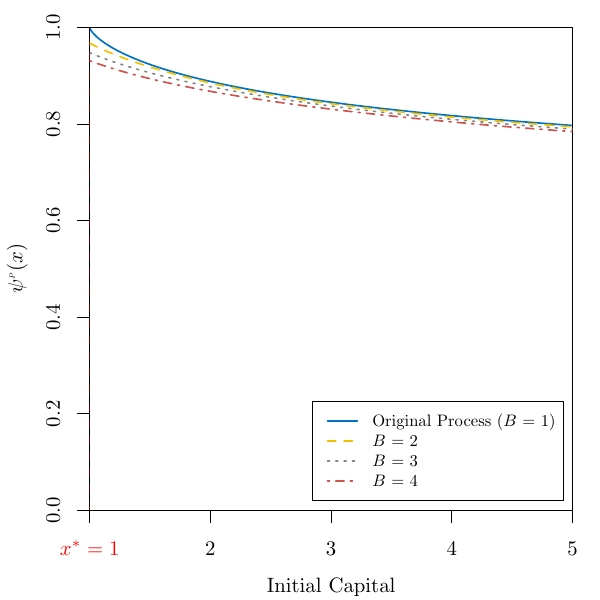}
		\caption{}
  		\label{TheTrappingTime-Subsection31-Figure1-b}
	\end{subfigure}
	\caption{(a) Trapping probability $\psi^{\scaleto{\text{ {\fontfamily{qcr}\selectfont P}}}{2.5pt}}(x)$ when $Z_{i} \sim Beta(0.8, 1)$, $a = 0.1$, $b = 4$, $c_{{\scaleto{S}{4pt}}} = 0.4$, $B=2$, $\lambda = 1$ and $x^{*} = 1$ for $c_{{\scaleto{T}{4pt}}} = 0.1, 1, 10, 100$ (b) Trapping probability $\psi^{\scaleto{\text{ {\fontfamily{qcr}\selectfont P}}}{2.5pt}}(x)$ when $Z_{i} \sim Beta(0.8, 1)$, $a = 0.1$, $b = 4$, $c_{{\scaleto{S}{4pt}}} = 0.4$, $c_{{\scaleto{T}{4pt}}} = 0.25$, $\lambda = 1$ and $x^{*} = 1$ for $B = 1, 2, 3, 4$.}
	\label{TheTrappingTime-Subsection31-Figure1}
\end{figure}

Figures \ref{TheTrappingTime-Subsection31-Figure1-a}  and \ref{TheTrappingTime-Subsection31-Figure1-b}  display the trapping probability $\psi^{\scaleto{\text{ {\fontfamily{qcr}\selectfont P}}}{2.5pt}}(x)$ for the capital process $X_{t}$. Not surprisingly, Figure \ref{TheTrappingTime-Subsection31-Figure1-a} shows the trapping probability is a decreasing function of both the capital transfer rate $c_{{\scaleto{T}{4pt}}}$ and the initial capital. In particular, it is worth noting the important role the capital transfer rate $c_{{\scaleto{T}{4pt}}}$ can play in attaining lower trapping probabilities for households with initial capital below the capital barrier level $B$ as very high capital transfer rates $c_{{\scaleto{T}{4pt}}}$ have the potential to level the likelihood of becoming poor for this particular group. However, high capital transfer rates $c_{{\scaleto{T}{4pt}}}$ seem to be less efficient for attaining lower trapping probabilities for households with initial capital above the capital barrier level $B$. This is due to the fact that households with initial capital above the capital barrier level $B$ are exposed to never receiving a capital cash transfer. Indeed, if they experience a large loss, they are likely to fall directly into the poverty trap without ever receiving a capital cash transfer. From Figure \ref{TheTrappingTime-Subsection31-Figure1-b}, we can also highlight the importance of the capital barrier level $B$ to reach lower values for the trapping probability. Although a higher capital transfer rate $c_{{\scaleto{T}{4pt}}}$ and a higher capital barrier level $B$ may lead to lower trapping probabilities, the sensitivity analyses shown in Appendices \ref{Appendix B: Effects of Underlying Factors on the Trapping Probability} and \ref{Appendix C: Effects of Underlying Factors on the Probability of Extreme Poverty} suggest the trapping probability is less sensitive to these parameters compared to the probability of extreme poverty.

\section{When and How Households Become Extremely Poor?} \label{WhenandHowHouseholdsBecomeExtremelyPoor?-Section4}

We define the random variable $\tau^{{\scaleto{\text{ {\fontfamily{qcr}\selectfont EP}}}{2.5pt}}}_{x}$ for $x \in (0,\infty)$ as the time of extreme poverty i.e. the moment at which a household with initial capital $x$ becomes extremely poor and $\psi^{{\scaleto{\text{ {\fontfamily{qcr}\selectfont EP}}}{2.5pt}}}\left(x\right)= \mathbbm{P}\left(\tau^{{\scaleto{\text{ {\fontfamily{qcr}\selectfont EP}}}{2.5pt}}}_{x}<\infty\right)$ as the probability of extreme poverty. Note that, we use the superscript ${\lq\lq\scaleto{\text{{\fontfamily{qcr}\selectfont EP}}}{4pt}}$\rq \rq \ to distinguish extreme poverty-related variables and functions. The probability of extreme poverty is quantified by an extreme poverty rate function $\omega\left(x\right)$, where $x$ represents the value of capital below the critical capital $x^{*}$. The extreme poverty rate function is defined in such a way in which the probability of extreme poverty increases when the capital deficit $\mid X_{s}-x^{*}\mid$ grows. Namely, for some capital $X_{s}<x^{*}$ and no prior extreme poverty event, the probability of extreme poverty on the time interval $\left[s,s+dt\right)$ is  given by $\omega\left(X_{s}\right)dt$. The expected discounted penalty function at extreme poverty is therefore given by

\vspace{0.3cm}

\begin{align}
    m^{\scaleto{\text{ {\fontfamily{qcr}\selectfont EP}}}{2.5pt}}_{\delta}(x)= \mathbb{E}\left[w^{\scaleto{\text{ {\fontfamily{qcr}\selectfont EP}}}{2.5pt}}(X_{\tau^{\scaleto{\text{ {\fontfamily{qcr}\selectfont EP -}}}{2.5pt}}_{x}}, \mid X_{\tau^{\scaleto{\text{ {\fontfamily{qcr}\selectfont EP}}}{2.5pt}}_{x}}-x^{*}\mid)e^{-\delta \tau^{\scaleto{\text{ {\fontfamily{qcr}\selectfont EP}}}{2.5pt}}_{x}} \mathbbm{1}_{\{\tau^{\scaleto{\text{ {\fontfamily{qcr}\selectfont EP}}}{2.5pt}}_{x} < \infty\}}\right],
    \label{WhenandHowHouseholdsBecomeExtremelyPoor?-Section4-Equation1}
\end{align}

\vspace{0.3cm}

where $w^{\scaleto{\text{ {\fontfamily{qcr}\selectfont EP}}}{2.5pt}}(x_{1}, x_{2})$, for $x_{1} < x^{*}$, $x_{2} < x^{*}$, is a non-negative penalty function of $x_{1}$, the accumulated capital prior to the time of extreme poverty, and $x_{2}$, the capital deficit at the time of extreme poverty. Note that, for the case of the expected discounted penalty function at extreme poverty, it is reasonable to consider the accumulated capital immediately before extreme poverty instead of the capital surplus, which was considered in \eqref{WhenandHowHouseholdsBecomePoor?-Section3-Equation2} for the expected discounted penalty function at the trapping time.

We point out that $m_{\delta}^{\scaleto{\text{ {\fontfamily{qcr}\selectfont EP}}}{2.5pt}}(x)$ has different sample paths for $x \geq B$, $x^{*} \leq x < B$ and $0 < x < x^{*}$. Hence, we distinguish the three situations by writing $m^{\scaleto{\text{ {\fontfamily{qcr}\selectfont EP}}}{2.5pt}}_{\delta}(x) = m^{\scaleto{\text{ {\fontfamily{qcr}\selectfont EP}}}{2.5pt}}_{\delta,u}(x)$ for  $x \geq B$, $m^{\scaleto{\text{ {\fontfamily{qcr}\selectfont EP}}}{2.5pt}}_{\delta}(x) = m^{\scaleto{\text{ {\fontfamily{qcr}\selectfont EP}}}{2.5pt}}_{\delta, m}(x)$ for $x^{*} \leq x < B$ and $m^{\scaleto{\text{ {\fontfamily{qcr}\selectfont EP}}}{2.5pt}}_{\delta}(x) = m^{\scaleto{\text{ {\fontfamily{qcr}\selectfont EP}}}{2.5pt}}_{\delta, l}(x)$ for $0 < x < x^{*}$. Similarly, we write $\psi^{\scaleto{\text{ {\fontfamily{qcr}\selectfont EP}}}{2.5pt}}_{u}(x) = \mathbb{P}(\tau^{\scaleto{\text{ {\fontfamily{qcr}\selectfont EP}}}{2.5pt}}_{x} < \infty)$ for $x \geq B$, $\psi^{\scaleto{\text{ {\fontfamily{qcr}\selectfont EP}}}{2.5pt}}_{m}(x) = \mathbb{P}(\tau^{\scaleto{\text{ {\fontfamily{qcr}\selectfont EP}}}{2.5pt}}_{x} < \infty)$ for $x^{*} \leq x < B$ and $\psi^{\scaleto{\text{ {\fontfamily{qcr}\selectfont EP}}}{2.5pt}}_{l}(x) = \mathbb{P}(\tau^{\scaleto{\text{ {\fontfamily{qcr}\selectfont EP}}}{2.5pt}}_{x} < \infty)$ for $0 < x < x^{*}$. 

Proceeding as in Section \ref{WhenandHowHouseholdsBecomePoor?-Section3}, one derives the following IEs for the expected discounted penalty function at extreme poverty in the following theorem

\vspace{0.3cm}

\begin{theorem}\label{WhenandHowHouseholdsBecomeExtremelyPoor?-Section4-Theorem1}
When $x \geq B$, we have

\vspace{0.3cm}

{\allowdisplaybreaks
\begin{align}
	m^{\scaleto{\text{ {\fontfamily{qcr}\selectfont EP}}}{2.5pt}}_{\delta,u}(x)&=\frac{\lambda}{r} (x-x^{*})^{\frac{\lambda + \delta}{r}} \int^{\infty}_{x} \frac{1}{(v_{u}-x^{*})^{\frac{\lambda + \delta}{r} + 1}} \left[\int^{x^{*}/v_{u}}_{0} m^{\scaleto{\text{ {\fontfamily{qcr}\selectfont EP}}}{2.5pt}}_{\delta, l} (v_{u}\cdot z) dG_{Z}(z) \right. \\ \\ 
	& \left. + \int^{B/v_{u}}_{x^{*}/v_{u}} m^{\scaleto{\text{ {\fontfamily{qcr}\selectfont EP}}}{2.5pt}}_{\delta, m} (v_{u}\cdot z) dG_{Z}(z) + \int^{1}_{B/v_{u}} m^{\scaleto{\text{ {\fontfamily{qcr}\selectfont EP}}}{2.5pt}}_{\delta, u} (v_{u}\cdot z) dG_{Z}(z) \right]dv_{u},
	\label{WhenandHowHouseholdsBecomeExtremelyPoor?-Section4-Equation2}
\end{align}
}

\vspace{0.3cm}

when $x^{*} \leq x < B$, we have

\vspace{0.3cm}

{\allowdisplaybreaks
\begin{align}
	m^{\scaleto{\text{ {\fontfamily{qcr}\selectfont EP}}}{2.5pt}}_{\delta,m}(x)&=\frac{\lambda}{r-c_{{\scaleto{T}{4pt}}}}(x+x^{**})^{\frac{\lambda + \delta}{r-c_{{\scaleto{T}{2pt}}}}} \int^{B}_{x}\frac{1}{(v_{m}+x^{**})^{\frac{\lambda + \delta}{r-c_{{\scaleto{T}{2pt}}}}+1}}\left[\int^{x^{*}/v_{m}}_{0} m^{\scaleto{\text{ {\fontfamily{qcr}\selectfont EP}}}{2.5pt}}_{\delta,l}(v_{m}\cdot z) dG_{Z}(z) \right. \\ \\
	& \left. + \int^{1}_{x^{*}/v_{m}} m^{\scaleto{\text{ {\fontfamily{qcr}\selectfont EP}}}{2.5pt}}_{\delta,m}(v_{m}\cdot z)  dG_{Z}(z) \right]dv_{m} \\ \\
	& 
	+ \frac{\lambda}{r}(B-x^{*})^{\frac{\lambda+\delta}{r}}\left(\frac{x+x^{**}}{B+x^{**}}\right)^{\frac{\lambda+\delta}{r-c_{{\scaleto{T}{2pt}}}}} \int^{\infty}_{B} \frac{1}{(v_{u}-x^{*})^{\frac{\lambda + \delta}{r}+1}}\left[\int^{x^{*}/v_{u}}_{0} m^{\scaleto{\text{ {\fontfamily{qcr}\selectfont EP}}}{2.5pt}}_{\delta,l}(v_{u}\cdot z) dG_{Z}(z) \right. \\ \\ 
	& \left. + \int^{B/v_{u}}_{x^{*}/v_{u}} m^{\scaleto{\text{ {\fontfamily{qcr}\selectfont EP}}}{2.5pt}}_{\delta,m}(v_{u}\cdot z) dG_{Z}(z) + \int^{1}_{B/v_{u}} m^{\scaleto{\text{ {\fontfamily{qcr}\selectfont EP}}}{2.5pt}}_{\delta,u}(v_{u}\cdot z) dG_{Z}(z) \right] dv_{u},
	\label{WhenandHowHouseholdsBecomeExtremelyPoor?-Section4-Equation3}
\end{align}
}

\vspace{0.3cm}

and when $0 < x < x^{*}$, we have

\vspace{0.3cm}

{\allowdisplaybreaks
\begin{align}
	m^{\scaleto{\text{ {\fontfamily{qcr}\selectfont EP}}}{2.5pt}}_{\delta,l}(x)&=-\frac{1}{c_{{\scaleto{T}{4pt}}}\left(x-B\right)^{\frac{\lambda + \delta}{c_{{\scaleto{T}{2pt}}}}}} \int^{x^{*}}_{x} \frac{1}{\left(v_{l}-B\right)^{1-\frac{\lambda+\delta}{c_{{\scaleto{T}{2pt}}}}}} e^{\frac{1}{c_{{\scaleto{T}{2pt}}}}\int^{v_{l}}_{x}\frac{\omega(u_{l})}{u_{l}-B}du_{l}}\omega(v_{l})w^{\scaleto{\text{ {\fontfamily{qcr}\selectfont EP}}}{2.5pt}}(v_{l}, x^{*}-v_{l})dv_{l} \\ \\ 
	&-\frac{\lambda}{c_{{\scaleto{T}{4pt}}}\left(x-B\right)^{\frac{\lambda + \delta}{c_{{\scaleto{T}{2pt}}}}}} \int^{x^{*}}_{x} \frac{1}{\left(v_{l}-B\right)^{1-\frac{\lambda+\delta}{c_{{\scaleto{T}{2pt}}}}}} e^{\frac{1}{c_{{\scaleto{T}{2pt}}}}\int^{v_{l}}_{x}\frac{\omega(u_{l})}{u_{l}-B}du_{l}}\int^{1}_{0} m^{\scaleto{\text{ {\fontfamily{qcr}\selectfont EP}}}{2.5pt}}_{\delta,l} (v_{l}\cdot z) dG_{z}(z) dv_{l} \\ \\
	&+\frac{\lambda}{r-c_{{\scaleto{T}{4pt}}}}\left(\frac{x^{*}-B}{x-B}\right)^{\frac{\lambda + \delta}{c_{{\scaleto{T}{2pt}}}}}(x^{*}+x^{**})^{\frac{\lambda + \delta}{r-c_{{\scaleto{T}{2pt}}}}} \int^{B}_{x^{*}}\frac{1}{\left(v_{m}+x^{**}\right)^{\frac{\lambda + \delta}{r-c_{{\scaleto{T}{2pt}}}}+1}}e^{\frac{1}{c_{{\scaleto{T}{2pt}}}}\int^{x^{*}}_{x}\frac{\omega(u_{l})}{u_{l}-B}du_{l}} \\ \\
	&\left[ \int^{x^{*}/v_{m}}_{0} m^{\scaleto{\text{ {\fontfamily{qcr}\selectfont EP}}}{2.5pt}}_{\delta,l} (v_{m} \cdot z) dG_{Z}(z) + \int^{1}_{x^{*}/v_{m}} m^{\scaleto{\text{ {\fontfamily{qcr}\selectfont EP}}}{2.5pt}}_{\delta,m}(v_{m}\cdot z) dG_{Z}(z)\right]dv_{m} \\ \\ &
	+ \frac{\lambda}{r}\left(\frac{x^{*}-B}{x-B}\right)^{\frac{\lambda + \delta}{c_{{\scaleto{T}{2pt}}}}}\left(\frac{x^{*}+x^{**}}{B+x^{**}}\right)^{\frac{\lambda + \delta}{r-c_{{\scaleto{T}{2pt}}}}}\left(B-x^{*}\right)^{\frac{\lambda + \delta}{r}}
	\int^{\infty}_{B}\frac{1}{(v_{u}-x^{*})^{\frac{\lambda + \delta}{r}+1}}e^{\frac{1}{c_{{\scaleto{T}{2pt}}}}\int^{x^{*}}_{x}\frac{\omega(u_{l})}{u_{l}-B}du_{l}} \\ \\
	& \left[\int^{x^{*}/v_{u}}_{0}m^{\scaleto{\text{ {\fontfamily{qcr}\selectfont EP}}}{2.5pt}}_{\delta,l}(v_{u}\cdot z)dG_{Z}(z) + \int^{B/v_{u}}_{x^{*}/v_{u}}m^{\scaleto{\text{ {\fontfamily{qcr}\selectfont EP}}}{2.5pt}}_{\delta,m}(v_{u}\cdot z)dG_{Z}(z) + \int^{1}_{B/v_{u}}m^{\scaleto{\text{ {\fontfamily{qcr}\selectfont EP}}}{2.5pt}}_{\delta,u}(v_{u}\cdot z)dG_{Z}(z)\right]dv_{u}.\\
	\label{WhenandHowHouseholdsBecomeExtremelyPoor?-Section4-Equation4}
\end{align}
}

\vspace{0.3cm}

\end{theorem}

See Appendix \ref{ProofofTheorem4.1} for proof of Theorem \ref{WhenandHowHouseholdsBecomeExtremelyPoor?-Section4-Theorem1}.

\vspace{0.3cm}

\begin{remark}
	We point out that the IEs \eqref{WhenandHowHouseholdsBecomeExtremelyPoor?-Section4-Equation2}, \eqref{WhenandHowHouseholdsBecomeExtremelyPoor?-Section4-Equation3} and \eqref{WhenandHowHouseholdsBecomeExtremelyPoor?-Section4-Equation4} allow us to consider the differentiability of the functions $m^{\scaleto{\text{ {\fontfamily{qcr}\selectfont EP}}}{2.5pt}}_{\delta,u}(x)$, $m^{\scaleto{\text{ {\fontfamily{qcr}\selectfont EP}}}{2.5pt}}_{\delta,m}(x)$ and $m^{\scaleto{\text{ {\fontfamily{qcr}\selectfont EP}}}{2.5pt}}_{\delta,l}(x)$. For instance, it is easy to see from \eqref{WhenandHowHouseholdsBecomeExtremelyPoor?-Section4-Equation2}, \eqref{WhenandHowHouseholdsBecomeExtremelyPoor?-Section4-Equation3} and \eqref{WhenandHowHouseholdsBecomeExtremelyPoor?-Section4-Equation4} that $m^{\scaleto{\text{ {\fontfamily{qcr}\selectfont EP}}}{2.5pt}}_{\delta,u}(x)$, $m^{\scaleto{\text{ {\fontfamily{qcr}\selectfont EP}}}{2.5pt}}_{\delta,m}(x)$ and $m^{\scaleto{\text{ {\fontfamily{qcr}\selectfont EP}}}{2.5pt}}_{\delta,l}(x)$ are differentiable in $(B,\infty)$, $(x^{*},B)$ and $(0,x^{*})$, respectively. Furthermore, they satisfy the following condition

\vspace{0.3cm}

\begin{align}
	m^{\scaleto{\text{ {\fontfamily{qcr}\selectfont EP}}}{2.5pt}}_{\delta,u}(B) = m^{\scaleto{\text{ {\fontfamily{qcr}\selectfont EP}}}{2.5pt}}_{\delta,m}(B^{-})\label{WhenandHowHouseholdsBecomeExtremelyPoor?-Section4-Equation5}
\end{align}

and 

\vspace{0.3cm}

\begin{align}
	m^{\scaleto{\text{ {\fontfamily{qcr}\selectfont EP}}}{2.5pt}}_{\delta,m}(x^{*}) = m^{\scaleto{\text{ {\fontfamily{qcr}\selectfont EP}}}{2.5pt}}_{\delta,l}(x^{*-}).	\label{WhenandHowHouseholdsBecomeExtremelyPoor?-Section4-Equation6}
\end{align}

\end{remark}

\vspace{0.3cm}

Now, by differentiating the IEs \eqref{WhenandHowHouseholdsBecomeExtremelyPoor?-Section4-Equation2}, \eqref{WhenandHowHouseholdsBecomeExtremelyPoor?-Section4-Equation3} and \eqref{WhenandHowHouseholdsBecomeExtremelyPoor?-Section4-Equation4}, we obtain the IDEs for $m^{\scaleto{\text{ {\fontfamily{qcr}\selectfont EP}}}{2.5pt}}_{\delta,u}(x)$, $m^{\scaleto{\text{ {\fontfamily{qcr}\selectfont EP}}}{2.5pt}}_{\delta,m}(x)$ and $m^{\scaleto{\text{ {\fontfamily{qcr}\selectfont EP}}}{2.5pt}}_{\delta,l}(x)$ in the following theorem

\vspace{0.3cm}

\begin{theorem}\label{WhenandHowHouseholdsBecomePoor?-Section3-Theorem2}
When $x \geq B$, we have

\vspace{0.3cm}

{\allowdisplaybreaks
\begin{align}
	r(x-x^{*})m^{\scaleto{\text{ {\fontfamily{qcr}\selectfont EP}}}{2.5pt}}_{\delta,u}(x)_{\delta,u}'(x)-(\lambda + \delta)m^{\scaleto{\text{ {\fontfamily{qcr}\selectfont EP}}}{2.5pt}}_{\delta,u}(x) & + \lambda \left[\int^{x^{*}/x}_{0} m^{\scaleto{\text{ {\fontfamily{qcr}\selectfont EP}}}{2.5pt}}_{\delta,l}(x\cdot z)dG_{Z}(z) \right. \\ \\ 
	& \left. + \int^{B/x}_{x^{*}/x} m^{\scaleto{\text{ {\fontfamily{qcr}\selectfont EP}}}{2.5pt}}_{\delta,m} (x\cdot z) dG_{Z}(z) + \int^{1}_{B/x} m^{\scaleto{\text{ {\fontfamily{qcr}\selectfont EP}}}{2.5pt}}_{\delta, u} (x\cdot z) dG_{Z}(z) \right] = 0,\\
	\label{WhenandHowHouseholdsBecomeExtremelyPoor?-Section4-Equation7}
\end{align}
}

\vspace{0.3cm}

when $x^{*} \leq x < B$, we have

\vspace{0.3cm}

{\allowdisplaybreaks
\begin{align}
	\left(r - c_{{\scaleto{T}{4pt}}}\right)\left(x + x^{**}\right)m'^{\scaleto{\text{ {\fontfamily{qcr}\selectfont EP}}}{2.5pt}}_{\delta,m}(x)-(\lambda + \delta)m^{\scaleto{\text{ {\fontfamily{qcr}\selectfont EP}}}{2.5pt}}_{\delta,m}(x) & + \lambda \left[\int^{x^{*}/x}_{0}m^{\scaleto{\text{ {\fontfamily{qcr}\selectfont EP}}}{2.5pt}}_{\delta,l}(x\cdot z) dG_{Z}(z) \right. \\ \\ 
	& \left. + \int^{1}_{x^{*}/x}m^{\scaleto{\text{ {\fontfamily{qcr}\selectfont EP}}}{2.5pt}}_{\delta,m}(x\cdot z)dG_{Z}(z)\right] = 0,
	\label{WhenandHowHouseholdsBecomeExtremelyPoor?-Section4-Equation8}
\end{align}
}

\vspace{0.3cm}

and when $0 < x < x^{*}$, we have

\vspace{0.3cm}

{\allowdisplaybreaks
\begin{align}
	c_{{\scaleto{T}{4pt}}}(x-B)m'^{\scaleto{\text{ {\fontfamily{qcr}\selectfont EP}}}{2.5pt}}_{\delta,l}(x)+[\lambda + \delta + \omega(x)]m^{\scaleto{\text{ {\fontfamily{qcr}\selectfont EP}}}{2.5pt}}_{\delta,l}(x)-\omega(x)w^{\scaleto{\text{ {\fontfamily{qcr}\selectfont EP}}}{2.5pt}}(x, x^{*}-x)-\lambda\int^{1}_{0}m^{\scaleto{\text{ {\fontfamily{qcr}\selectfont EP}}}{2.5pt}}_{\delta,l}(x\cdot z)dG_{Z}(z)=0.\\
	\label{WhenandHowHouseholdsBecomeExtremelyPoor?-Section4-Equation9}
\end{align}
}

In addition, the boundary conditions for $m^{\scaleto{\text{ {\fontfamily{qcr}\selectfont EP}}}{2.5pt}}_{\delta,u}(x)$, $m^{\scaleto{\text{ {\fontfamily{qcr}\selectfont EP}}}{2.5pt}}_{\delta,m}(x)$ and $m^{\scaleto{\text{ {\fontfamily{qcr}\selectfont EP}}}{2.5pt}}_{\delta,l}(x)$ are given by \eqref{WhenandHowHouseholdsBecomeExtremelyPoor?-Section4-Equation5}, \eqref{WhenandHowHouseholdsBecomeExtremelyPoor?-Section4-Equation6} and $\lim\limits_{x\to\infty} m^{\scaleto{\text{ {\fontfamily{qcr}\selectfont EP}}}{2.5pt}}_{\delta,u}(x) = 0$.

\end{theorem}

\vspace{0.3cm}

\begin{remark}
Equation \eqref{WhenandHowHouseholdsBecomeExtremelyPoor?-Section4-Equation9} for $m^{\scaleto{\text{ {\fontfamily{qcr}\selectfont EP}}}{2.5pt}}_{\delta,l}(x)$ is independent of $m^{\scaleto{\text{ {\fontfamily{qcr}\selectfont EP}}}{2.5pt}}_{\delta,u}(x)$ and $m^{\scaleto{\text{ {\fontfamily{qcr}\selectfont EP}}}{2.5pt}}_{\delta,m}(x)$. However, $m^{\scaleto{\text{ {\fontfamily{qcr}\selectfont EP}}}{2.5pt}}_{\delta,l}(x)$ is subject to the boundary condition \eqref{WhenandHowHouseholdsBecomeExtremelyPoor?-Section4-Equation6} which is involved with $m^{\scaleto{\text{ {\fontfamily{qcr}\selectfont EP}}}{2.5pt}}_{\delta,m}(x)$. At the same time, $m^{\scaleto{\text{ {\fontfamily{qcr}\selectfont EP}}}{2.5pt}}_{\delta,m}(x)$ is subject to the boundary condition \eqref{WhenandHowHouseholdsBecomeExtremelyPoor?-Section4-Equation5} which is involved with $m^{\scaleto{\text{ {\fontfamily{qcr}\selectfont EP}}}{2.5pt}}_{\delta,u}(x)$. Furthermore, it is easy to see from \eqref{WhenandHowHouseholdsBecomeExtremelyPoor?-Section4-Equation7}, \eqref{WhenandHowHouseholdsBecomeExtremelyPoor?-Section4-Equation8} and \eqref{WhenandHowHouseholdsBecomeExtremelyPoor?-Section4-Equation9} that $m^{\scaleto{\text{ {\fontfamily{qcr}\selectfont EP}}}{2.5pt}}_{\delta,u}(x)$, $m^{\scaleto{\text{ {\fontfamily{qcr}\selectfont EP}}}{2.5pt}}_{\delta,m}(x)$ and $m^{\scaleto{\text{ {\fontfamily{qcr}\selectfont EP}}}{2.5pt}}_{\delta,l}(x)$ satisfy 

\vspace{0.3cm}

\begin{align}
	m'^{\scaleto{\text{ {\fontfamily{qcr}\selectfont EP}}}{2.5pt}}_{\delta,u}(B) = m'^{\scaleto{\text{ {\fontfamily{qcr}\selectfont EP}}}{2.5pt}}_{\delta,m}(B^{-})	\label{WhenandHowHouseholdsBecomeExtremelyPoor?-Section4-Equation10}
\end{align}

and 

\vspace{0.3cm}

\begin{align}
	m'^{\scaleto{\text{ {\fontfamily{qcr}\selectfont EP}}}{2.5pt}}_{\delta,m}(x^{*}) = m'^{\scaleto{\text{ {\fontfamily{qcr}\selectfont EP}}}{2.5pt}}_{\delta,l}(x^{*-}).	\label{WhenandHowHouseholdsBecomeExtremelyPoor?-Section4-Equation11}
\end{align}

\end{remark}

\vspace{0.3cm}

\subsection{The Time of Extreme Poverty} \label{TheTimeofExtremePoverty-Subsection41}

Focusing again in studying the Laplace transform of the random variable of interest (the time of extreme poverty) we note that by specifying the penalty function such that $w^{\scaleto{\text{ {\fontfamily{qcr}\selectfont EP}}}{2.5pt}}(x_{1}, x_{2}) = 1$, $m^{\scaleto{\text{ {\fontfamily{qcr}\selectfont EP}}}{2.5pt}}_{\delta}(x)$ becomes the Laplace transform of the time of extreme poverty, also interpreted as the expected present value of a unit payment due at the time of extreme poverty. Thus, equation \eqref{WhenandHowHouseholdsBecomeExtremelyPoor?-Section4-Equation9} can then be written such that when $0 < x <  x^{*}$,

\vspace{0.3cm}

\begin{align}
	0&=c_{{\scaleto{T}{4pt}}}(x-B)m'^{\scaleto{\text{ {\fontfamily{qcr}\selectfont EP}}}{2.5pt}}_{\delta,l}(x)+[\lambda + \delta + \omega(x)]m^{\scaleto{\text{ {\fontfamily{qcr}\selectfont EP}}}{2.5pt}}_{\delta,l}(x)-\omega(x)-\lambda\int^{1}_{0}m^{\scaleto{\text{ {\fontfamily{qcr}\selectfont EP}}}{2.5pt}}_{\delta,l}(x\cdot z)dG_{Z}(z). 
	\label{TheTimeofExtremePoverty-Subsection41-Equation1}
\end{align}

\vspace{0.3cm}

\subsubsection{Examples of Extreme Poverty Rate Functions} \label{ExamplesofExtremePovertyRateFunctions-Subsection411}

\paragraph{Constant extreme poverty rate functions.} Let $\omega_{1}\left(x\right)\equiv \omega_{c} \cdot \mathbbm{1}_{\{x<x^{*}\}} $ with $\omega_{c}>0$. This is the simplest case of extreme poverty rate function and it could be interpreted as the situation in which the events of extreme poverty occur at discrete times. For instance, let $\xi_{1}, \xi_{2},...$ be i.i.d. exponential random variables with mean $\frac{1}{\omega_{c}}$ and $\Xi_{k}=\xi_{1} + \xi_{2}+...+\xi_{k}$ denote the $kth$ event of extreme poverty (e.g. unexpected loss of assets or health), with $k=1,2,...$. In this context, extreme poverty occurs when at such an event of extreme poverty the capital level is below $x^{*}$ \citep{Article:Albrecher2013a}. 

\vspace{0.3cm}

\begin{proposition}\label{ExamplesofExtremePovertyRateFunctions-Subsection411-Proposition1}

Consider a household capital process with initial capital $x\ge x^{*}$, capital growth rate $r$, capital barrier level $B$, capital transfer rate $c_{{\scaleto{T}{4pt}}}$, intensity $\lambda > 0$ and remaining proportions of capital with distribution $Beta(\alpha, 1)$ where $\alpha >0$; that is, $Z_{i}\sim Beta(\alpha, 1)$. For a constant extreme poverty rate function $\omega_{1}\left(x\right)\equiv \omega_{c} \cdot \mathbbm{1}_{\{x<x^{*}\}}$, with $\omega_{c} > 0$, the Laplace transform of the time of extreme poverty is given by

\vspace{0.3cm}

\begin{align}
	m^{\scaleto{\text{ {\fontfamily{qcr}\selectfont EP}}}{2.5pt}}_{\delta}\left(x\right)&=
		\begin{cases}
		\begin{aligned}
		 A^{\scaleto{\text{ {\fontfamily{qcr}\selectfont EP}}}{2.5pt}}_{2,u}{y^{\scaleto{\text{ {\fontfamily{qcr}\selectfont EP}}}{2.5pt}}_{u}(x)}^{-b^{\scaleto{\text{ {\fontfamily{qcr}\selectfont EP}}}{2pt}}_{u}} { }_{2} F_{1}\left(b^{\scaleto{\text{ {\fontfamily{qcr}\selectfont EP}}}{2.5pt}}_{u}, b^{\scaleto{\text{ {\fontfamily{qcr}\selectfont EP}}}{2.5pt}}_{u}-c^{\scaleto{\text{ {\fontfamily{qcr}\selectfont EP}}}{2.5pt}}_{u}+1 ; b^{\scaleto{\text{ {\fontfamily{qcr}\selectfont EP}}}{2.5pt}}_{u}-a^{\scaleto{\text{ {\fontfamily{qcr}\selectfont EP}}}{2.5pt}}_{u}+1 ; {y^{\scaleto{\text{ {\fontfamily{qcr}\selectfont EP}}}{2.5pt}}_{u}(x)}^{-1}\right) \hspace{1.5cm} \textit{for} \hspace{0.3cm} x \geq B,\end{aligned} \\ \\
		\begin{aligned}
		A^{\scaleto{\text{ {\fontfamily{qcr}\selectfont EP}}}{2.5pt}}_{1,m}&{y^{\scaleto{\text{ {\fontfamily{qcr}\selectfont EP}}}{2.5pt}}_{m}(x)}^{-a^{\scaleto{\text{ {\fontfamily{qcr}\selectfont EP}}}{2pt}}_{m}} { }_{2} F_{1}\left(a^{\scaleto{\text{ {\fontfamily{qcr}\selectfont EP}}}{2.5pt}}_{m}, a^{\scaleto{\text{ {\fontfamily{qcr}\selectfont EP}}}{2.5pt}}_{m}-c^{\scaleto{\text{ {\fontfamily{qcr}\selectfont EP}}}{2.5pt}}_{m}+1 ; a^{\scaleto{\text{ {\fontfamily{qcr}\selectfont EP}}}{2.5pt}}_{m}-b^{\scaleto{\text{ {\fontfamily{qcr}\selectfont EP}}}{2.5pt}}_{m}+1 ; {y^{\scaleto{\text{ {\fontfamily{qcr}\selectfont EP}}}{2.5pt}}_{m}(x)}^{-1}\right)\\ \\ & +A^{\scaleto{\text{ {\fontfamily{qcr}\selectfont EP}}}{2.5pt}}_{2,m}{y^{\scaleto{\text{ {\fontfamily{qcr}\selectfont EP}}}{2.5pt}}_{m}(x)}^{-b^{\scaleto{\text{ {\fontfamily{qcr}\selectfont EP}}}{2pt}}_{m}} { }_{2} F_{1}\left(b^{\scaleto{\text{ {\fontfamily{qcr}\selectfont EP}}}{2.5pt}}_{m}, b^{\scaleto{\text{ {\fontfamily{qcr}\selectfont EP}}}{2.5pt}}_{m}-c^{\scaleto{\text{ {\fontfamily{qcr}\selectfont EP}}}{2.5pt}}_{m}+1 ; b^{\scaleto{\text{ {\fontfamily{qcr}\selectfont EP}}}{2.5pt}}_{m}-a^{\scaleto{\text{ {\fontfamily{qcr}\selectfont EP}}}{2.5pt}}_{m}+1 ; {y^{\scaleto{\text{ {\fontfamily{qcr}\selectfont EP}}}{2.5pt}}_{m}(x)}^{-1}\right)  \hspace{0.1cm} \textit{for}  \hspace{0.3cm} x^{*} \leq x < B,\end{aligned}\\ \\
	\begin{aligned}
		 \frac{\omega_{c}}{\delta + \omega_{c}} + A^{\scaleto{\text{ {\fontfamily{qcr}\selectfont EP}}}{2.5pt}}_{1,l} { }_{2} F_{1}\left(a^{\scaleto{\text{ {\fontfamily{qcr}\selectfont EP}}}{2.5pt}}_{l}, b^{\scaleto{\text{ {\fontfamily{qcr}\selectfont EP}}}{2.5pt}}_{l}; c^{\scaleto{\text{ {\fontfamily{qcr}\selectfont EP}}}{2.5pt}}_{l} ; {y^{\scaleto{\text{ {\fontfamily{qcr}\selectfont EP}}}{2.5pt}}_{l}(x)}\right) \hspace{4.85cm} \textit{for} \hspace{0.3cm} 0< x < x^{*},\end{aligned}
	\end{cases}
\label{ExamplesofExtremePovertyRateFunctions-Subsection411-Equation1}
\end{align}

\vspace{0.3cm}

where $\delta \ge 0$ is the force of interest for valuation, ${ }_{2} F_{1}\left(\cdot \right)$ is Gauss\rq s Hypergeometric Function as defined in \eqref{Appendix A: Mathematical Proofs-Equation10}, $y^{\scaleto{\text{ {\fontfamily{qcr}\selectfont EP}}}{2.5pt}}_{u}(x)=\frac{x}{x^{*}}$, $a^{\scaleto{\text{ {\fontfamily{qcr}\selectfont EP}}}{2.5pt}}_{u}=\frac{-(\delta + \lambda - \alpha r) - \sqrt{(\delta + \lambda -\alpha r)^{2}+4 r \alpha \delta}}{2r}$, $b^{\scaleto{\text{ {\fontfamily{qcr}\selectfont EP}}}{2.5pt}}_{u}=\frac{-(\delta + \lambda - \alpha r) + \sqrt{(\delta + \lambda -\alpha r)^{2}+4 r \alpha \delta}}{2r}$, $y^{\scaleto{\text{ {\fontfamily{qcr}\selectfont EP}}}{2.5pt}}_{m}(x)=-\frac{x}{x^{**}}$, with $x^{**}=\frac{c_{{\scaleto{T}{2.5pt}}}B-rx^{*}}{r-c_{{\scaleto{T}{2.5pt}}}}$, $a^{\scaleto{\text{ {\fontfamily{qcr}\selectfont EP}}}{2.5pt}}_{m}= \frac{-(\delta + \lambda - \alpha \left(r-c_{{\scaleto{T}{2.5pt}}}\right)) - \sqrt{(\delta + \lambda -\alpha \left(r-c_{{\scaleto{T}{2.5pt}}}\right))^{2}+4 \left(r - c_{{\scaleto{T}{2.5pt}}}\right) \alpha \delta}}{2\left(r - c_{{\scaleto{T}{2.5pt}}}\right)}$, $b^{\scaleto{\text{ {\fontfamily{qcr}\selectfont EP}}}{2.5pt}}_{m}=\frac{-(\delta + \lambda - \alpha \left(r- c_{{\scaleto{T}{2.5pt}}}\right)) + \sqrt{(\delta + \lambda -\alpha \left(r - c_{{\scaleto{T}{2.5pt}}}\right))^{2}+4 \left(r - c_{{\scaleto{T}{2.5pt}}}\right) \alpha \delta}}{2\left(r - c_{{\scaleto{T}{2.5pt}}}\right)}$, $a^{\scaleto{\text{ {\fontfamily{qcr}\selectfont EP}}}{2.5pt}}_{l}=\frac{\alpha c_{{\scaleto{T}{2.5pt}}} + \lambda + \delta + \omega_{c} - \sqrt{(\alpha c_{{\scaleto{T}{2.5pt}}} + \lambda +\delta + \omega_{c})^{2}-4 \alpha c_{{\scaleto{T}{2.5pt}}}\left(\delta + \omega_{c}\right)}}{2 c_{{\scaleto{T}{2.5pt}}}}$, $y^{\scaleto{\text{ {\fontfamily{qcr}\selectfont EP}}}{2.5pt}}_{l}(x)=\frac{x}{B}$, $b^{\scaleto{\text{ {\fontfamily{qcr}\selectfont EP}}}{2.5pt}}_{l}=\frac{\alpha c_{{\scaleto{T}{2.5pt}}} + \lambda + \delta + \omega_{c} + \sqrt{(\alpha c_{{\scaleto{T}{2.5pt}}} + \lambda +\delta + \omega_{c})^{2}-4 \alpha c_{{\scaleto{T}{2.5pt}}}\left(\delta + \omega_{c}\right)}}{2 c_{{\scaleto{T}{2.5pt}}}}$, $c^{\scaleto{\text{ {\fontfamily{qcr}\selectfont EP}}}{2.5pt}}_{u}=c^{\scaleto{\text{ {\fontfamily{qcr}\selectfont EP}}}{2.5pt}}_{m}=c^{\scaleto{\text{ {\fontfamily{qcr}\selectfont EP}}}{2.5pt}}_{l}=\alpha$ and the constants $A^{\scaleto{\text{ {\fontfamily{qcr}\selectfont EP}}}{2.5pt}}_{2,u}$, $A^{\scaleto{\text{ {\fontfamily{qcr}\selectfont EP}}}{2.5pt}}_{1,m}$, $A^{\scaleto{\text{ {\fontfamily{qcr}\selectfont EP}}}{2.5pt}}_{2,m}$ and $A^{\scaleto{\text{ {\fontfamily{qcr}\selectfont EP}}}{2.5pt}}_{1,l}$ are obtained as explained in Appendix \ref{ProofofProposition4.1}.

\end{proposition}

\vspace{0.3cm}

See Appendix \ref{ProofofProposition4.1} for proof of Proposition \ref{ExamplesofExtremePovertyRateFunctions-Subsection411-Proposition1}.

\vspace{0.3cm}

\begin{remark}
As for the trapping time, the Laplace transform of the time of extreme poverty approaches the probability of extreme poverty as $\delta$ tends to zero, i.e. 
    
\vspace{0.3cm}
    
\begin{align}
        \lim _{\delta \downarrow 0} m^{\scaleto{\text{ {\fontfamily{qcr}\selectfont EP}}}{2.5pt}}_{\delta}(x) =\mathbb{P}(\tau^{\scaleto{\text{ {\fontfamily{qcr}	\selectfont EP}}}{2.5pt}}_{x}<\infty)\equiv\psi^{\scaleto{\text{ {\fontfamily{qcr}\selectfont EP}}}{2.5pt}}(x),
        \label{TheTrappingTime-Subsection21-Equation9}
\end{align}
    
\vspace{0.3cm}
    
for $\frac{\lambda}{r} < \alpha$.    
     
 \end{remark}   

\vspace{0.3cm}

\begin{figure}[H]
	\begin{subfigure}[b]{0.5\linewidth}
  		\includegraphics[width=7.5cm, height=7.5cm]{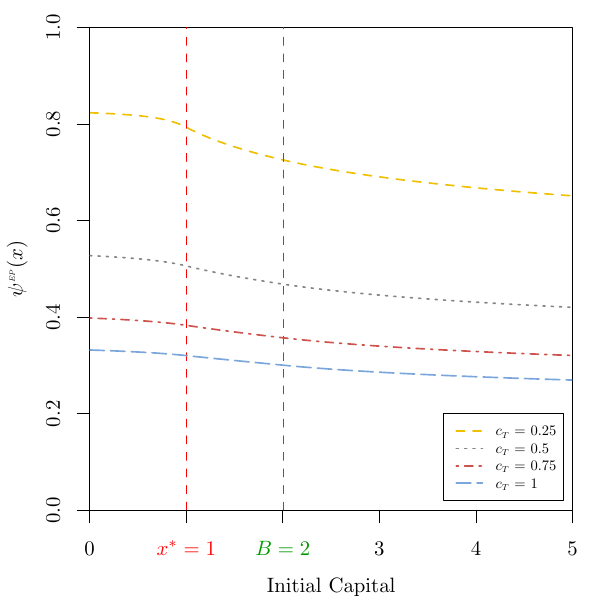}
		\caption{}
  		\label{ExamplesofExtremePovertyRateFunctions-Subsubsection4111-Figure1-a}
	\end{subfigure}
	\begin{subfigure}[b]{0.5\linewidth}
  		\includegraphics[width=7.5cm, height=7.5cm]{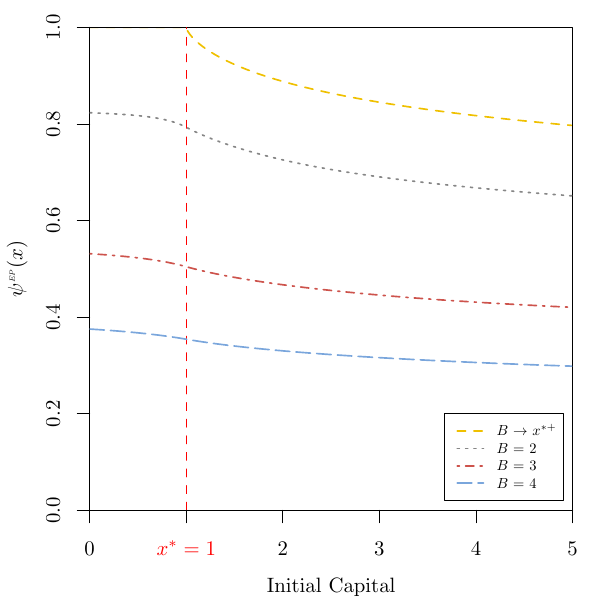}
		\caption{}
  		\label{ExamplesofExtremePovertyRateFunctions-Subsubsection4111-Figure1-b}
	\end{subfigure}
	\caption{(a) Probability of extreme poverty $\psi^{\scaleto{\text{ {\fontfamily{qcr}\selectfont EP}}}{2.5pt}}(x)$ when $Z_{i} \sim Beta(0.8, 1)$, $a = 0.1$, $b = 4$, $c_{{\scaleto{S}{4pt}}} = 0.4$, $B=2$, $\lambda = 1$, $x^{*} = 1$ and $\omega_{1}(x)=0.02$ for $c_{{\scaleto{T}{4pt}}} = 0.25, 0.5, 0.75, 1$ (b) Probability of extreme poverty $\psi^{\scaleto{\text{ {\fontfamily{qcr}\selectfont EP}}}{2.5pt}}(x)$ when $Z_{i} \sim Beta(0.8, 1)$, $a = 0.1$, $b = 4$, $c_{{\scaleto{S}{4pt}}} = 0.4$, $c_{{\scaleto{T}{4pt}}} = 0.25$, $\lambda = 1$, $x^{*} = 1$ and $\omega_{1}(x)=0.02$ for $B \rightarrow x^{*+}$ and $B = 2, 3, 4$.}
	\label{ExamplesofExtremePovertyRateFunctions-Subsubsection4111-Figure1}
\end{figure}

\begin{figure}[H]
		\centering
  		\includegraphics[width=7.5cm, height=7.5cm]{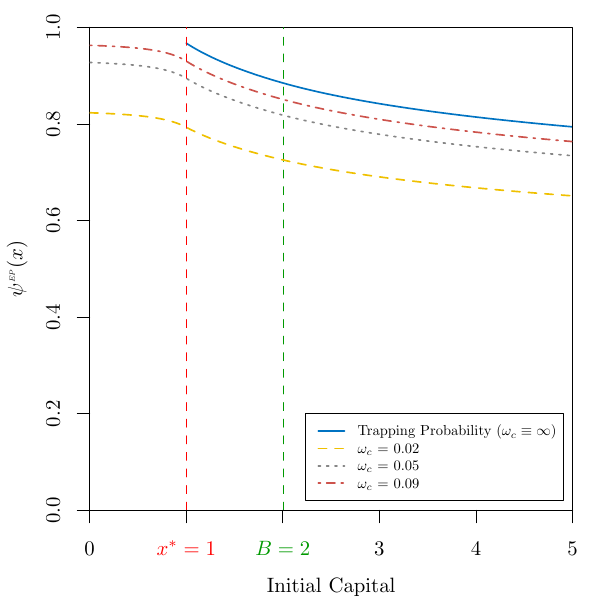}
	\caption{Probability of extreme poverty $\psi^{\scaleto{\text{ {\fontfamily{qcr}\selectfont EP}}}{2.5pt}}(x)$ when $Z_{i} \sim Beta(0.8, 1)$, $a = 0.1$, $b = 4$, $c_{{\scaleto{S}{4pt}}} = 0.4$, $c_{{\scaleto{T}{4pt}}} = 0.25$, $B=2$, $\lambda = 1$, $x^{*} = 1$ and $\omega_{1}(x)=\omega_{c}$ for $\omega_{c} = 0.02, 0.05, 0.09$.}
	 \label{ExamplesofExtremePovertyRateFunctions-Subsubsection4111-Figure2}
\end{figure}

Figure \ref{ExamplesofExtremePovertyRateFunctions-Subsubsection4111-Figure1} shows the probability of extreme poverty $\psi^{\scaleto{\text{ {\fontfamily{qcr}\selectfont EP}}}{2.5pt}}(x)$ for the capital process $X_{t}$ for a constant extreme poverty rate function. As shown in Figure \ref{TheTrappingTime-Subsection31-Figure1} for the case of the trapping probability, Figure \ref{ExamplesofExtremePovertyRateFunctions-Subsubsection4111-Figure1-a} and \ref{ExamplesofExtremePovertyRateFunctions-Subsubsection4111-Figure1-b}  reveal the probability of extreme poverty is also a decreasing function of the cash transfer rate $c_{{\scaleto{T}{4pt}}}$, the capital barrier level $B$ and the initial capital. In addition, in line with the definition of the extreme poverty rate function, Figure \ref{ExamplesofExtremePovertyRateFunctions-Subsubsection4111-Figure2} demonstrates the probability of extreme poverty is an increasing function of the extreme poverty rate function. Furthermore, Figure \ref{ExamplesofExtremePovertyRateFunctions-Subsubsection4111-Figure2} also plots the trapping probability obtained in Section \ref{WhenandHowHouseholdsBecomePoor?-Section3} for reference, which is given by the particular case when $\omega_{c} \equiv \infty$ and therefore represents an upper bound for the probability of extreme poverty. Appendix \ref{Appendix C: Effects of Underlying Factors on the Probability of Extreme Poverty} provides a sensitivity analysis for the probability of extreme poverty with a constant extreme poverty rate function. 

\vspace{0.3cm}

\paragraph{Exponential extreme poverty rate functions.} Let now $\omega_{2}\left(x\right) = \frac{\beta}{x} \cdot \mathbbm{1}_{\{x<x^{*}\}}$, for some $\beta >0$. In this case, the extreme poverty rates take fairly flat values for lower deficit levels and reach higher values when the capital level gets close to zero. Such a function could be considered as the analogous version of the exponential bankruptcy rate function studied in \cite{Article:Albrecher2013b}.

\vspace{0.3cm}

\begin{remark}
	In general, it is not straightforward to obtain the solution of \eqref{Appendix A: Mathematical Proofs-Equation27} for more general extreme poverty rates $\omega(x)$, as functions of $\omega(x)$ appear both in the coefficients of the homogeneous equation and in the inhomogeneous term. Thus, for the particular case of exponential extreme poverty rate functions $\omega_{2}\left(x\right) = \frac{\beta}{x} \cdot \mathbbm{1}_{\{x<x^{*}\}}$, we will only discuss the probability of extreme poverty. 
	 
\end{remark}

\vspace{0.3cm}

\begin{proposition}\label{ExamplesofExtremePovertyRateFunctions-Subsection411-Proposition2}

Consider a household capital process with initial capital $x\ge x^{*}$, capital growth rate $r$, capital barrier level $B$, capital transfer rate $c_{{\scaleto{T}{4pt}}}$, intensity $\lambda > 0$ and remaining proportions of capital with distribution $Beta(\alpha, 1)$ where $\alpha >0$; that is, $Z_{i}\sim Beta(\alpha, 1)$. For an exponential extreme poverty rate function $\omega_{2}\left(x\right) = \frac{\beta}{x} \cdot \mathbbm{1}_{\{x<x^{*}\}}$, with $\beta > 0$, the probability of extreme poverty is given by

\vspace{0.3cm}

\begin{align}
	\psi^{\scaleto{\text{ {\fontfamily{qcr}\selectfont EP}}}{2.5pt}}\left(x\right)&=
		\begin{cases}
		\begin{aligned}
		 A^{\scaleto{\text{ {\fontfamily{qcr}\selectfont EP}}}{2.5pt}}_{2,u}{\left(\frac{x}{x^{*}}\right)}^{\frac{\lambda}{r}-\alpha} { }_{2} F_{1}\left(\alpha - \frac{\lambda}{r}, 1- \frac{\lambda}{r} ;1 + \alpha - \frac{\lambda}{r} ; \frac{x^{*}}{x}\right) \hspace{3cm} \textit{for} \hspace{0.3cm} x \geq B,\end{aligned} \\ \\
		\begin{aligned}
		A^{\scaleto{\text{ {\fontfamily{qcr}\selectfont EP}}}{2.5pt}}_{1,m}&{y^{\scaleto{\text{ {\fontfamily{qcr}\selectfont EP}}}{2.5pt}}_{m}(x)}^{-a^{\scaleto{\text{ {\fontfamily{qcr}\selectfont EP}}}{2pt}}_{m}} { }_{2} F_{1}\left(a^{\scaleto{\text{ {\fontfamily{qcr}\selectfont EP}}}{2.5pt}}_{m}, a^{\scaleto{\text{ {\fontfamily{qcr}\selectfont EP}}}{2.5pt}}_{m}-c^{\scaleto{\text{ {\fontfamily{qcr}\selectfont EP}}}{2.5pt}}_{m}+1 ; a^{\scaleto{\text{ {\fontfamily{qcr}\selectfont EP}}}{2.5pt}}_{m}-b^{\scaleto{\text{ {\fontfamily{qcr}\selectfont EP}}}{2.5pt}}_{m}+1 ; {y^{\scaleto{\text{ {\fontfamily{qcr}\selectfont EP}}}{2.5pt}}_{m}(x)}^{-1}\right)\\ \\ & +A^{\scaleto{\text{ {\fontfamily{qcr}\selectfont EP}}}{2.5pt}}_{2,m}{y^{\scaleto{\text{ {\fontfamily{qcr}\selectfont EP}}}{2.5pt}}_{m}(x)}^{-b^{\scaleto{\text{ {\fontfamily{qcr}\selectfont EP}}}{2pt}}_{m}} { }_{2} F_{1}\left(b^{\scaleto{\text{ {\fontfamily{qcr}\selectfont EP}}}{2.5pt}}_{m}, b^{\scaleto{\text{ {\fontfamily{qcr}\selectfont EP}}}{2.5pt}}_{m}-c^{\scaleto{\text{ {\fontfamily{qcr}\selectfont EP}}}{2.5pt}}_{m}+1 ; b^{\scaleto{\text{ {\fontfamily{qcr}\selectfont EP}}}{2.5pt}}_{m}-a^{\scaleto{\text{ {\fontfamily{qcr}\selectfont EP}}}{2.5pt}}_{m}+1 ; {y^{\scaleto{\text{ {\fontfamily{qcr}\selectfont EP}}}{2.5pt}}_{m}(x)}^{-1}\right)  \hspace{0.1cm} \textit{for}  \hspace{0.3cm} x^{*} \leq x < B,\end{aligned}\\ \\
	\begin{aligned}
		 1+ A^{\scaleto{\text{ {\fontfamily{qcr}\selectfont EP}}}{2.5pt}}_{2,l}{y^{\scaleto{\text{ {\fontfamily{qcr}\selectfont EP}}}{2.5pt}}_{l}(x)}^{2 - \alpha - c^{\scaleto{\text{ {\fontfamily{qcr}\selectfont EP}}}{2pt}}_{l}} { }_{2} F_{1}\left(a^{\scaleto{\text{ {\fontfamily{qcr}\selectfont EP}}}{2.5pt}}_{l}-c^{\scaleto{\text{ {\fontfamily{qcr}\selectfont EP}}}{2.5pt}}_{l}+1, b^{\scaleto{\text{ {\fontfamily{qcr}\selectfont EP}}}{2.5pt}}_{l}-c^{\scaleto{\text{ {\fontfamily{qcr}\selectfont EP}}}{2.5pt}}_{l}+1 ; 2-c^{\scaleto{\text{ {\fontfamily{qcr}\selectfont EP}}}{2.5pt}}_{l} ; {y^{\scaleto{\text{ {\fontfamily{qcr}\selectfont EP}}}{2.5pt}}_{l}(x)}\right) \hspace{0.2cm} \textit{for} \hspace{0.3cm} 0< x < x^{*},\end{aligned}
	\end{cases}
\label{ExamplesofExtremePovertyRateFunctions-Subsection411-Equation1}
\end{align}

\vspace{0.3cm}

where ${ }_{2} F_{1}\left(\cdot \right)$ is Gauss\rq s Hypergeometric Function as defined in \eqref{Appendix A: Mathematical Proofs-Equation10}, $y^{\scaleto{\text{ {\fontfamily{qcr}\selectfont EP}}}{2.5pt}}_{m}(x)=-\frac{x}{x^{**}}$, with $x^{**}=\frac{c_{{\scaleto{T}{2.5pt}}}B-rx^{*}}{r-c_{{\scaleto{T}{2.5pt}}}}$, $a^{\scaleto{\text{ {\fontfamily{qcr}\selectfont EP}}}{2.5pt}}_{m}= \frac{-(\lambda - \alpha \left(r-c_{{\scaleto{T}{2.5pt}}}\right)) - \sqrt{(\lambda -\alpha \left(r-c_{{\scaleto{T}{2.5pt}}}\right))^{2}}}{2\left(r - c_{{\scaleto{T}{2.5pt}}}\right)}$, $b^{\scaleto{\text{ {\fontfamily{qcr}\selectfont EP}}}{2.5pt}}_{m}=\frac{-(\lambda - \alpha \left(r- c_{{\scaleto{T}{2.5pt}}}\right)) + \sqrt{(\lambda -\alpha \left(r - c_{{\scaleto{T}{2.5pt}}}\right))^{2}}}{2\left(r - c_{{\scaleto{T}{2.5pt}}}\right)}$, $c^{\scaleto{\text{ {\fontfamily{qcr}\selectfont EP}}}{2.5pt}}_{m}=\alpha$, $y^{\scaleto{\text{ {\fontfamily{qcr}\selectfont EP}}}{2.5pt}}_{l}(x)=\frac{x}{B}$, $a^{\scaleto{\text{ {\fontfamily{qcr}\selectfont EP}}}{2.5pt}}_{l}=1 - \alpha$, $b^{\scaleto{\text{ {\fontfamily{qcr}\selectfont EP}}}{2.5pt}}_{l}=\frac{c_{{\scaleto{T}{2.5pt}}} + \lambda}{c_{{\scaleto{T}{2.5pt}}}}$ and $c^{\scaleto{\text{ {\fontfamily{qcr}\selectfont EP}}}{2.5pt}}_{l}=-\frac{Bc_{{\scaleto{T}{2.5pt}}}\left(\alpha - 2\right) + \beta}{Bc_{{\scaleto{T}{2.5pt}}}}$ for $\alpha > \frac{\lambda}{r}$ and the constants $A^{\scaleto{\text{ {\fontfamily{qcr}\selectfont EP}}}{2.5pt}}_{2,u}$, $A^{\scaleto{\text{ {\fontfamily{qcr}\selectfont EP}}}{2.5pt}}_{1,m}$, $A^{\scaleto{\text{ {\fontfamily{qcr}\selectfont EP}}}{2.5pt}}_{2,m}$ and $A^{\scaleto{\text{ {\fontfamily{qcr}\selectfont EP}}}{2.5pt}}_{2,l}$ are obtained as explained in Appendix \ref{ProofofProposition4.2}.
\end{proposition}

The mathematical proof of Proposition \ref{ExamplesofExtremePovertyRateFunctions-Subsection411-Proposition2} is given in Appendix \ref{ProofofProposition4.2}.

\vspace{0.3cm}

\begin{figure}[H]
	\begin{subfigure}[b]{0.5\linewidth}
  		\includegraphics[width=7.5cm, height=7.5cm]{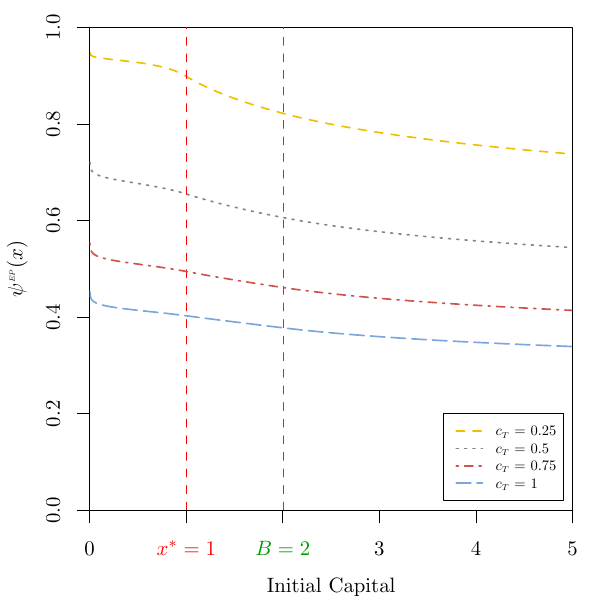}
		\caption{}
  		\label{ExamplesofExtremePovertyRateFunctions-Subsubsection4112-Figure1-a}
	\end{subfigure}
	\begin{subfigure}[b]{0.5\linewidth}
  		\includegraphics[width=7.5cm, height=7.5cm]{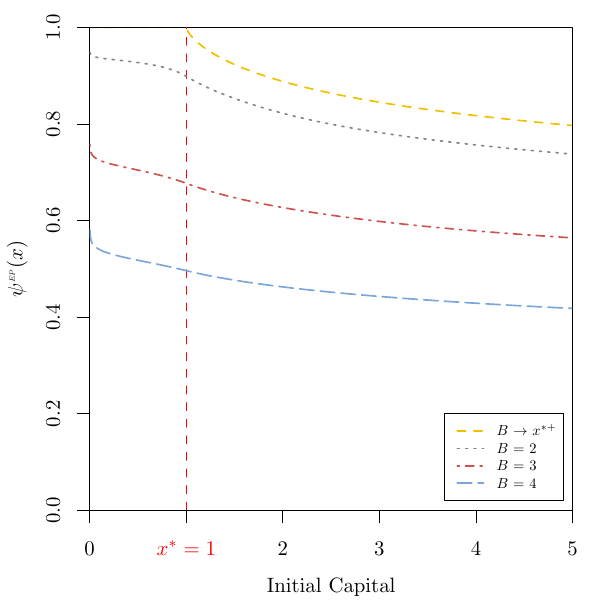}
		\caption{}
  		\label{ExamplesofExtremePovertyRateFunctions-Subsubsection4112-Figure1-b}
	\end{subfigure}
	\caption{(a) Probability of extreme poverty $\psi^{\scaleto{\text{ {\fontfamily{qcr}\selectfont EP}}}{2.5pt}}(x)$ when $Z_{i} \sim Beta(0.8, 1)$, $a = 0.1$, $b = 4$, $c_{{\scaleto{S}{4pt}}} = 0.4$, $B=2$, $\lambda = 1$, $x^{*} = 1$ and $\omega_{2}(x)=\frac{0.02}{x}$ for $c_{{\scaleto{T}{4pt}}} = 0.25, 0.5, 0.75, 1$ (b) Probability of extreme poverty $\psi^{\scaleto{\text{ {\fontfamily{qcr}\selectfont EP}}}{2.5pt}}(x)$ when $Z_{i} \sim Beta(0.8, 1)$, $a = 0.1$, $b = 4$, $c_{{\scaleto{S}{4pt}}} = 0.4$, $c_{{\scaleto{T}{4pt}}} = 0.25$, $\lambda = 1$, $x^{*} = 1$ and $\omega_{2}(x)=\frac{0.02}{x}$ for $B \rightarrow x^{*+}$ and $B = 2, 3, 4$.}
	\label{ExamplesofExtremePovertyRateFunctions-Subsubsection4112-Figure1}
\end{figure}

Figures \ref{ExamplesofExtremePovertyRateFunctions-Subsubsection4112-Figure1} and \ref{ExamplesofExtremePovertyRateFunctions-Subsubsection4112-Figure2} display the probability of extreme poverty when dealing with an exponential extreme poverty rate function. Evidently, under this setup, the probability of extreme poverty attains higher values compared to that under which a constant extreme poverty rate is considered. This can be verified by comparing Figures \ref{ExamplesofExtremePovertyRateFunctions-Subsubsection4111-Figure1-a} and \ref{ExamplesofExtremePovertyRateFunctions-Subsubsection4112-Figure1-a}, for several cash transfer rates $c_{{\scaleto{T}{4pt}}}$, Figures \ref{ExamplesofExtremePovertyRateFunctions-Subsubsection4111-Figure1-b} and \ref{ExamplesofExtremePovertyRateFunctions-Subsubsection4112-Figure1-b}, for different capital barrier levels $B$, and Figures \ref{ExamplesofExtremePovertyRateFunctions-Subsubsection4111-Figure2} and \ref{ExamplesofExtremePovertyRateFunctions-Subsubsection4112-Figure2}, for different values of the extreme poverty rate, respectively. This result is not particularly surprising because of the fact that the exponential extreme poverty rate takes higher values for higher capital deficits while the constant extreme poverty rate remains flat for all capital levels. Appendix \ref{Appendix C: Effects of Underlying Factors on the Probability of Extreme Poverty} also presents a sensitivity analysis of the probability of extreme poverty for an exponential extreme poverty rate function.

\vspace{0.3cm}

\begin{figure}[H]
		\centering
  		\includegraphics[width=7.5cm, height=7.5cm]{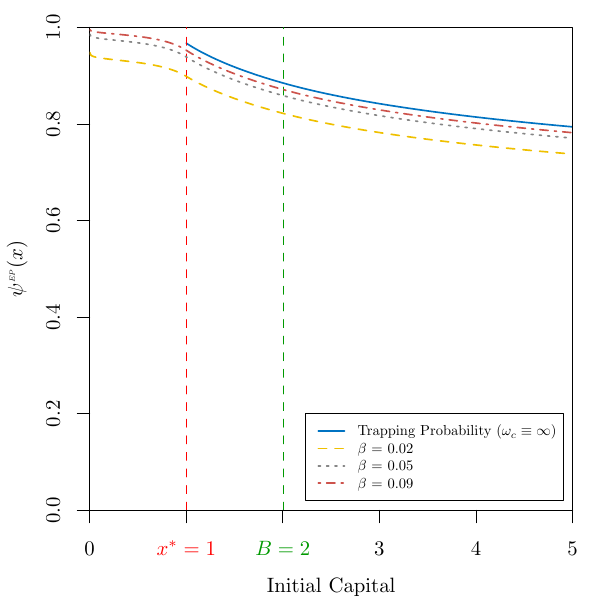}
  		\label{ExamplesofExtremePovertyRateFunctions-Subsubsection4112-Figure2}
	\caption{Probability of extreme poverty $\psi^{\scaleto{\text{ {\fontfamily{qcr}\selectfont EP}}}{2.5pt}}(x)$ when $Z_{i} \sim Beta(0.8, 1)$, $a = 0.1$, $b = 4$, $c_{{\scaleto{S}{4pt}}} = 0.4$, $c_{{\scaleto{T}{4pt}}} = 0.25$, $B=2$, $\lambda = 1$, $x^{*} = 1$ and $\omega_{2}(x)=\frac{\beta}{x}$ for $\beta = 0.02, 0.05, 0.09$.}
	\label{ExamplesofExtremePovertyRateFunctions-Subsubsection4112-Figure2}
\end{figure}

As mentioned previously, Appendix \ref{Appendix C: Effects of Underlying Factors on the Probability of Extreme Poverty} shows how sensitive the probability of extreme poverty is with respect to all the underlying parameters (for both constant and exponential extreme poverty rate functions). In particular, the sensitivity with respect to the cash transfer rate $c_{{\scaleto{T}{4pt}}}$ and the capital barrier level $B$ is worth noting. These results accentuate the importance of selecting an appropriate cash transfer rate $c_{{\scaleto{T}{4pt}}}$ (i.e. an adequate frequency or intensity of the capital cash transfers) and a suitable capital barrier level $B$ (i.e. an opportune targeting), when designing the social protection strategy for achieving extreme poverty reduction.

\begin{figure}[H]
	\begin{subfigure}[b]{0.5\linewidth}
  		\includegraphics[width=7.5cm, height=7.5cm]{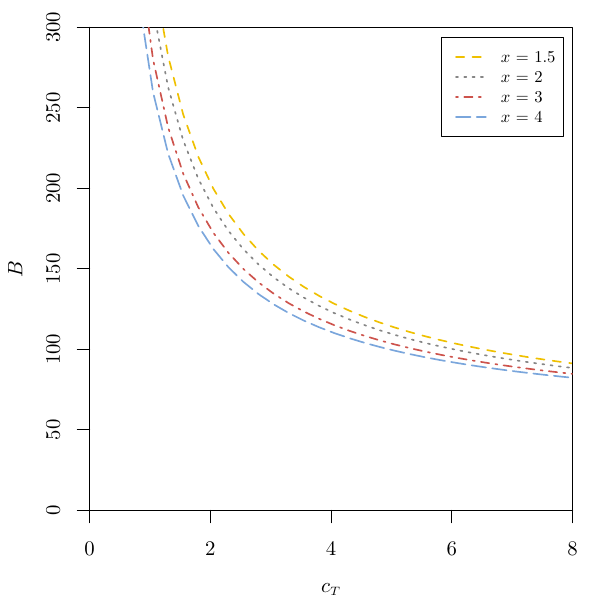}
		\caption{}
  		\label{WhenandHowHouseholdsBecomeExtremelyPoor?-Section4-Figure1-a}
	\end{subfigure}
	\begin{subfigure}[b]{0.5\linewidth}
  		\includegraphics[width=7.5cm, height=7.5cm]{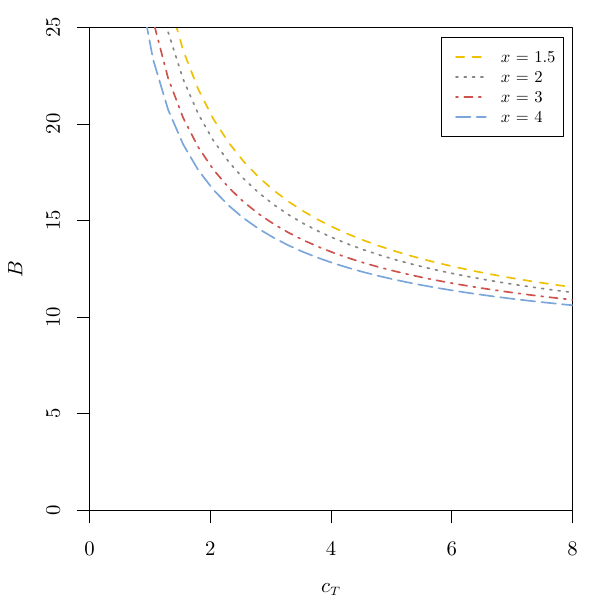}
		\caption{}
  		\label{WhenandHowHouseholdsBecomeExtremelyPoor?-Section4-Figure1-b}
	\end{subfigure}
	\caption{(a) Cash transfer rate $c_{{\scaleto{T}{4pt}}}$ and capital barrier level $B$ required to attain a given target trapping probability of $\psi^{\scaleto{\text{ {\fontfamily{qcr}\selectfont P}}}{2.5pt}}(x) = 0.01$ when $Z_{i} \sim Beta(1.25, 1)$, $a = 0.1$, $b = 4$, $c_{{\scaleto{S}{4pt}}} = 0.4$, $\lambda = 1$ and $x^{*} = 1$ for initial capital $x = 1.5, 2, 3, 4$ (b) Cash transfer rate $c_{{\scaleto{T}{4pt}}}$ and capital barrier level $B$ required to attain a given target probability of extreme poverty of $\psi^{\scaleto{\text{ {\fontfamily{qcr}\selectfont EP}}}{2.5pt}}(x) = 0.01$ when $Z_{i} \sim Beta(1.25, 1)$, $a = 0.1$, $b = 4$, $c_{{\scaleto{S}{4pt}}} = 0.4$, $\lambda = 1$, $x^{*} = 1$ and $\omega_{c} = 0.09$ for initial capital $x = 1.5, 2, 3, 4$. }
	\label{WhenandHowHouseholdsBecomeExtremelyPoor?-Section4-Figure1}
\end{figure}

Figures \ref{WhenandHowHouseholdsBecomeExtremelyPoor?-Section4-Figure1-a} and \ref{WhenandHowHouseholdsBecomeExtremelyPoor?-Section4-Figure1-b} provide an example of the cash transfer rate $c_{{\scaleto{T}{4pt}}}$ and the capital barrier level $B$ required to attain a given target trapping probability and probability of extreme poverty (for a constant extreme poverty rate function), respectively. Clearly, policymakers must decide between reducing the intensity of the capital cash transfers (lowering the capital cash transfer rate $c_{{\scaleto{T}{4pt}}}$) to a wider group of households (increasing the capital barrier level $B$) or increasing the intensity of the capital cash transfers (rising the capital cash transfer rate $c_{{\scaleto{T}{4pt}}}$) to a narrower group of households (lowering the capital barrier level $B$) in order to achieve the target probabilities, showing an evident trade-off between these two parameters.

\vspace{0.3cm}

\section{Monte Carlo Simulation} \label{MonteCarloSimulation-Section5}

In general, it is not straightforward to derive explicit formulas for both the trapping probability and the probability of extreme poverty when more general cases are considered. Monte Carlo simulation is an alternative way to produce estimates for both quantities and is particularly useful when dealing with cases for which closed-form formulas are not available. In this section, following \cite{Article:Albrecher2013b}, we introduce a simple and efficient methodology that allows to generate fairly accurate approximations for the probability of extreme poverty.

\subsection{Methodology} \label{Methodology-Subsection}

Following \cite{Article:Albrecher2013b}, we note that for any capital level $x \in (0, \infty)$ it holds that

\vspace{0.3cm}

\begin{align}
    \psi^{\scaleto{\text{ {\fontfamily{qcr}\selectfont EP}}}{2.5pt}}(x)=1-\mathbb{E}\left[e^{-\int_0^{\infty} \omega\left(X_{t}\right) \mathbbm{1}_{\{X_{t}<x^{*}\}} d t} \mid X_{0}=x\right],
    \label{Methodology-Subsection-Equation1}
\end{align}

\vspace{0.3cm}

as extreme poverty can only be avoided if there is no event of the Poisson process with level-dependent intensity $\omega\left(\cdot\right)$ during the time the capital process spends below the critical capital $x^{*}$. The above expectation can then be computed by conditioning on the simulated sample path. Concretely, conditioning on the remaining proportions $\Theta_{i}$, with

\vspace{0.3cm}

\begin{align}
\Psi^{\scaleto{\text{ {\fontfamily{qcr}\selectfont EP}}}{2.5pt}}(\omega, x) \mid\left(T_1, \Theta_1\right),\left(T_2, \Theta_2\right) \ldots & = \int_0^{\infty} \omega\left(X_{t}\right) \cdot \mathbbm{1}_{\left\{X_{t}<x^{*}\right\}} d t \\
& =-\sum_{i=0}^{\infty} \mathbbm{1}_{\left\{X_{T_{i}<x^{*}}\right\}} \int_{T_i}^{\min \left(T_{i+1}, T_i + \tau_{x^{{\scaleto{*}{2pt}}}}\left(X_{T_{i}}\right) \right)} \omega\left(X_{s}\right) d s \label{Methodology-Subsection-Equation2}
\end{align}

\vspace{0.3cm}

with $T_{0}=0$, we can write

\vspace{0.3cm}

\begin{align}
\psi^{\scaleto{\text{ {\fontfamily{qcr}\selectfont EP}}}{2.5pt}}(\omega, x)=\mathbb{E}_{\left(T_{1}, \Theta_{1}\right),\left(T_{2}, \Theta_{2}\right) \ldots}\left[1-e^{\Psi^{\scaleto{\text{ {\fontfamily{qcr}\selectfont EP}}}{2.5pt}}(\omega, x) \mid\left(T_{1}, \Theta_{1}\right),\left(T_{2}, \Theta_{2}\right) \ldots}\right]. \label{Methodology-Subsection-Equation3}
\end{align}

\vspace{0.3cm}

\begin{figure}[H]
	\begin{center}
  		\includegraphics[width=9cm, height=9cm]{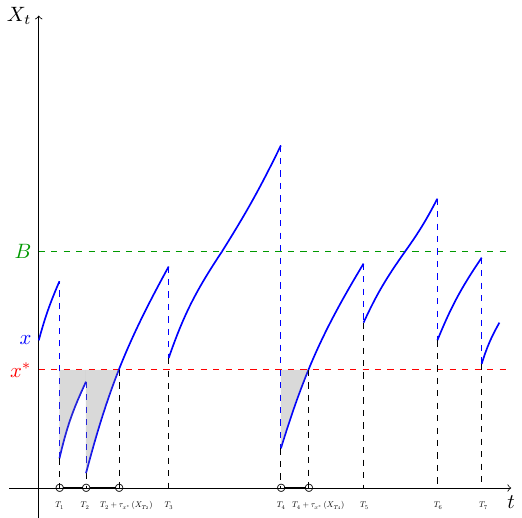}
		\caption{Computation of $\Psi^{\scaleto{\text{ {\fontfamily{qcr}\selectfont EP}}}{2.5pt}}\left(\omega, x\right)$ conditional on a realised sample path.}
		\label{MonteCarloSimulation-Section5-Figure1}
	\end{center}
\end{figure}

\vspace{0.3cm}

\begin{figure}[H]
	\begin{subfigure}[b]{0.5\linewidth}
  		\includegraphics[width=7.5cm, height=7.5cm]{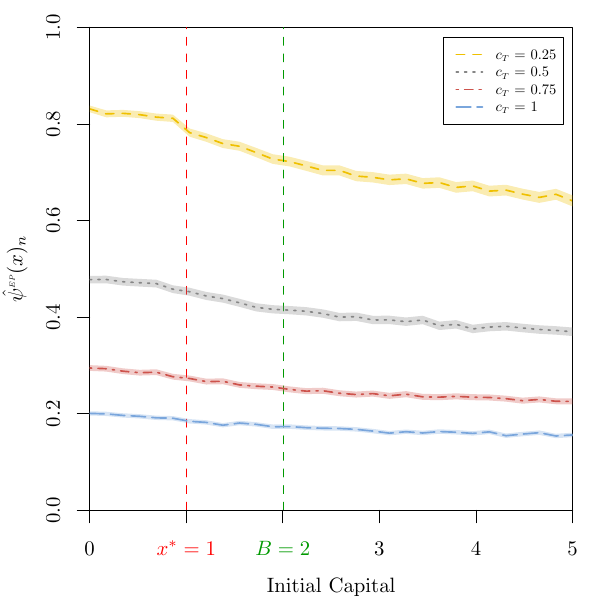}
		\caption{}
  		\label{MonteCarloSimulation-Section5-Figure2-a}
	\end{subfigure}
	\begin{subfigure}[b]{0.5\linewidth}
  		\includegraphics[width=7.5cm, height=7.5cm]{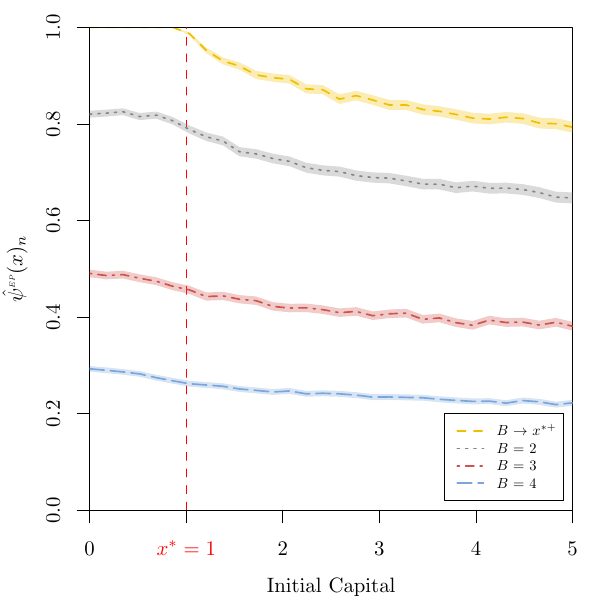}
		\caption{}
  		\label{MonteCarloSimulation-Section5-Figure2-b}
	\end{subfigure}
	\caption{(a) Probability of extreme poverty $\hat{\psi}^{\scaleto{\text{ {\fontfamily{qcr}\selectfont EP}}}{2.5pt}}(x)_{n}$ when $n=10,000$, $Z_{i} \sim Beta(0.8, 1)$, $a = 0.1$, $b = 4$, $c_{{\scaleto{S}{4pt}}} = 0.4$, $B=2$, $\lambda = 1$, $x^{*} = 1$ and $\omega_{1}(x)=0.02$ for $c_{{\scaleto{T}{4pt}}} = 0.25, 0.5, 0.75, 1$ (b) Probability of extreme poverty $\psi^{\scaleto{\text{ {\fontfamily{qcr}\selectfont EP}}}{2.5pt}}(x)_{n}$ when $n=10,000$, $Z_{i} \sim Beta(0.8, 1)$, $a = 0.1$, $b = 4$, $c_{{\scaleto{S}{4pt}}} = 0.4$, $c_{{\scaleto{T}{4pt}}} = 0.25$, $\lambda = 1$, $x^{*} = 1$ and $\omega_{1}(x)=0.02$ for $B \rightarrow x^{*+}$ and $B = 2, 3, 4$.}
	\label{MonteCarloSimulation-Section5-Figure2}
\end{figure}

\vspace{0.3cm}

In particular, for the two choices $\omega_{1}(x)=\omega_{c}$, $\omega_{c}>0$, and $\omega_{2}(x)=\frac{\beta}{x}$, $\beta > 0$, \eqref{Methodology-Subsection-Equation2} reads

\vspace{0.3cm}

{\allowdisplaybreaks
\begin{align}
\Psi^{\scaleto{\text{ {\fontfamily{qcr}\selectfont EP}}}{2.5pt}}(\omega_{1}, x) \mid\left(T_1, \Theta_1\right),\left(T_2, \Theta_2\right) \ldots & = -\sum_{i=0}^{\infty} \mathbbm{1}_{\left\{X_{T_{i}<x^{*}}\right\}}  \int_{T_i}^{\min \left(T_{i+1}, T_i + \tau_{x^{{\scaleto{*}{2pt}}}} \left(X_{T_{i}}\right) \right)} \omega_{c} \ ds \\ \\
& = -\omega_{c} \sum_{i=0}^{\infty} \mathbbm{1}_{\left\{X_{T_{i}<x^{*}}\right\}} \left[ \min \left(T_{i+1} - T_{i}, \tau_{x^{{\scaleto{*}{2pt}}}} \left(X_{T_{i}}\right) \right)\right] \label{Methodology-Subsection-Equation4}
\end{align}
}

\vspace{0.3cm}

and

\vspace{0.3cm}

\small

{\allowdisplaybreaks
\begin{align}
&\Psi^{\scaleto{\text{ {\fontfamily{qcr}\selectfont EP}}}{2.5pt}}(\omega_{2}, x) \mid\left(T_1, \Theta_1\right),\left(T_2, \Theta_2\right) \ldots = -\sum_{i=0}^{\infty} \mathbbm{1}_{\left\{X_{T_{i}<x^{*}}\right\}} \int_{T_i}^{\min \left(T_{i+1}, T_i + \tau_{x^{{\scaleto{*}{2pt}}}}\left(X_{T_{i}}\right) \right)} \frac{\beta}{\left(X_{T_{i}}-B\right)e^{c_{{\scaleto{T}{2.5pt}}} \left(T_{i} - s\right)} + B} ds\\ \\
& = -\frac{\beta}{c_{{\scaleto{T}{2.5pt}}} B}\sum_{i=0}^{\infty} \mathbbm{1}_{\left\{X_{T_{i}<x^{*}}\right\}} \left[ c_{{\scaleto{T}{2.5pt}}} \min \left(T_{i+1} - T_{i}, \tau_{x^{{\scaleto{*}{2pt}}}}\left(X_{T_{i}}\right) \right) + ln\left(B + \left(X_{T_{i}}-B\right)e^{c_{{\scaleto{T}{2.5pt}}}\left[T_{i} - min\left(T_{i+1}, T_{i} +\tau_{x^{{\scaleto{*}{2pt}}}} \left(X_{T_{i}}\right)\right)\right] }\right)- ln\left(X_{T_{i}}\right)\right], \\ \label{Methodology-Subsection-Equation5}
\end{align}
}

\normalsize

\vspace{0.3cm}

respectively. Figure \ref{MonteCarloSimulation-Section5-Figure1} depicts a particular path, and the shaded area refers to $\Psi^{\scaleto{\text{ {\fontfamily{qcr}\selectfont EP}}}{2.5pt}}(\omega, x) \mid\left(T_1, \Theta_1\right),\left(T_2, \Theta_2\right) \ldots$ as in \eqref{Methodology-Subsection-Equation2}. 

\begin{figure}[H]
	\centering
  	\includegraphics[width=7.5cm, height=7.5cm]{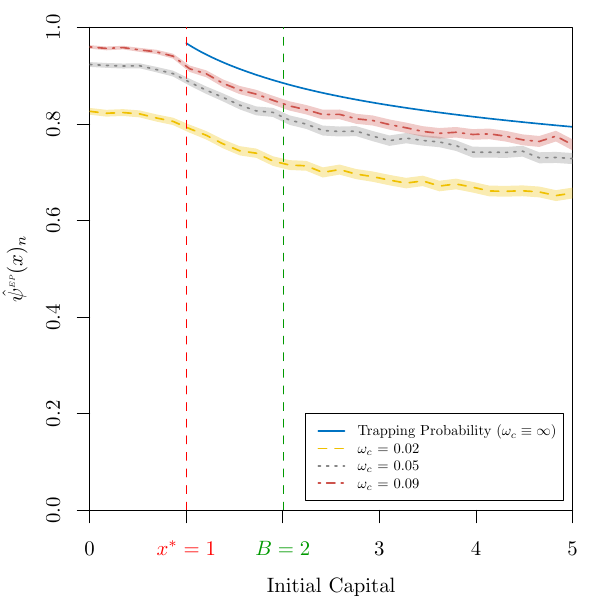}
	\caption{Probability of extreme poverty $\hat{\psi}^{\scaleto{\text{ {\fontfamily{qcr}\selectfont EP}}}{2.5pt}}(x)_{n}$ when $n=10,000$, $Z_{i} \sim Beta(0.8, 1)$, $a = 0.1$, $b = 4$, $c_{{\scaleto{S}{4pt}}} = 0.4$, $c_{{\scaleto{T}{4pt}}} = 0.25$, $B=2$, $\lambda = 1$, $x^{*} = 1$ and $\omega_{1}(x)=\omega_{c}$ for $\omega_{c} = 0.02, 0.05, 0.09$.}
	\label{MonteCarloSimulation-Section5-Figure3}
\end{figure}

In the following simulations, $n$ capital process paths are generated and for the $k$th such path, the function $\Psi^{\scaleto{\text{ {\fontfamily{qcr}\selectfont EP}}}{2.5pt}}(\omega, x)_{k} \mid\left(T_1, \Theta_1\right),\left(T_2, \Theta_2\right) \ldots$ is computed as per \eqref{Methodology-Subsection-Equation4} and \eqref{Methodology-Subsection-Equation5}. The estimator of the probability of extreme poverty is given by 

\vspace{0.3cm}

\begin{align}
\hat{\psi}^{\scaleto{\text{ {\fontfamily{qcr}\selectfont EP}}}{2.5pt}}(x)_n=\frac{1}{n} \sum_{k=1}^n\left(1-e^{\Psi^{\scaleto{\text{ {\fontfamily{qcr}\selectfont EP}}}{2.5pt}}(\omega, x)_k}\right), \label{Methodology-Subsection-Equation6}
\end{align}

\vspace{0.3cm}

and the two sided $99\%$ confidence interval of the estimator can be written as 

\vspace{0.3cm}

\begin{align}
\left(\max \left[\hat{\psi}^{\scaleto{\text{ {\fontfamily{qcr}\selectfont EP}}}{2.5pt}}(x)_n-\frac{2.81}{\sqrt{n}} \sigma_n, 0\right], \min \left[\hat{\psi}^{\scaleto{\text{ {\fontfamily{qcr}\selectfont EP}}}{2.5pt}}(x)_n+\frac{2.81}{\sqrt{n}} \sigma_n, 1\right]\right), \label{Methodology-Subsection-Equation7}
\end{align}

\vspace{0.3cm}

with $\sigma_n=\sqrt{\frac{1}{n-1} \sum_{k=1}^n\left(1-e^{\Psi^{\scaleto{\text{ {\fontfamily{qcr}\selectfont EP}}}{2.5pt}}(\omega, x)_k}-\hat{\psi}^{\scaleto{\text{ {\fontfamily{qcr}\selectfont EP}}}{2.5pt}}(x)_{n}\right)^2}$, such that the bounds of the confidence interval converge to $\hat{\psi}^{\scaleto{\text{ {\fontfamily{qcr}\selectfont EP}}}{2.5pt}}(x)_n$ for $n \rightarrow \infty$.

\begin{figure}[H]
	\begin{subfigure}[b]{0.5\linewidth}
  		\includegraphics[width=7.5cm, height=7.5cm]{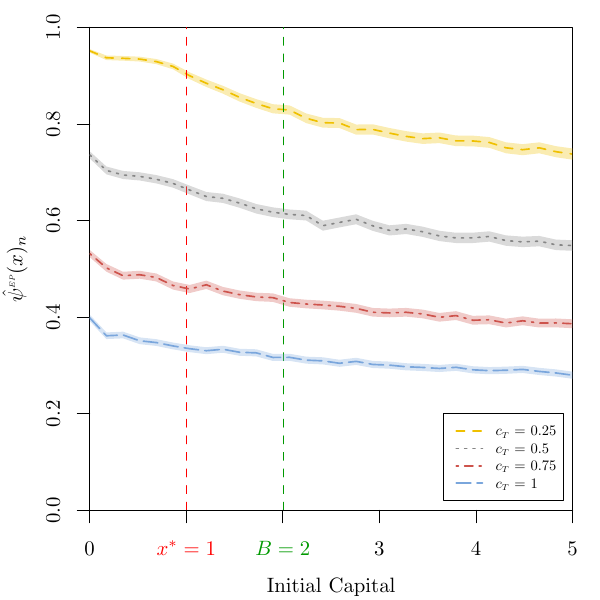}
		\caption{}
  		\label{MonteCarloSimulation-Section5-Figure4-a}
	\end{subfigure}
	\begin{subfigure}[b]{0.5\linewidth}
  		\includegraphics[width=7.5cm, height=7.5cm]{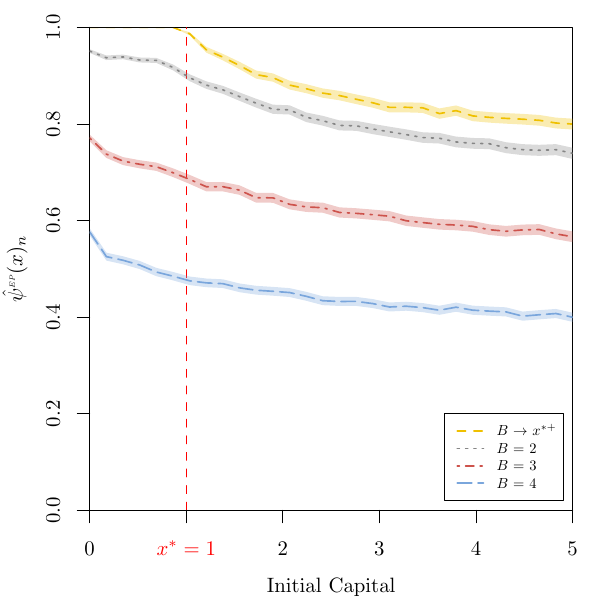}
		\caption{}
  		\label{MonteCarloSimulation-Section5-Figure4-b}
	\end{subfigure}
	\caption{(a) Probability of extreme poverty $\hat{\psi}^{\scaleto{\text{ {\fontfamily{qcr}\selectfont EP}}}{2.5pt}}(x)_{n}$ when $n=10,000$, $Z_{i} \sim Beta(0.8, 1)$, $a = 0.1$, $b = 4$, $c_{{\scaleto{S}{4pt}}} = 0.4$, $B=2$, $\lambda = 1$, $x^{*} = 1$ and $\omega_{2}(x)=\frac{0.02}{x}$ for $c_{{\scaleto{T}{4pt}}} = 0.25, 0.5, 0.75, 1$ (b) Probability of extreme poverty $\psi^{\scaleto{\text{ {\fontfamily{qcr}\selectfont EP}}}{2.5pt}}(x)$ when $Z_{i} \sim Beta(0.8, 1)$, $a = 0.1$, $b = 4$, $c_{{\scaleto{S}{4pt}}} = 0.4$, $c_{{\scaleto{T}{4pt}}} = 0.25$, $\lambda = 1$, $x^{*} = 1$ and $\omega_{2}(x)=\frac{0.02}{x}$ for $B \rightarrow x^{*+}$ and $B = 2, 3, 4$.}
	\label{MonteCarloSimulation-Section5-Figure4}
\end{figure}

\begin{figure}[H]
	\centering
  	\includegraphics[width=7.5cm, height=7.5cm]{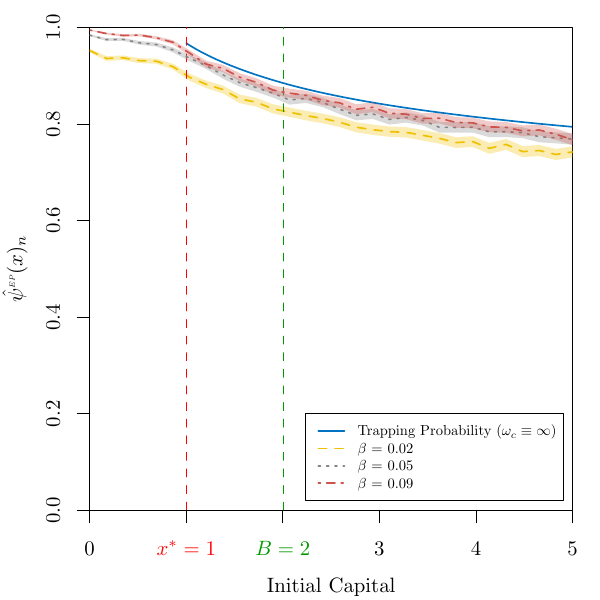}
	\caption{Probability of extreme poverty $\hat{\psi}^{\scaleto{\text{ {\fontfamily{qcr}\selectfont EP}}}{2.5pt}}(x)_{n}$ when $n=10,000$, $Z_{i} \sim Beta(0.8, 1)$, $a = 0.1$, $b = 4$, $c_{{\scaleto{S}{4pt}}} = 0.4$, $c_{{\scaleto{T}{4pt}}} = 0.25$, $B=2$, $\lambda = 1$, $x^{*} = 1$ and $\omega_{2}(x)=\frac{\beta}{x}$ for $\beta = 0.02, 0.05, 0.09$.}
	\label{MonteCarloSimulation-Section5-Figure5}
\end{figure}

Figures \ref{MonteCarloSimulation-Section5-Figure2}, \ref{MonteCarloSimulation-Section5-Figure3}, \ref{MonteCarloSimulation-Section5-Figure4} and \ref{MonteCarloSimulation-Section5-Figure5} provide an example of the Monte Carlo methodology discussed in this Section \ref{MonteCarloSimulation-Section5}. A comparison of Figure \ref{ExamplesofExtremePovertyRateFunctions-Subsubsection4111-Figure1} with Figure \ref{MonteCarloSimulation-Section5-Figure2}, Figure \ref{ExamplesofExtremePovertyRateFunctions-Subsubsection4111-Figure2} with Figure \ref{MonteCarloSimulation-Section5-Figure3}, Figure \ref{ExamplesofExtremePovertyRateFunctions-Subsubsection4112-Figure1} with Figure \ref{MonteCarloSimulation-Section5-Figure4} and Figure \ref{ExamplesofExtremePovertyRateFunctions-Subsubsection4112-Figure2} with Figure \ref{MonteCarloSimulation-Section5-Figure5}, respectively, provides insight into the ability of this method to produce approximations of the probability of extreme poverty when considering more general cases. Although, in general, Monte Carlo simulations produce fairly accurate approximations, it is especially important to note that, for some cases of selected parameters, Monte Carlo simulations may lead to less accurate approximations. Comparing Figures \ref{ExamplesofExtremePovertyRateFunctions-Subsubsection4111-Figure1-a} and \ref{MonteCarloSimulation-Section5-Figure2-a}, and Figures \ref{ExamplesofExtremePovertyRateFunctions-Subsubsection4111-Figure1-b} and \ref{MonteCarloSimulation-Section5-Figure2-b}, for higher capital cash transfer rates $c_{{\scaleto{T}{4pt}}}$ and higher capital barrier levels $B$, respectively, leads to a clear evidence of this imprecision. In this particular case, the differences between the closed-form formula and the Monte Carlo approximates are mainly due to the fact that for high capital cash transfer rates $c_{{\scaleto{T}{4pt}}}$ and capital barrier levels $B$, the capital trajectory will grow rapidly up to the capital barrier level $B$, even in those cases where capital levels close to zero are reached, whereas for the closed-form formula, this would almost certainly be considered as an event of extreme poverty. Nevertheless, it is also worth noting that even though there are the aforementioned discrepancies, Monte Carlo estimates are able to capture the main trend in the probability of extreme poverty.

As mentioned previously, the proposed methodology could be of great advantage when dealing with dynamics for which closed-form formulas are not available. For instance, one could produce approximates of the probability of extreme poverty for situations under which the remaining proportions of capital after experiencing a loss are non $Beta(\alpha,1)-$distributed; i.e. when the random variables $Z_{i}$ follow another distribution with support in $(0,1]$.

\section{Conclusion} \label{Conclusion-Section6}

Using standard techniques from actuarial science and, in particular, from ruin theory, this study analyses the efficiency of regular unconditional cash transfer (UCT) programmes in achieving one of the global public\rq s priority: ending poverty in all its forms everywhere. Introducing an alternative version of the household\rq s capital model originally proposed in \cite{Article:Kovacevic2011}, where we consider a particular group of households are entitled to benefit from capital cash transfers and, adopting ideas from the Omega risk process, first introduced in \cite{Article:Albrecher2011}, this article focuses in studying two main random variables: the trapping time and the time of extreme poverty. While the trapping time has been previously studied for more common risk processes (see, for example, \cite{Article:Flores-Contro2021} and \cite{Article:Flores-Contro2024}), for the best of our knowledge, this is the first work that considers the trapping time and the time of extreme poverty under the dynamics of a household\rq s capital process that incorporates capital cash transfers. Furthermore, for the particular case of the time of extreme poverty, this work also introduces the concept of the extreme poverty rate function for the first time. This article analyses the behavior of two main risk measures associated to these random times: the trapping probability and the probability of extreme poverty. 

From a ruin-theoretic perspective, our main contribution is obtaining closed-form solutions for both risk measures, which is considered to be the ideal situation when working with ruin probabilities \citep{Book:Asmussen2010}. We derive explicit formulas for both the trapping probability and the probability of extreme poverty assuming the proportion of the remaining capital of a household after experiencing a loss is $Beta(\alpha,1)-$distributed. Moreover, for the particular case of the probability of extreme poverty, we also consider two examples of extreme poverty rate functions for which closed-from solutions for the probability of extreme poverty are available: constant and exponential extreme poverty rate functions. Nevertheless, explicit formulas are generally not straightforward to obtain for more general cases. Hence, following \cite{Article:Albrecher2013b}, in Section \ref{MonteCarloSimulation-Section5} we also illustrate how to produce approximates of the probability of extreme poverty via an efficient Monte Carlo simulation method.

Numerical examples presented in Sections \ref{WhenandHowHouseholdsBecomePoor?-Section3} and \ref{WhenandHowHouseholdsBecomeExtremelyPoor?-Section4} indicate that regular UCT programmes are an efficient social protection strategy to keep households out of poverty and extreme poverty, as their trapping probability and the probability of extreme poverty, respectively, decrease when they are part of such strategy. In particular, the role played by both the capital cash transfer rate $c_{{\scaleto{T}{4pt}}}$ and the capital barrier level $B$ for attaining lower probabilities is outlined. Our findings can provide policy makers with a mathematically sound starting point for designing UCT programmes. That is, our model, for instance, could provide insights during  the planning phase of an UCT programme to policy makers about the impact on the probability of (extreme) household impoverishment when targeting a particular group of households (depending on the selection of the capital barrier level $B$). Moreover, the sensitivity of the probability of (extreme) household impoverishment to the frequency or intensity of the capital cash transfers (depending on the choice of the capital cash transfer rate $c_{{\scaleto{T}{4pt}}}$) can also be assessed with our results. Furthermore, it is important to note that our analyses show that the probability of extreme poverty appears to be more sensitive to changes in these parameters, compared to the trapping probability, therefore suggesting that policy makers should specially watch out on these parameters when designing social protection strategies aimed at reducing extreme poverty. 

From the point of view of development economics, previous empirical studies are in line with our findings.  Furthermore, our work presents an alternative approach to analyse cash transfer programmes and may represent a point of departure for applying knowledge of another discipline, such as actuarial science, in development economics.

It is important to highlight some of the limitations of our study. For example, due to the construction of the model, our analysis does not capture the direct effect of an UCT programme on a household\rq s consumption. Recently,  \cite{Article:Habimana2021} show how Rwanda's UCT programme (VUP-Direct Support) increases a household\rq s total and food consumption. In the same way, in its current form, the capital model is unable to incorporate the rationale behind conditional cash transfer (CCT) programmes, as it does not track any beneficiary actions such as: enrollment and attendance of children and adolescents in school, use of health services and uptake of food and nutritional supplements \citep{Article:Cruz2017}. Alternative versions of the proposed model should address these issues.

Finally, future research should also consider the cost of an UCT programme. This cost could be estimated, for instance, by computing the total expected discounted value of capital cash transfers made to a household. This concept would be analogous to other well-known quantities previously studied in ruin theory, such as the expected discounted capital injections \citep{Article:Albrecher2014}. These quantities could, for example, be useful for estimating the required capital cash transfer rate $c_{{\scaleto{T}{4pt}}}$ and capital barrier level $B$ such that, for a given social protection budget, the trapping probability or probability of extreme poverty is minimised.

\setcitestyle{numbers} 

\bibliographystyle{chicago} 

\bibliography{main}

\begin{thebibliography}{}

\bibitem[\protect\citeauthoryear{Abramowitz and Stegun}{Abramowitz and
  Stegun}{1972}]{Book:Abramowitz1964}
Abramowitz, M. and I.~A. Stegun (1972).
\newblock {\em Handbook of Mathematical Functions with Formulas, Graphs, and
  Mathematical Tables}.
\newblock Washington, D.C.: U.S. Department of Commerce.

\bibitem[\protect\citeauthoryear{Adams}{Adams}{2003}]{Book:Adams2003}
Adams, R.~H. (2003).
\newblock {\em Economic Growth, Inequality and Poverty: Findings from a New
  Data Set}.
\newblock Number 2972 in Policy Research Working Paper. Washington, D.C.: World
  Bank Publications.

\bibitem[\protect\citeauthoryear{Aker}{Aker}{2013}]{Article:Aker2013}
Aker, J.~C. (2013).
\newblock {Cash or Coupons? Testing the Impacts of Cash Versus Vouchers in the
  Democratic Republic of Congo}.
\newblock {\em Center for Global Development Working Paper\/}~(320).

\bibitem[\protect\citeauthoryear{Albrecher, Cheung, and Thonhauser}{Albrecher
  et~al.}{2013}]{Article:Albrecher2013a}
Albrecher, H., E.~C.~K. Cheung, and S.~Thonhauser (2013).
\newblock {Randomized Observation Periods for the Compound Poisson Risk Model:
  The Discounted Penalty Function}.
\newblock {\em Scandinavian Actuarial Journal\/}~(6), 424--452.

\bibitem[\protect\citeauthoryear{Albrecher, Gerber, and Shiu}{Albrecher
  et~al.}{2011}]{Article:Albrecher2011}
Albrecher, H., H.~U. Gerber, and E.~S.~W. Shiu (2011).
\newblock {The Optimal Dividend Barrier in the Gamma–Omega Model}.
\newblock {\em European Actuarial Journal\/}~{\em 1\/}(1), 43--55.

\bibitem[\protect\citeauthoryear{Albrecher and Hipp}{Albrecher and
  Hipp}{2007}]{Article:Albrecher2007}
Albrecher, H. and C.~Hipp (2007).
\newblock {Lundberg\rq s Risk Process with Tax}.
\newblock {\em Blätter der DGVFM\/}~{\em 28\/}(1), 13--28.

\bibitem[\protect\citeauthoryear{Albrecher and Ivanovs}{Albrecher and
  Ivanovs}{2014}]{Article:Albrecher2014}
Albrecher, H. and J.~Ivanovs (2014).
\newblock {Power Identities for Lévy Risk Models Under Taxation and Capital
  Injections}.
\newblock {\em Stochastic Systems\/}~{\em 4\/}(1), 157--172.

\bibitem[\protect\citeauthoryear{Albrecher and Lautscham}{Albrecher and
  Lautscham}{2013}]{Article:Albrecher2013b}
Albrecher, H. and V.~Lautscham (2013).
\newblock {From Ruin to Bankruptcy for Compound Poisson Surplus Processes}.
\newblock {\em ASTIN Bulletin\/}~{\em 43\/}(2), 213--243.

\bibitem[\protect\citeauthoryear{Ambler and De~Brauw}{Ambler and
  De~Brauw}{2017}]{Book:Ambler2017}
Ambler, K. and A.~De~Brauw (2017).
\newblock {\em {The Impacts of Cash Transfers on Women\rq s Empowerment:
  Learning from Pakistan\rq s BISP Program}}.
\newblock Number 1702 in Social Protection and Labor Discssion. Washington,
  D.C.: World Bank Group.

\bibitem[\protect\citeauthoryear{Asmussen and Albrecher}{Asmussen and
  Albrecher}{2010}]{Book:Asmussen2010}
Asmussen, S. and H.~Albrecher (2010).
\newblock {\em Ruin Probabilities}.
\newblock Singapore: World Scientific.

\bibitem[\protect\citeauthoryear{Aza{\"\i}s and Genadot}{Aza{\"\i}s and
  Genadot}{2015}]{Article:Azais2015}
Aza{\"\i}s, R. and A.~Genadot (2015).
\newblock {Semi-Parametric Inference for the Absorption Features of a
  Growth-Fragmentation Model}.
\newblock {\em TEST\/}~{\em 24\/}(2), 341--360.

\bibitem[\protect\citeauthoryear{Baird, Ferreira, Özler, and Woolcock}{Baird
  et~al.}{2014}]{Article:Baird2014}
Baird, S., F.~H.~G. Ferreira, B.~Özler, and M.~Woolcock (2014).
\newblock {Conditional, Unconditional and Everything in Between: A Systematic
  Review of the Effects of Cash Transfer Programmes on Schooling Outcomes}.
\newblock {\em Journal of Development Effectiveness\/}~{\em 6\/}(1), 1--43.

\bibitem[\protect\citeauthoryear{Bjerk}{Bjerk}{2007}]{Article:Bjerk2007}
Bjerk, D. (2007).
\newblock {Measuring the Relationship Between Youth Criminal Participation and
  Household Economic Resources}.
\newblock {\em Journal of Quantitative Criminology\/}~{\em 23\/}(1), 23--39.

\bibitem[\protect\citeauthoryear{Blattman and Niehaus}{Blattman and
  Niehaus}{2014}]{Article:Blattman2014}
Blattman, C. and P.~Niehaus (2014).
\newblock {Show Them The Money: Why Giving Cash Helps Alleviate Poverty}.
\newblock {\em Foreign Affairs\/}~{\em 93\/}(3), 117--126.

\bibitem[\protect\citeauthoryear{Brooks-Gunn and Duncan}{Brooks-Gunn and
  Duncan}{1997}]{Article:Brooks1997}
Brooks-Gunn, J. and G.~J. Duncan (1997).
\newblock {The Effects of Poverty on Children}.
\newblock {\em The Future of Children\/}~{\em 7\/}(2), 55--71.

\bibitem[\protect\citeauthoryear{Case, Lubotsky, and Paxson}{Case
  et~al.}{2002}]{Article:Case2002}
Case, A., D.~Lubotsky, and C.~Paxson (2002).
\newblock {Economic Status and Health in Childhood: The Origins of the
  Gradient}.
\newblock {\em American Economic Review\/}~{\em 92\/}(5), 1308--1334.

\bibitem[\protect\citeauthoryear{{Children\rq s Defense Fund
  (U.S.)}}{{Children\rq s Defense Fund
  (U.S.)}}{1994}]{Book:ChildrensDefenseFund1994}
{Children\rq s Defense Fund (U.S.)} (1994).
\newblock {\em Wasting America\rq s Future: The Children\rq s Defense Fund
  Report on the Costs of Child Poverty}.
\newblock Boston: Beacon Press.

\bibitem[\protect\citeauthoryear{Cruz, Moura, and Soares~Neto}{Cruz
  et~al.}{2017}]{Article:Cruz2017}
Cruz, R. C. d.~S., L.~B. A.~d. Moura, and J.~J. Soares~Neto (2017).
\newblock {Conditional Cash Transfers and the Creation of Equal Opportunities
  of Health for Children in Low and Middle-Income Countries: A Literature
  Review}.
\newblock {\em International Journal for Equity in Health\/}~{\em 16\/}(1),
  161.

\bibitem[\protect\citeauthoryear{Cui and Nguyen}{Cui and
  Nguyen}{2016}]{Article:Cui2016}
Cui, Z. and D.~Nguyen (2016).
\newblock {Omega Diffusion Risk Model with Surplus-Dependent Tax and Capital
  Injections}.
\newblock {\em Insurance: Mathematics and Economics\/}~{\em 68}, 150--161.

\bibitem[\protect\citeauthoryear{Davis}{Davis}{1984}]{Article:Davis1984}
Davis, M. H.~A. (1984).
\newblock {Piecewise-Deterministic Markov Processes: A General Class of
  Non-Diffusion Stochastic Models}.
\newblock {\em Journal of the Royal Statistical Society: Series B
  (Methodological)\/}~{\em 46\/}(3), 353--388.

\bibitem[\protect\citeauthoryear{Duflo}{Duflo}{2012}]{Article:Duflo2012}
Duflo, E. (2012).
\newblock {Women Empowerment and Economic Development}.
\newblock {\em Journal of Economic Literature\/}~{\em 50\/}(4), 1051--1079.

\bibitem[\protect\citeauthoryear{Emran, Robano, and Smith}{Emran
  et~al.}{2014}]{Article:Emran2014}
Emran, M.~S., V.~Robano, and S.~C. Smith (2014).
\newblock {Assessing the Frontiers of Ultrapoverty Reduction: Evidence from
  Challenging the Frontiers of Poverty Reduction/Targeting the Ultra-Poor, an
  Innovative Program in Bangladesh}.
\newblock {\em Economic Development and Cultural Change\/}~{\em 62\/}(2),
  339--380.

\bibitem[\protect\citeauthoryear{Flores-Contró}{Flores-Contró}{2024}]{Article:Flores-Contro2024}
Flores-Contró, J.~M. (2024).
\newblock {The Gerber-Shiu Expected Discounted Penalty Function: An Application
  to Poverty Trapping}.
\newblock {\em Working Paper\/}.

\bibitem[\protect\citeauthoryear{Flores-Contró, Henshaw, Loke, Arnold, and
  Constantinescu}{Flores-Contró et~al.}{2021}]{Article:Flores-Contro2021}
Flores-Contró, J.~M., K.~Henshaw, S.~H. Loke, S.~Arnold, and C.~D.
  Constantinescu (2021).
\newblock {Subsidising Inclusive Insurance to Reduce Poverty}.
\newblock {\em Working Paper\/}.

\bibitem[\protect\citeauthoryear{Gao and He}{Gao and
  He}{2019}]{Article:Gao2019}
Gao, Z. and J.~He (2019).
\newblock {The Gerber-Shiu Function for the Compound Poisson Omega Model with a
  Three-Step Premium Rate}.
\newblock {\em Communications in Statistics - Theory and Methods\/}~{\em
  48\/}(24), 6019--6037.

\bibitem[\protect\citeauthoryear{Gao, He, Zhao, and Wang}{Gao
  et~al.}{2022}]{Article:Gao2022}
Gao, Z., J.~He, Z.~Zhao, and B.~Wang (2022).
\newblock {Omega Model for a Jump-Diffusion Process with a Two-Step Premium
  Rate and a Threshold Dividend Strategy}.
\newblock {\em Methodology and Computing in Applied Probability\/}~{\em
  24\/}(1), 233--258.

\bibitem[\protect\citeauthoryear{Garcia and Moore}{Garcia and
  Moore}{2012}]{Book:WorldBank2012}
Garcia, M. and C.~M.~T. Moore (2012).
\newblock {\em {The Cash Dividend: The Rise of Cash Transfer Programs in
  Sub-Saharan Africa}}.
\newblock Directions in Development - Human Development. Washington, D.C.: The
  World Bank.

\bibitem[\protect\citeauthoryear{Gauss}{Gauss}{1866}]{Article:Gauss1812}
Gauss, C.~F. (1866).
\newblock {Disquisitiones Generales Circa Seriem Infinitam}.
\newblock {\em Ges. Werke Gottingen\/}~{\em 2}, 437--45.

\bibitem[\protect\citeauthoryear{Gentilini}{Gentilini}{2022}]{Book:Gentilini2022}
Gentilini, U. (2022).
\newblock {\em Cash Transfers in Pandemic Times: Evidence, Practices, and
  Implications from the Largest Scale Up in History}.
\newblock Washington, D.C.: World Bank.

\bibitem[\protect\citeauthoryear{Gerber, Shiu, and Yang}{Gerber
  et~al.}{2012}]{Article:Gerber2012}
Gerber, H.~U., E.~Shiu, and H.~Yang (2012).
\newblock {The Omega Model: From Bankruptcy to Occupation Times in the Red}.
\newblock {\em European Actuarial Journal\/}~{\em 2\/}(2), 259--272.

\bibitem[\protect\citeauthoryear{Gerber and Shiu}{Gerber and
  Shiu}{1998}]{Article:Gerber1998}
Gerber, H.~U. and E.~S.~W. Shiu (1998).
\newblock {On the Time Value of Ruin}.
\newblock {\em North American Actuarial Journal\/}~{\em 2\/}(1), 48--72.

\bibitem[\protect\citeauthoryear{Habimana, Haughton, Nkurunziza, and
  Haughton}{Habimana et~al.}{2021}]{Article:Habimana2021}
Habimana, D., J.~Haughton, J.~Nkurunziza, and D.~M.~A. Haughton (2021).
\newblock {Measuring the Impact of Unconditional Cash Transfers on Consumption
  and Poverty in Rwanda}.
\newblock {\em World Development Perspectives\/}~{\em 23}, 100341.

\bibitem[\protect\citeauthoryear{Handa and Davis}{Handa and
  Davis}{2006}]{Article:Handa2006}
Handa, S. and B.~Davis (2006).
\newblock {The Experience of Conditional Cash Transfers in Latin America and
  the Caribbean}.
\newblock {\em Development Policy Review\/}~{\em 24\/}(5), 513--536.

\bibitem[\protect\citeauthoryear{Handa, Natali, Seidenfeld, Tembo, and
  Davis}{Handa et~al.}{2016}]{Book:Handa2016}
Handa, S., L.~Natali, D.~Seidenfeld, G.~Tembo, and B.~Davis (2016).
\newblock {\em {Can Unconditional Cash Transfers Lead to Sustainable Poverty
  Reduction? Evidence from Two Government-Led Programmes in Zambia}}.
\newblock Number 2016-21 in Innocenti Working Paper. Florence: UNICEF Office of
  Research.

\bibitem[\protect\citeauthoryear{Haushofer and Shapiro}{Haushofer and
  Shapiro}{2016}]{Article:Haushofer2016}
Haushofer, J. and J.~Shapiro (2016).
\newblock {The Short-Term Impact of Unconditional Cash Transfers to the Poor:
  Experimental Evidence from Kenya}.
\newblock {\em The Quarterly Journal of Economics\/}~{\em 131\/}(4),
  1973--2042.

\bibitem[\protect\citeauthoryear{He, Gao, and Wang}{He
  et~al.}{2019}]{Article:He2019}
He, J., Z.~Gao, and B.~Wang (2019).
\newblock {Omega Model for a Jump–Diffusion Process with a Two-Step Premium
  Rate}.
\newblock {\em Journal of the Korean Statistical Society\/}~{\em 48\/}(3),
  426--438.

\bibitem[\protect\citeauthoryear{He, Kawai, Shimizu, and Yamazaki}{He
  et~al.}{2023}]{Article:He2023}
He, Y., R.~Kawai, Y.~Shimizu, and K.~Yamazaki (2023).
\newblock {The Gerber-Shiu Discounted Penalty Function: A Review from Practical
  Perspectives}.
\newblock {\em Insurance: Mathematics and Economics\/}~{\em 109}, 1--28.

\bibitem[\protect\citeauthoryear{Henshaw, Ramirez, Flores-Contró, Thomann,
  Loke, and Constantinescu}{Henshaw et~al.}{2023}]{Article:Henshaw2023}
Henshaw, K., J.~M. Ramirez, J.~M. Flores-Contró, E.~A. Thomann, S.~H. Loke,
  and C.~D. Constantinescu (2023).
\newblock {On the Impact of Insurance on Households Susceptible to Random
  Proportional Losses: An Analysis of Poverty Trapping}.
\newblock {\em Working Paper\/}.

\bibitem[\protect\citeauthoryear{Jensen, Barrett, and Mude}{Jensen
  et~al.}{2017}]{Article:Jensen2017}
Jensen, N.~D., C.~B. Barrett, and A.~G. Mude (2017).
\newblock {Cash Transfers and Index Insurance: A Comparative Impact Analysis
  from Northern Kenya}.
\newblock {\em Journal of Development Economics\/}~{\em 129}, 14--28.

\bibitem[\protect\citeauthoryear{Jolliffe, Mahler, Lakner, Atamanov, and
  Tetteh~Baah}{Jolliffe et~al.}{2022}]{Book:Jolliffe2022}
Jolliffe, D.~M., D.~G. Mahler, C.~Lakner, A.~Atamanov, and S.~K. Tetteh~Baah
  (2022).
\newblock {\em {Assessing the Impact of the 2017 PPPs on the International
  Poverty Line and Global Poverty}}.
\newblock Number WPS 9941 in Policy Research Working Paper. Washington, D.C.:
  World Bank Group.

\bibitem[\protect\citeauthoryear{Kaszubowski}{Kaszubowski}{2019}]{Article:Kaszubowski2019}
Kaszubowski, A. (2019).
\newblock {Omega Bankruptcy for Different Lévy Models}.
\newblock {\em Śląski Przegląd Statystyczny\/}~{\em 23\/}(17), 31--57.

\bibitem[\protect\citeauthoryear{Kovacevic and Pflug}{Kovacevic and
  Pflug}{2011}]{Article:Kovacevic2011}
Kovacevic, R.~M. and G.~C. Pflug (2011).
\newblock {Does Insurance Help to Escape the Poverty Trap? — A Ruin Theoretic
  Approach}.
\newblock {\em Journal of Risk and Insurance\/}~{\em 78\/}(4), 1003--1027.

\bibitem[\protect\citeauthoryear{Kyprianou}{Kyprianou}{2013}]{Book:Kyprianou2013}
Kyprianou, A.~E. (2013).
\newblock {\em Gerber–Shiu Risk Theory}.
\newblock Switzerland: Springer International Publishing.

\bibitem[\protect\citeauthoryear{McLaughlin and Rank}{McLaughlin and
  Rank}{2018}]{Article:McLaughlin2018}
McLaughlin, M. and M.~R. Rank (2018).
\newblock {Estimating the Economic Cost of Childhood Poverty in the United
  States}.
\newblock {\em Social Work Research\/}~{\em 42\/}(2), 73--83.

\bibitem[\protect\citeauthoryear{Pega, Pabayo, Benny, Lee, Lhachimi, and
  Liu}{Pega et~al.}{2022}]{Article:Pega2022}
Pega, F., R.~Pabayo, C.~Benny, E.~Y. Lee, S.~K. Lhachimi, and S.~Y. Liu (2022).
\newblock {Unconditional Cash Transfers for Reducing Poverty and
  Vulnerabilities: Effect on Use of Health Services and Health Outcomes in
  Low‐ and Middle‐Income Countries}.
\newblock {\em Cochrane Database of Systematic Reviews\/}~(3).

\bibitem[\protect\citeauthoryear{Rank, Eppard, and Bullock}{Rank
  et~al.}{2021}]{Book:Rank2021}
Rank, M.~R., L.~M. Eppard, and H.~E. Bullock (2021).
\newblock {\em {Poorly Understood: What America Gets Wrong About Poverty}}.
\newblock New York: Oxford University Press.

\bibitem[\protect\citeauthoryear{Rank, Hirschl, and Foster}{Rank
  et~al.}{2014}]{Book:Rank2014}
Rank, M.~R., T.~A. Hirschl, and K.~A. Foster (2014).
\newblock {\em {Chasing the American Dream: Understanding What Shapes Our
  Fortunes}}.
\newblock New York: Oxford University Press.

\bibitem[\protect\citeauthoryear{Ravallion}{Ravallion}{2016}]{Book:Ravallion2016}
Ravallion, M. (2016).
\newblock {\em {The Economics of Poverty: History, Measurement, and Policy}}.
\newblock New York: Oxford University Press.

\bibitem[\protect\citeauthoryear{Seaborn}{Seaborn}{1991}]{Book:Seaborn1991}
Seaborn, J.~B. (1991).
\newblock {\em Hypergeometric Functions and Their Applications}.
\newblock New York: Springer-Verlag.

\bibitem[\protect\citeauthoryear{Slater}{Slater}{1960}]{Book:Slater1960}
Slater, L.~J. (1960).
\newblock {\em Confluent Hypergeometric Functions}.
\newblock New York: Cambridge University Press.

\bibitem[\protect\citeauthoryear{Sumner, Hoy, and Ortiz-Juarez}{Sumner
  et~al.}{2020}]{Article:Hoy2020}
Sumner, A., C.~Hoy, and E.~Ortiz-Juarez (2020).
\newblock {Estimates of the Impact of COVID-19 on Global Poverty}.
\newblock {\em UNU-WIDER\/}~(43).

\bibitem[\protect\citeauthoryear{{United Nations}}{{United
  Nations}}{2015}]{Book:UnitedNations2015}
{United Nations} (2015).
\newblock {\em {Transforming Our World: The 2030 Agenda for Sustainable
  Development}}.
\newblock Number A/RES/70/1. New York: United Nations.

\bibitem[\protect\citeauthoryear{{United Nations and World Summit for Social
  Development}}{{United Nations and World Summit for Social
  Development}}{1996}]{Book:UnitedNations1996}
{United Nations and World Summit for Social Development} (1996).
\newblock {\em {Report of the World Summit for Social Development: Copenhagen,
  6-12 March 1995}}.
\newblock Number 96.IV.8 in A/CONF. 166. New York: United Nations Publication.

\bibitem[\protect\citeauthoryear{Vera-Cossio, Hoffmann, Pecha, Gallego,
  Stampini, Vargas, Medina, and Álvarez}{Vera-Cossio
  et~al.}{2023}]{Book:Vera-Cossio2023}
Vera-Cossio, D.~A., B.~Hoffmann, C.~Pecha, J.~Gallego, M.~Stampini, D.~Vargas,
  M.~P. Medina, and E.~Álvarez (2023).
\newblock {\em {Re-thinking Social Protection: From Poverty Alleviation to
  Building Resilience in Middle-Income Households}}.
\newblock Number IDB-WP-1412 in IDB Working Paper. Inter-American Development
  Bank.

\bibitem[\protect\citeauthoryear{Wang, Wang, and Zhang}{Wang
  et~al.}{2016}]{Article:Wang2016}
Wang, X., W.~Wang, and C.~Zhang (2016).
\newblock {Ornstein-Uhlenback Type Omega Model}.
\newblock {\em Frontiers of Mathematics in China\/}~{\em 11\/}(3), 737--751.

\bibitem[\protect\citeauthoryear{{World Bank}}{{World
  Bank}}{2018}]{Book:WorldBank2018}
{World Bank} (2018).
\newblock {\em {Poverty and Shared Prosperity 2018: Piecing Together the
  Poverty Puzzle}}.
\newblock Washington, D.C.: World Bank Group.

\end{thebibliography}

\setcitestyle{authoryear}


\section*{Appendices}
\addcontentsline{toc}{section}{Appendices}
\renewcommand{\thesubsection}{\Alph{subsection}}
\setcounter{subsection}{0}

\numberwithin{equation}{subsubsection}

\subsection{Mathematical Proofs}\label{Appendix A: Mathematical Proofs}


\subsubsection{Proof of Theorem \ref{WhenandHowHouseholdsBecomePoor?-Section3-Theorem1}} \label{ProofofTheorem3.1}

For $x\geq B$, the capital immediately before the first capital loss is $h_{r}(t,x) = (x-x^{*})e^{rt}+x^{*}$. Hence, by conditioning on the time and the remaining proportion of the first capital loss and discounting the expected values to time 0 at the force of interest $\delta$, when $x \geq B$ we obtain

\vspace{0.3cm}

{\allowdisplaybreaks
\begin{align}
	m^{\scaleto{\text{ {\fontfamily{qcr}\selectfont P}}}{2.5pt}}_{\delta,u}(x)&=\int^{\infty}_{0} \lambda e^{-(\lambda + \delta)t} \left[\int^{1}_{B/h_{r}(t,x)}m^{\scaleto{\text{ {\fontfamily{qcr}\selectfont P}}}{2.5pt}}_{\delta,u}(h_{r}(t,x)\cdot z)dG_{Z}(z) +\int^{B/h_{r}(t,x)}_{x^{*}/h_{r}(t,x)}m^{\scaleto{\text{ {\fontfamily{qcr}\selectfont P}}}{2.5pt}}_{\delta,l}(h_{r}(t,x)\cdot z)dG_{Z}(z) \right. \\ \\
	&\left.+ \int^{x^{*}/h_{r}(t,x)}_{0}w^{\scaleto{\text{ {\fontfamily{qcr}\selectfont P}}}{2.5pt}}(h_{r}(t,x) - x^{*}, x^{*}-h_{r}(t,x)\cdot z)dG_{Z}(z) \right]dt.
	\label{Appendix A: Mathematical Proofs-Equation1}
\end{align}
}

\vspace{0.3cm}

The above equation for $m^{\scaleto{\text{ {\fontfamily{qcr}\selectfont P}}}{2.5pt}}_{\delta,u}(x)$ involves $m^{\scaleto{\text{ {\fontfamily{qcr}\selectfont P}}}{2.5pt}}_{\delta,l}(x)$ for $x^{*} \leq x < B$. When the initial capital is below the capital barrier level $B$, the capital growth is driven by both the capital growth rate $r$ and the capital transfer rate $c_{{\scaleto{T}{4pt}}}$ before the capital returns to the capital barrier level $B$. Thus, for $x^{*}\leq x < B$, let $\tau_{{{\scaleto{B}{3pt}}}}:=\tau_{{{\scaleto{B}{3pt}}}}(x)$ be the solution to 

\vspace{0.3cm}

{\allowdisplaybreaks
\begin{align}
	h_{r-c_{{\scaleto{T}{2.5pt}}}}(t,x)=(x+x^{**})e^{(r-c_{{\scaleto{T}{2.5pt}}})t}-x^{**}=B,
	\label{Appendix A: Mathematical Proofs-Equation2}
\end{align}
}

\vspace{0.3cm}

with $x^{**}=\frac{c_{{\scaleto{T}{2.5pt}}}B-rx^{*}}{r-c_{{\scaleto{T}{2.5pt}}}}$. Namely, $\tau_{{{\scaleto{B}{3pt}}}}:=\tau_{{{\scaleto{B}{3pt}}}}(x)=\frac{1}{r-c_{{\scaleto{T}{2.5pt}}}}ln\left(\frac{B+x^{**}}{x+x^{**}}\right)$, which is the time when the capital returns to the capital barrier level $B$ if no capital loss occurs prior to time $\tau_{{{\scaleto{B}{3pt}}}}$. Furthermore, $h_{r-c_{{\scaleto{T}{2.5pt}}}}(t,x)<B$ for $t<\tau_{{{\scaleto{B}{3pt}}}}$ and $h_{r-c_{{\scaleto{T}{2.5pt}}}}(\tau_{{{\scaleto{B}{3pt}}}},x)=B$. Moreover, $h_{r-c_{{\scaleto{T}{2.5pt}}}}(t,x)$ is the capital at time $t \leq \tau_{{{\scaleto{B}{3pt}}}}$ if no capital loss occurs prior to time $\tau_{{{\scaleto{B}{3pt}}}}$. Thus, by conditioning on the time and the remaining proportion of the first capital loss and discounting the expected values to time 0 at the force of interest $\delta$, when $x^{*} \leq x < B$ we obtain

\vspace{0.3cm}

{\allowdisplaybreaks
\begin{align}
	m^{\scaleto{\text{ {\fontfamily{qcr}\selectfont P}}}{2.5pt}}_{\delta,l}(x)&=\int^{\tau_{{{\scaleto{B}{2pt}}}}}_{0}\lambda e^{-(\lambda+\delta)t}\left[\int^{1}_{x^{*}/h_{\scaleto{r-c_{{\scaleto{T}{2pt}}}}{4pt}}(t,x)}m^{\scaleto{\text{ {\fontfamily{qcr}\selectfont P}}}{2.5pt}}_{\delta,l}(h_{r-c_{{\scaleto{T}{2.5pt}}}}(t,x) \cdot z)dG_{Z}(z) \right.\\ \\
	&\left.+ \int^{x^{*}/h_{\scaleto{r-c_{{\scaleto{T}{2pt}}}}{4pt}}(t,x)}_{0} w^{\scaleto{\text{ {\fontfamily{qcr}\selectfont P}}}{2.5pt}}\left(h_{r-c_{{\scaleto{T}{2.5pt}}}}(t,x)-x^{*},x^{*}-h_{r-c_{{\scaleto{T}{2.5pt}}}}(t,x) \cdot z\right)dG_{Z}(z)\right]dt \\ \\
	& + \int^{\infty}_{\tau_{{{\scaleto{B}{2pt}}}}}\lambda e^{-(\lambda+\delta)t}\left[\int^{1}_{B/h_{r}(t-\tau_{{{\scaleto{B}{2pt}}}},B)}m^{\scaleto{\text{ {\fontfamily{qcr}\selectfont P}}}{2.5pt}}_{\delta,u}(h_{r}(t-\tau_{{{\scaleto{B}{3pt}}}},B)\cdot z)dG_{Z}(z) \right. \\ \\
	& + \int^{B/h_{r}(t-\tau_{{{\scaleto{B}{2pt}}}},B)}_{x^{*}/h_{r}(t-\tau_{{{\scaleto{B}{2pt}}}},B)}m^{\scaleto{\text{ {\fontfamily{qcr}\selectfont P}}}{2.5pt}}_{\delta,l}(h_{r}(t-\tau_{{{\scaleto{B}{3pt}}}},B)\cdot z)dG_{Z}(z) \\ \\
	& \left. + \int^{x^{*}/h_{r}(t-\tau_{{{\scaleto{B}{2pt}}}},B)}_{0}w^{\scaleto{\text{ {\fontfamily{qcr}\selectfont P}}}{2.5pt}}(h_{r}(t-\tau_{{{\scaleto{B}{3pt}}}},B) - x^{*}, x^{*}- h_{r}(t-\tau_{{{\scaleto{B}{3pt}}}},B)\cdot z)dG_{Z}(z) \right]dt.
	\label{Appendix A: Mathematical Proofs-Equation3}
\end{align}
}

\vspace{0.3cm}

Now, changing variables $u=h_{r}(t,x)$ in \eqref{Appendix A: Mathematical Proofs-Equation1}, we obtain \eqref{WhenandHowHouseholdsBecomePoor?-Section3-Equation3}. Moreover, first changing variables $u=h_{r-c_{{\scaleto{T}{2.5pt}}}}(t,x)$ in the integrals with respect to $t$ from $0$ to $\tau_{{{\scaleto{B}{3pt}}}}$ in \eqref{Appendix A: Mathematical Proofs-Equation3}, and then changing variables $v=h_{r}(t-\tau_{{{\scaleto{B}{3pt}}}},B)$ in the integrals with respect to $t$ from $\tau_{{{\scaleto{B}{3pt}}}}$ to $\infty$ in \eqref{Appendix A: Mathematical Proofs-Equation3}, we obtain \eqref{WhenandHowHouseholdsBecomePoor?-Section3-Equation4}.


\subsubsection{Proof of Proposition \ref{TheTrappingTime-Subsection21-Proposition1}} \label{ProofofProposition3.1.1}

When $Z_{i}\sim Beta(\alpha, 1)$, i.e. $g_{Z}(z)=\alpha z^{\alpha - 1}\mathbbm{1}_{\{0 < z < 1\}}$ with $\alpha>0$, equation \eqref{TheTrappingTime-Subsection31-Equation1} and \eqref{TheTrappingTime-Subsection31-Equation2} can be written such that when $x \geq B$,

\vspace{0.3cm}

{\allowdisplaybreaks
\begin{align}
        0&= r(x-x^{*})m'^{\scaleto{\text{ {\fontfamily{qcr}\selectfont P}}}{2.5pt}}_{\delta,u}(x)
        -(\lambda + \delta)m^{\scaleto{\text{ {\fontfamily{qcr}\selectfont P}}}{2.5pt}}_{\delta,u}(x) +\lambda \left[ \int^{1}_{B/x}m^{\scaleto{\text{ {\fontfamily{qcr}\selectfont P}}}{2.5pt}}_{\delta,u}(x\cdot z)\alpha z^{\alpha - 1} dz \right. \\ \\ 
        & \left. + \int^{B/x}_{x^{*}/x} m^{\scaleto{\text{ {\fontfamily{qcr}\selectfont P}}}{2.5pt}}_{\delta,l}(x\cdot z)\alpha z^{\alpha - 1} dz + \left(\frac{x^{*}}{x}\right)^{\alpha} \right],  \label{Appendix A: Mathematical Proofs-Equation4}
\end{align}
}
        
\vspace{0.3cm}

and when $x^{*} \leq x <  B$,

\vspace{0.3cm}

{\allowdisplaybreaks
\begin{align}
        0&= (r - c_{{\scaleto{T}{4pt}}})(x+x^{**}) m'^{\scaleto{\text{ {\fontfamily{qcr}\selectfont P}}}{2.5pt}}_{\delta,l}(x) - (\lambda + \delta) m^{\scaleto{\text{ {\fontfamily{qcr}\selectfont P}}}{2.5pt}}_{\delta,l}(x) + \lambda \left[\int^{1}_{x^{*}/x}m^{\scaleto{\text{ {\fontfamily{qcr}\selectfont P}}}{2.5pt}}_{\delta,l}(x\cdot z) \alpha z^{\alpha - 1} dz + \left(\frac{x^{*}}{x}\right)^{\alpha} \right].\\ \\
        \label{Appendix A: Mathematical Proofs-Equation5}
\end{align}
}

\vspace{0.3cm}

Applying the operator $\frac{d}{dx}$ to both sides of \eqref{Appendix A: Mathematical Proofs-Equation4} and \eqref{Appendix A: Mathematical Proofs-Equation5}, together with a number of algebraic manipulations, yields to the following second order Ordinary Differential Equations (ODEs),

\vspace{0.3cm}

{\allowdisplaybreaks
\begin{align}
        x \geq B: 0 &= r(x^{2}-xx^{*})m''^{\scaleto{\text{ {\fontfamily{qcr}\selectfont P}}}{2.5pt}}_{\delta,u}(x)+\left[(r(1+\alpha)-\delta-\lambda)x-r\alpha x^{*}\right]m'^{\scaleto{\text{ {\fontfamily{qcr}\selectfont P}}}{2.5pt}}_{\delta,u}(x)-\alpha \delta m^{\scaleto{\text{ {\fontfamily{qcr}\selectfont P}}}{2.5pt}}_{\delta,u}(x)\quad \label{Appendix A: Mathematical Proofs-Equation6} 
\end{align}
}
        
\vspace{0.3cm}

and

\vspace{0.3cm}

{\allowdisplaybreaks
\begin{align}
        x^{*} \leq x <  B: 0&= (r - c_{{\scaleto{T}{4pt}}})(x^{2}+xx^{**}) m''^{\scaleto{\text{ {\fontfamily{qcr}\selectfont P}}}{2.5pt}}_{\delta,l}(x) \\ \\ 
        &+ \left[ \left(\left(r - c_{{\scaleto{T}{4pt}}}\right) \left( 1 + \alpha\right) - \delta - \lambda \right) x + \alpha \left(r - c_{{\scaleto{T}{4pt}}}\right)x^{**} \right] m'^{\scaleto{\text{ {\fontfamily{qcr}\selectfont P}}}{2.5pt}}_{\delta,l}(x) - \alpha\delta m^{\scaleto{\text{ {\fontfamily{qcr}\selectfont P}}}{2.5pt}}_{\delta,l}(x).
        \label{Appendix A: Mathematical Proofs-Equation7}
\end{align}
}

\vspace{0.3cm}

Letting $f^{\scaleto{\text{ {\fontfamily{qcr}\selectfont P}}}{2.5pt}}_{i}(y^{\scaleto{\text{ {\fontfamily{qcr}\selectfont P}}}{2.5pt}}_{i}):=m^{\scaleto{\text{ {\fontfamily{qcr}\selectfont P}}}{2.5pt}}_{\delta,i}(x)$  for $i=u,l$, such that $y^{\scaleto{\text{ {\fontfamily{qcr}\selectfont P}}}{2.5pt}}_{u}$ and $y^{\scaleto{\text{ {\fontfamily{qcr}\selectfont P}}}{2.5pt}}_{l}$ are associated with the change of variables $y^{\scaleto{\text{ {\fontfamily{qcr}\selectfont P}}}{2.5pt}}_{u}:=y^{\scaleto{\text{ {\fontfamily{qcr}\selectfont P}}}{2.5pt}}_{u}(x)=\frac{x}{x^{*}}$ and $y^{\scaleto{\text{ {\fontfamily{qcr}\selectfont P}}}{2.5pt}}_{l}:=y^{\scaleto{\text{ {\fontfamily{qcr}\selectfont P}}}{2.5pt}}_{l}(x)=-\frac{x}{x^{**}}$, respectively, equations \eqref{Appendix A: Mathematical Proofs-Equation6} and \eqref{Appendix A: Mathematical Proofs-Equation7} reduce to Gauss\rq s Hypergeometric Differential Equation \citep{Book:Slater1960}

\vspace{0.3cm}

{\allowdisplaybreaks
\begin{align}
    0 & = y^{\scaleto{\text{ {\fontfamily{qcr}\selectfont P}}}{2.5pt}}_{i}(1-y^{\scaleto{\text{ {\fontfamily{qcr}\selectfont P}}}{2.5pt}}_{i})\cdot f''^{\scaleto{\text{ {\fontfamily{qcr}\selectfont P}}}{2.5pt}}_{i}(y^{\scaleto{\text{ {\fontfamily{qcr}\selectfont P}}}{2.5pt}}_{i}) + [c^{\scaleto{\text{ {\fontfamily{qcr}\selectfont P}}}{2.5pt}}_{i} - (1+a^{\scaleto{\text{ {\fontfamily{qcr}\selectfont P}}}{2.5pt}}_{i}+b^{\scaleto{\text{ {\fontfamily{qcr}\selectfont P}}}{2.5pt}}_{i})y^{\scaleto{\text{ {\fontfamily{qcr}\selectfont P}}}{2.5pt}}_{i}] f'^{\scaleto{\text{ {\fontfamily{qcr}\selectfont P}}}{2.5pt}}_{i}(y^{\scaleto{\text{ {\fontfamily{qcr}\selectfont P}}}{2.5pt}}_{i}) - a^{\scaleto{\text{ {\fontfamily{qcr}\selectfont P}}}{2.5pt}}_{i}b^{\scaleto{\text{ {\fontfamily{qcr}\selectfont P}}}{2.5pt}}_{i} f^{\scaleto{\text{ {\fontfamily{qcr}\selectfont P}}}{2.5pt}}_{i}(y^{\scaleto{\text{ {\fontfamily{qcr}\selectfont P}}}{2.5pt}}_{i}),
    \label{Appendix A: Mathematical Proofs-Equation8}
\end{align}
}

\vspace{0.3cm}

for $a^{\scaleto{\text{ {\fontfamily{qcr}\selectfont P}}}{2.5pt}}_{l}=\frac{-(\delta + \lambda - \alpha \left(r-c_{{\scaleto{T}{2.5pt}}}\right)) - \sqrt{(\delta + \lambda -\alpha \left(r-c_{{\scaleto{T}{2.5pt}}}\right))^{2}+4 \left(r - c_{{\scaleto{T}{2.5pt}}}\right) \alpha \delta}}{2\left(r - c_{{\scaleto{T}{2.5pt}}}\right)}$, $b^{\scaleto{\text{ {\fontfamily{qcr}\selectfont P}}}{2.5pt}}_{l}=\frac{-(\delta + \lambda - \alpha \left(r- c_{{\scaleto{T}{2.5pt}}}\right)) + \sqrt{(\delta + \lambda -\alpha \left(r - c_{{\scaleto{T}{2.5pt}}}\right))^{2}+4 \left(r - c_{{\scaleto{T}{2.5pt}}}\right) \alpha \delta}}{2\left(r - c_{{\scaleto{T}{2.5pt}}}\right)}$, $a^{\scaleto{\text{ {\fontfamily{qcr}\selectfont P}}}{2.5pt}}_{u}=\frac{-(\delta + \lambda - \alpha r) - \sqrt{(\delta + \lambda -\alpha r)^{2}+4 r \alpha \delta}}{2r}$, $b^{\scaleto{\text{ {\fontfamily{qcr}\selectfont P}}}{2.5pt}}_{u}=\frac{-(\delta + \lambda - \alpha r) + \sqrt{(\delta + \lambda -\alpha r)^{2}+4 r \alpha \delta}}{2r}$ and $c^{\scaleto{\text{ {\fontfamily{qcr}\selectfont P}}}{2.5pt}}_{u}=c^{\scaleto{\text{ {\fontfamily{qcr}\selectfont P}}}{2.5pt}}_{l}=\alpha$ with regular singular points at $y^{\scaleto{\text{ {\fontfamily{qcr}\selectfont P}}}{2.5pt}}_{i}=0, 1, \infty$ (corresponding to $x=-\infty, -x^{**}, 0,x^{*}$ and $\infty$). A general solution of \eqref{Appendix A: Mathematical Proofs-Equation8} in the neighborhood of the singular point $y^{\scaleto{\text{ {\fontfamily{qcr}\selectfont P}}}{2.5pt}}_{i}=\infty$ is given by

\vspace{0.3cm}

\normalsize

{\allowdisplaybreaks
\begin{align}
    f^{\scaleto{\text{ {\fontfamily{qcr}\selectfont P}}}{2.5pt}}_{i}(y^{\scaleto{\text{ {\fontfamily{qcr}\selectfont P}}}{2.5pt}}_{i})&:=m^{\scaleto{\text{ {\fontfamily{qcr}\selectfont P}}}{2.5pt}}_{\delta,i}(x)= A^{\scaleto{\text{ {\fontfamily{qcr}\selectfont P}}}{2.5pt}}_{1,i}{y^{\scaleto{\text{ {\fontfamily{qcr}\selectfont P}}}{2.5pt}}_{i}(x)}^{-a^{\scaleto{\text{ {\fontfamily{qcr}\selectfont P}}}{2.5pt}}_{i}} { }_{2} F_{1}\left(a^{\scaleto{\text{ {\fontfamily{qcr}\selectfont P}}}{2.5pt}}_{i}, a^{\scaleto{\text{ {\fontfamily{qcr}\selectfont P}}}{2.5pt}}_{i}-c^{\scaleto{\text{ {\fontfamily{qcr}\selectfont P}}}{2.5pt}}_{i}+1 ; a^{\scaleto{\text{ {\fontfamily{qcr}\selectfont P}}}{2.5pt}}_{i}-b^{\scaleto{\text{ {\fontfamily{qcr}\selectfont P}}}{2.5pt}}_{i}+1 ; {y^{\scaleto{\text{ {\fontfamily{qcr}\selectfont P}}}{2.5pt}}_{i}(x)}^{-1}\right)\\ \\ 
    &+A^{\scaleto{\text{ {\fontfamily{qcr}\selectfont P}}}{2.5pt}}_{2,i}{y^{\scaleto{\text{ {\fontfamily{qcr}\selectfont P}}}{2.5pt}}_{i}(x)}^{-b^{\scaleto{\text{ {\fontfamily{qcr}\selectfont P}}}{2.5pt}}_{i}} { }_{2} F_{1}\left(b^{\scaleto{\text{ {\fontfamily{qcr}\selectfont P}}}{2.5pt}}_{i}, b^{\scaleto{\text{ {\fontfamily{qcr}\selectfont P}}}{2.5pt}}_{i}-c^{\scaleto{\text{ {\fontfamily{qcr}\selectfont P}}}{2.5pt}}_{i}+1 ; b^{\scaleto{\text{ {\fontfamily{qcr}\selectfont P}}}{2.5pt}}_{i}-a^{\scaleto{\text{ {\fontfamily{qcr}\selectfont P}}}{2.5pt}}_{i}+1 ; {y^{\scaleto{\text{ {\fontfamily{qcr}\selectfont P}}}{2.5pt}}_{i}(x)}^{-1}\right),\\
    \label{Appendix A: Mathematical Proofs-Equation9}
\end{align}
}

\normalsize

\vspace{-0.5cm}

for arbitrary constants $A^{\scaleto{\text{ {\fontfamily{qcr}\selectfont P}}}{2.5pt}}_{1,i},A^{\scaleto{\text{ {\fontfamily{qcr}\selectfont P}}}{2.5pt}}_{2,i} \in \mathbb {R}$ (see for example, equations (15.5.7) and (15.5.8) of \cite{Book:Abramowitz1964}). Here, 

\vspace{0.3cm}

{\allowdisplaybreaks
\begin{align}
    { }_{2} F_{1}(a, b ; c ; z)=\sum_{n=0}^{\infty} \frac{(a)_{n}(b)_{n}}{(c)_{n}} \frac{z^{n}}{n !}
    \label{Appendix A: Mathematical Proofs-Equation10}
\end{align}
}

\vspace{0.3cm}

is Gauss\rq s Hypergeometric Function \citep{Article:Gauss1812} and $(a)_{n}=\frac{\Gamma(a+n)}{\Gamma(n)}$ denotes the Pochhammer symbol \citep{Book:Seaborn1991}. 

To determine the constants $A^{\scaleto{\text{ {\fontfamily{qcr}\selectfont P}}}{2.5pt}}_{1,i}$ and $A^{\scaleto{\text{ {\fontfamily{qcr}\selectfont P}}}{2.5pt}}_{2,i}$ we use the boundary conditions at $x^*$ and at $\infty$. In addition, we use \eqref{WhenandHowHouseholdsBecomePoor?-Section3-Equation6}, \eqref{WhenandHowHouseholdsBecomePoor?-Section3-Equation9} and the differential properties of Gauss\rq s Hypergeometric Function

\vspace{0.3cm}

{\allowdisplaybreaks
\begin{align}
    \frac{d}{d z}{ }_2 F_1(a, b ; c ; z)=\frac{a b}{c}{ }_2 F_1(a+1, b+1 ; c+1 ; z).
   \label{Appendix A: Mathematical Proofs-Equation11}
\end{align}
}

\vspace{0.3cm}

The boundary condition $\lim\limits_{x\to\infty} m^{\scaleto{\text{ {\fontfamily{qcr}\selectfont P}}}{2.5pt}}_{\delta,u}(x) = 0$, by definition of $m^{\scaleto{\text{ {\fontfamily{qcr}\selectfont P}}}{2.5pt}}_{\delta}(x)$ in \eqref{WhenandHowHouseholdsBecomePoor?-Section3-Equation2}, thus implies that $A^{\scaleto{\text{ {\fontfamily{qcr}\selectfont P}}}{2.5pt}}_{1,u}=0$. Moreover, letting $x=x^{*}$ in \eqref{TheTrappingTime-Subsection31-Equation2} yields

\vspace{0.3cm}

{\allowdisplaybreaks
\begin{align}
    m^{\scaleto{\text{ {\fontfamily{qcr}\selectfont P}}}{2.5pt}}_{\delta,l}(x^{*})=\frac{1}{\lambda + \delta}\left[c_{{\scaleto{T}{4pt}}}\left(B - x^{*}\right) m'^{\scaleto{\text{ {\fontfamily{qcr}\selectfont P}}}{2.5pt}}_{\delta,l}(x^{*}) + \lambda\right].
    \label{Appendix A: Mathematical Proofs-Equation14}
\end{align}
}

\vspace{0.3cm}

Hence, this yields to

\footnotesize

\vspace{0.3cm}

{\allowdisplaybreaks
\begin{align}
    A^{\scaleto{\text{ {\fontfamily{qcr}\selectfont P}}}{2.5pt}}_{2,u}&= \left[\lambda y^{\scaleto{\text{ {\fontfamily{qcr}\selectfont P}}}{2.5pt}}_{u}\left(B\right)^{b^{\scaleto{\text{ {\fontfamily{qcr}\selectfont P}}}{2pt}}_{u}}x^{*}y^{\scaleto{\text{ {\fontfamily{qcr}\selectfont P}}}{2.5pt}}_{l}\left(B\right)^{-(a^{\scaleto{\text{ {\fontfamily{qcr}\selectfont P}}}{2pt}}_{l}+b^{\scaleto{\text{ {\fontfamily{qcr}\selectfont P}}}{2pt}}_{l})}y^{\scaleto{\text{ {\fontfamily{qcr}\selectfont P}}}{2.5pt}}_{l}\left(x^{*}\right)^{a^{\scaleto{\text{ {\fontfamily{qcr}\selectfont P}}}{2pt}}_{l}} \left(a^{\scaleto{\text{ {\fontfamily{qcr}\selectfont P}}}{2.5pt}}_{l} { }_{2} \tilde{F}_{1}\left(a^{\scaleto{\text{ {\fontfamily{qcr}\selectfont P}}}{2.5pt}}_{l}+1, a^{\scaleto{\text{ {\fontfamily{qcr}\selectfont P}}}{2.5pt}}_{l}-c^{\scaleto{\text{ {\fontfamily{qcr}\selectfont P}}}{2.5pt}}_{l}+1 ; a^{\scaleto{\text{ {\fontfamily{qcr}\selectfont P}}}{2.5pt}}_{l}-b^{\scaleto{\text{ {\fontfamily{qcr}\selectfont P}}}{2.5pt}}_{l}+1 ; y^{\scaleto{\text{ {\fontfamily{qcr}\selectfont P}}}{2.5pt}}_{l}(B)^{-1}\right)\right. \right. \\ \\
    & { }_{2} \tilde{F}_{1}\left(b^{\scaleto{\text{ {\fontfamily{qcr}\selectfont P}}}{2.5pt}}_{l}, b^{\scaleto{\text{ {\fontfamily{qcr}\selectfont P}}}{2.5pt}}_{l}-c^{\scaleto{\text{ {\fontfamily{qcr}\selectfont P}}}{2.5pt}}_{l}+1 ; b^{\scaleto{\text{ {\fontfamily{qcr}\selectfont P}}}{2.5pt}}_{l}-a^{\scaleto{\text{ {\fontfamily{qcr}\selectfont P}}}{2.5pt}}_{l}+1 ; y^{\scaleto{\text{ {\fontfamily{qcr}\selectfont P}}}{2.5pt}}_{l}(B)^{-1}\right) - b^{\scaleto{\text{ {\fontfamily{qcr}\selectfont P}}}{2.5pt}}_{l} { }_{2} \tilde{F}_{1}\left(a^{\scaleto{\text{ {\fontfamily{qcr}\selectfont P}}}{2.5pt}}_{l}, a^{\scaleto{\text{ {\fontfamily{qcr}\selectfont P}}}{2.5pt}}_{l}-c^{\scaleto{\text{ {\fontfamily{qcr}\selectfont P}}}{2.5pt}}_{l}+1 ; a^{\scaleto{\text{ {\fontfamily{qcr}\selectfont P}}}{2.5pt}}_{l}-b^{\scaleto{\text{ {\fontfamily{qcr}\selectfont P}}}{2.5pt}}_{l}+1 ; y^{\scaleto{\text{ {\fontfamily{qcr}\selectfont P}}}{2.5pt}}_{l}(B)^{-1}\right)\\ \\
    & \left. \left. { }_{2} \tilde{F}_{1}\left(b^{\scaleto{\text{ {\fontfamily{qcr}\selectfont P}}}{2.5pt}}_{l}+1, b^{\scaleto{\text{ {\fontfamily{qcr}\selectfont P}}}{2.5pt}}_{l}-c^{\scaleto{\text{ {\fontfamily{qcr}\selectfont P}}}{2.5pt}}_{l}+1 ; b^{\scaleto{\text{ {\fontfamily{qcr}\selectfont P}}}{2.5pt}}_{l}-a^{\scaleto{\text{ {\fontfamily{qcr}\selectfont P}}}{2.5pt}}_{l}+1 ; y^{\scaleto{\text{ {\fontfamily{qcr}\selectfont P}}}{2.5pt}}_{l}(B)^{-1}\right)\right) \right]/\\ \\
    & \left[\Gamma\left(1-a^{\scaleto{\text{ {\fontfamily{qcr}\selectfont P}}}{2.5pt}}_{u}+b^{\scaleto{\text{ {\fontfamily{qcr}\selectfont P}}}{2.5pt}}_{u}\right) \left(y^{\scaleto{\text{ {\fontfamily{qcr}\selectfont P}}}{2.5pt}}_{l}\left(B\right)^{-b^{\scaleto{\text{ {\fontfamily{qcr}\selectfont P}}}{2pt}}_{l}}\left(\left(\delta+\lambda\right)x^{*} { }_{2} \tilde{F}_{1}\left(a^{\scaleto{\text{ {\fontfamily{qcr}\selectfont P}}}{2.5pt}}_{l}, a^{\scaleto{\text{ {\fontfamily{qcr}\selectfont P}}}{2.5pt}}_{l}-c^{\scaleto{\text{ {\fontfamily{qcr}\selectfont P}}}{2.5pt}}_{l}+1 ; a^{\scaleto{\text{ {\fontfamily{qcr}\selectfont P}}}{2.5pt}}_{l}-b^{\scaleto{\text{ {\fontfamily{qcr}\selectfont P}}}{2.5pt}}_{l}+1 ; y^{\scaleto{\text{ {\fontfamily{qcr}\selectfont P}}}{2.5pt}}_{l}(x^{*})^{-1}\right) \right. \right. \right. \\ \\
    & \left. + c_{{\scaleto{T}{3pt}}} a^{\scaleto{\text{ {\fontfamily{qcr}\selectfont P}}}{2.5pt}}_{l} \left(B - x^{*}\right) { }_{2} \tilde{F}_{1}\left(a^{\scaleto{\text{ {\fontfamily{qcr}\selectfont P}}}{2.5pt}}_{l} + 1, a^{\scaleto{\text{ {\fontfamily{qcr}\selectfont P}}}{2.5pt}}_{l}-c^{\scaleto{\text{ {\fontfamily{qcr}\selectfont P}}}{2.5pt}}_{l}+1 ; a^{\scaleto{\text{ {\fontfamily{qcr}\selectfont P}}}{2.5pt}}_{l}-b^{\scaleto{\text{ {\fontfamily{qcr}\selectfont P}}}{2.5pt}}_{l}+1 ; y^{\scaleto{\text{ {\fontfamily{qcr}\selectfont P}}}{2.5pt}}_{l}(x^{*})^{-1}\right)\right)\\ \\
    & \left(b^{\scaleto{\text{ {\fontfamily{qcr}\selectfont P}}}{2.5pt}}_{u} { }_{2} \tilde{F}_{1}\left(b^{\scaleto{\text{ {\fontfamily{qcr}\selectfont P}}}{2.5pt}}_{u} + 1, b^{\scaleto{\text{ {\fontfamily{qcr}\selectfont P}}}{2.5pt}}_{u}-c^{\scaleto{\text{ {\fontfamily{qcr}\selectfont P}}}{2.5pt}}_{u}+1 ; b^{\scaleto{\text{ {\fontfamily{qcr}\selectfont P}}}{2.5pt}}_{u}-a^{\scaleto{\text{ {\fontfamily{qcr}\selectfont P}}}{2.5pt}}_{u}+1 ; y^{\scaleto{\text{ {\fontfamily{qcr}\selectfont P}}}{2.5pt}}_{u}(B)^{-1}\right) { }_{2} \tilde{F}_{1}\left(b^{\scaleto{\text{ {\fontfamily{qcr}\selectfont P}}}{2.5pt}}_{l}, b^{\scaleto{\text{ {\fontfamily{qcr}\selectfont P}}}{2.5pt}}_{l}-c^{\scaleto{\text{ {\fontfamily{qcr}\selectfont P}}}{2.5pt}}_{l}+1 ; b^{\scaleto{\text{ {\fontfamily{qcr}\selectfont P}}}{2.5pt}}_{l}-a^{\scaleto{\text{ {\fontfamily{qcr}\selectfont P}}}{2.5pt}}_{l}+1 ; y^{\scaleto{\text{ {\fontfamily{qcr}\selectfont P}}}{2.5pt}}_{l}(B)^{-1}\right) \right. \\ \\
    & \left. - b^{\scaleto{\text{ {\fontfamily{qcr}\selectfont P}}}{2.5pt}}_{l} { }_{2} \tilde{F}_{1}\left(b^{\scaleto{\text{ {\fontfamily{qcr}\selectfont P}}}{2.5pt}}_{u}, b^{\scaleto{\text{ {\fontfamily{qcr}\selectfont P}}}{2.5pt}}_{u}-c^{\scaleto{\text{ {\fontfamily{qcr}\selectfont P}}}{2.5pt}}_{u}+1 ; b^{\scaleto{\text{ {\fontfamily{qcr}\selectfont P}}}{2.5pt}}_{u}-a^{\scaleto{\text{ {\fontfamily{qcr}\selectfont P}}}{2.5pt}}_{u}+1 ; y^{\scaleto{\text{ {\fontfamily{qcr}\selectfont P}}}{2.5pt}}_{u}(B)^{-1}\right) { }_{2} \tilde{F}_{1}\left(b^{\scaleto{\text{ {\fontfamily{qcr}\selectfont P}}}{2.5pt}}_{l} + 1, b^{\scaleto{\text{ {\fontfamily{qcr}\selectfont P}}}{2.5pt}}_{l}-c^{\scaleto{\text{ {\fontfamily{qcr}\selectfont P}}}{2.5pt}}_{l}+1 ; b^{\scaleto{\text{ {\fontfamily{qcr}\selectfont P}}}{2.5pt}}_{l}-a^{\scaleto{\text{ {\fontfamily{qcr}\selectfont P}}}{2.5pt}}_{l}+1 ; y^{\scaleto{\text{ {\fontfamily{qcr}\selectfont P}}}{2.5pt}}_{l}(B)^{-1}\right)\right) \\ \\
    & - y^{\scaleto{\text{ {\fontfamily{qcr}\selectfont P}}}{2.5pt}}_{l}\left(B\right)^{-a^{\scaleto{\text{ {\fontfamily{qcr}\selectfont P}}}{2pt}}_{l}} y^{\scaleto{\text{ {\fontfamily{qcr}\selectfont P}}}{2.5pt}}_{l}\left(x^{*}\right)^{a^{\scaleto{\text{ {\fontfamily{qcr}\selectfont P}}}{2pt}}_{l}-b^{\scaleto{\text{ {\fontfamily{qcr}\selectfont P}}}{2pt}}_{l}} \left(b^{\scaleto{\text{ {\fontfamily{qcr}\selectfont P}}}{2.5pt}}_{u} { }_{2} \tilde{F}_{1}\left(b^{\scaleto{\text{ {\fontfamily{qcr}\selectfont P}}}{2.5pt}}_{u} + 1, b^{\scaleto{\text{ {\fontfamily{qcr}\selectfont P}}}{2.5pt}}_{u}-c^{\scaleto{\text{ {\fontfamily{qcr}\selectfont P}}}{2.5pt}}_{u}+1 ; b^{\scaleto{\text{ {\fontfamily{qcr}\selectfont P}}}{2.5pt}}_{u}-a^{\scaleto{\text{ {\fontfamily{qcr}\selectfont P}}}{2.5pt}}_{u}+1 ; y^{\scaleto{\text{ {\fontfamily{qcr}\selectfont P}}}{2.5pt}}_{u}(B)^{-1}\right) \right. \\ \\
    & { }_{2} \tilde{F}_{1}\left(a^{\scaleto{\text{ {\fontfamily{qcr}\selectfont P}}}{2.5pt}}_{l}, a^{\scaleto{\text{ {\fontfamily{qcr}\selectfont P}}}{2.5pt}}_{l}-c^{\scaleto{\text{ {\fontfamily{qcr}\selectfont P}}}{2.5pt}}_{l}+1 ; a^{\scaleto{\text{ {\fontfamily{qcr}\selectfont P}}}{2.5pt}}_{l}-b^{\scaleto{\text{ {\fontfamily{qcr}\selectfont P}}}{2.5pt}}_{l}+1 ; y^{\scaleto{\text{ {\fontfamily{qcr}\selectfont P}}}{2.5pt}}_{l}(B)^{-1}\right) - a^{\scaleto{\text{ {\fontfamily{qcr}\selectfont P}}}{2.5pt}}_{l} { }_{2} \tilde{F}_{1}\left(b^{\scaleto{\text{ {\fontfamily{qcr}\selectfont P}}}{2.5pt}}_{u}, b^{\scaleto{\text{ {\fontfamily{qcr}\selectfont P}}}{2.5pt}}_{u}-c^{\scaleto{\text{ {\fontfamily{qcr}\selectfont P}}}{2.5pt}}_{u}+1 ; b^{\scaleto{\text{ {\fontfamily{qcr}\selectfont P}}}{2.5pt}}_{u}-a^{\scaleto{\text{ {\fontfamily{qcr}\selectfont P}}}{2.5pt}}_{u}+1 ; y^{\scaleto{\text{ {\fontfamily{qcr}\selectfont P}}}{2.5pt}}_{u}(B)^{-1}\right)\\ \\
    & \left. { }_{2} \tilde{F}_{1}\left(1 + a^{\scaleto{\text{ {\fontfamily{qcr}\selectfont P}}}{2.5pt}}_{l}, a^{\scaleto{\text{ {\fontfamily{qcr}\selectfont P}}}{2.5pt}}_{l}-c^{\scaleto{\text{ {\fontfamily{qcr}\selectfont P}}}{2.5pt}}_{l}+1 ; a^{\scaleto{\text{ {\fontfamily{qcr}\selectfont P}}}{2.5pt}}_{l}-b^{\scaleto{\text{ {\fontfamily{qcr}\selectfont P}}}{2.5pt}}_{l}+1 ; y^{\scaleto{\text{ {\fontfamily{qcr}\selectfont P}}}{2.5pt}}_{l}(B)^{-1}\right)\right)\left(\left(\delta + \lambda\right)x^{*} { }_{2} \tilde{F}_{1}\left(b^{\scaleto{\text{ {\fontfamily{qcr}\selectfont P}}}{2.5pt}}_{l}, b^{\scaleto{\text{ {\fontfamily{qcr}\selectfont P}}}{2.5pt}}_{l}-c^{\scaleto{\text{ {\fontfamily{qcr}\selectfont P}}}{2.5pt}}_{l}+1 ; b^{\scaleto{\text{ {\fontfamily{qcr}\selectfont P}}}{2.5pt}}_{l}-a^{\scaleto{\text{ {\fontfamily{qcr}\selectfont P}}}{2.5pt}}_{l}+1 ; y^{\scaleto{\text{ {\fontfamily{qcr}\selectfont P}}}{2.5pt}}_{l}(x^{*})^{-1}\right) \right.\\ \\
    & \left. \left. \left. + c_{{\scaleto{T}{3pt}}} b^{\scaleto{\text{ {\fontfamily{qcr}\selectfont P}}}{2.5pt}}_{l} \left(B - x^{*}\right) { }_{2} \tilde{F}_{1}\left(b^{\scaleto{\text{ {\fontfamily{qcr}\selectfont P}}}{2.5pt}}_{l} + 1, b^{\scaleto{\text{ {\fontfamily{qcr}\selectfont P}}}{2.5pt}}_{l}-c^{\scaleto{\text{ {\fontfamily{qcr}\selectfont P}}}{2.5pt}}_{l}+1 ; b^{\scaleto{\text{ {\fontfamily{qcr}\selectfont P}}}{2.5pt}}_{l}-a^{\scaleto{\text{ {\fontfamily{qcr}\selectfont P}}}{2.5pt}}_{l}+1 ; y^{\scaleto{\text{ {\fontfamily{qcr}\selectfont P}}}{2.5pt}}_{l}(x^{*})^{-1}\right) \right) \right)\right],
    \label{Appendix A: Mathematical Proofs-Equation15}
\end{align}
}

\vspace{0.3cm}

{\allowdisplaybreaks
\begin{align}
    A^{\scaleto{\text{ {\fontfamily{qcr}\selectfont P}}}{2.5pt}}_{1,l}&= \left[ \lambda x^{*} y^{\scaleto{\text{ {\fontfamily{qcr}\selectfont P}}}{2.5pt}}_{l}\left(x^{*}\right)^{a^{\scaleto{\text{ {\fontfamily{qcr}\selectfont P}}}{2pt}}_{l}} \left(1 + 1/\left(-1 + \left(y^{\scaleto{\text{ {\fontfamily{qcr}\selectfont P}}}{2.5pt}}_{l}\left(B\right)^{a^{\scaleto{\text{ {\fontfamily{qcr}\selectfont P}}}{2pt}}_{l}-b^{\scaleto{\text{ {\fontfamily{qcr}\selectfont P}}}{2pt}}_{l}} y^{\scaleto{\text{ {\fontfamily{qcr}\selectfont P}}}{2.5pt}}_{l}\left(x^{*}\right)^{b^{\scaleto{\text{ {\fontfamily{qcr}\selectfont P}}}{2pt}}_{l}-a^{\scaleto{\text{ {\fontfamily{qcr}\selectfont P}}}{2pt}}_{l}}\left(\left(\delta + \lambda\right) x^{*} { }_{2} \tilde{F}_{1}\left(a^{\scaleto{\text{ {\fontfamily{qcr}\selectfont P}}}{2.5pt}}_{l}, a^{\scaleto{\text{ {\fontfamily{qcr}\selectfont P}}}{2.5pt}}_{l}-c^{\scaleto{\text{ {\fontfamily{qcr}\selectfont P}}}{2.5pt}}_{l}+1 ; a^{\scaleto{\text{ {\fontfamily{qcr}\selectfont P}}}{2.5pt}}_{l}-b^{\scaleto{\text{ {\fontfamily{qcr}\selectfont P}}}{2.5pt}}_{l}+1 ; y^{\scaleto{\text{ {\fontfamily{qcr}\selectfont P}}}{2.5pt}}_{l}(x^{*})^{-1}\right) \right. \right.  \right.  \right. \right.\\ \\
    & \left. + c_{{\scaleto{T}{3pt}}} a^{\scaleto{\text{ {\fontfamily{qcr}\selectfont P}}}{2.5pt}}_{l} \left(B - x^{*}\right) { }_{2} \tilde{F}_{1}\left(a^{\scaleto{\text{ {\fontfamily{qcr}\selectfont P}}}{2.5pt}}_{l} + 1, a^{\scaleto{\text{ {\fontfamily{qcr}\selectfont P}}}{2.5pt}}_{l}-c^{\scaleto{\text{ {\fontfamily{qcr}\selectfont P}}}{2.5pt}}_{l}+1 ; a^{\scaleto{\text{ {\fontfamily{qcr}\selectfont P}}}{2.5pt}}_{l}-b^{\scaleto{\text{ {\fontfamily{qcr}\selectfont P}}}{2.5pt}}_{l}+1 ; y^{\scaleto{\text{ {\fontfamily{qcr}\selectfont P}}}{2.5pt}}_{l}(x^{*})^{-1}\right) \right)\left(b^{\scaleto{\text{ {\fontfamily{qcr}\selectfont P}}}{2.5pt}}_{u} { }_{2} \tilde{F}_{1}\left(b^{\scaleto{\text{ {\fontfamily{qcr}\selectfont P}}}{2.5pt}}_{u} + 1, b^{\scaleto{\text{ {\fontfamily{qcr}\selectfont P}}}{2.5pt}}_{u}-c^{\scaleto{\text{ {\fontfamily{qcr}\selectfont P}}}{2.5pt}}_{u}+1 ; b^{\scaleto{\text{ {\fontfamily{qcr}\selectfont P}}}{2.5pt}}_{u}-a^{\scaleto{\text{ {\fontfamily{qcr}\selectfont P}}}{2.5pt}}_{u}+1 ; y^{\scaleto{\text{ {\fontfamily{qcr}\selectfont P}}}{2.5pt}}_{u}(B)^{-1}\right) \right. \\ \\
    & { }_{2} \tilde{F}_{1}\left(b^{\scaleto{\text{ {\fontfamily{qcr}\selectfont P}}}{2.5pt}}_{l}, b^{\scaleto{\text{ {\fontfamily{qcr}\selectfont P}}}{2.5pt}}_{l}-c^{\scaleto{\text{ {\fontfamily{qcr}\selectfont P}}}{2.5pt}}_{l}+1 ; b^{\scaleto{\text{ {\fontfamily{qcr}\selectfont P}}}{2.5pt}}_{l}-a^{\scaleto{\text{ {\fontfamily{qcr}\selectfont P}}}{2.5pt}}_{l}+1 ; y^{\scaleto{\text{ {\fontfamily{qcr}\selectfont P}}}{2.5pt}}_{l}(B)^{-1}\right) - b^{\scaleto{\text{ {\fontfamily{qcr}\selectfont P}}}{2.5pt}}_{l} { }_{2} \tilde{F}_{1}\left(b^{\scaleto{\text{ {\fontfamily{qcr}\selectfont P}}}{2.5pt}}_{u}, b^{\scaleto{\text{ {\fontfamily{qcr}\selectfont P}}}{2.5pt}}_{u}-c^{\scaleto{\text{ {\fontfamily{qcr}\selectfont P}}}{2.5pt}}_{u}+1 ; b^{\scaleto{\text{ {\fontfamily{qcr}\selectfont P}}}{2.5pt}}_{u}-a^{\scaleto{\text{ {\fontfamily{qcr}\selectfont P}}}{2.5pt}}_{u}+1 ; y^{\scaleto{\text{ {\fontfamily{qcr}\selectfont P}}}{2.5pt}}_{u}(B)^{-1}\right) \\ \\
    &  \left. \left. { }_{2} \tilde{F}_{1}\left(b^{\scaleto{\text{ {\fontfamily{qcr}\selectfont P}}}{2.5pt}}_{l} + 1, b^{\scaleto{\text{ {\fontfamily{qcr}\selectfont P}}}{2.5pt}}_{l}-c^{\scaleto{\text{ {\fontfamily{qcr}\selectfont P}}}{2.5pt}}_{l}+1 ; b^{\scaleto{\text{ {\fontfamily{qcr}\selectfont P}}}{2.5pt}}_{l}-a^{\scaleto{\text{ {\fontfamily{qcr}\selectfont P}}}{2.5pt}}_{l}+1 ; y^{\scaleto{\text{ {\fontfamily{qcr}\selectfont P}}}{2.5pt}}_{l}(B)^{-1}\right)\right)\right)/\left(\left(b^{\scaleto{\text{ {\fontfamily{qcr}\selectfont P}}}{2.5pt}}_{u}{ }_{2} \tilde{F}_{1}\left(b^{\scaleto{\text{ {\fontfamily{qcr}\selectfont P}}}{2.5pt}}_{u} + 1, b^{\scaleto{\text{ {\fontfamily{qcr}\selectfont P}}}{2.5pt}}_{u}-c^{\scaleto{\text{ {\fontfamily{qcr}\selectfont P}}}{2.5pt}}_{u}+1 ; b^{\scaleto{\text{ {\fontfamily{qcr}\selectfont P}}}{2.5pt}}_{u}-a^{\scaleto{\text{ {\fontfamily{qcr}\selectfont P}}}{2.5pt}}_{u}+1 ; y^{\scaleto{\text{ {\fontfamily{qcr}\selectfont P}}}{2.5pt}}_{u}(B)^{-1}\right) \right. \right. \\ \\
    & { }_{2} \tilde{F}_{1}\left(a^{\scaleto{\text{ {\fontfamily{qcr}\selectfont P}}}{2.5pt}}_{l}, a^{\scaleto{\text{ {\fontfamily{qcr}\selectfont P}}}{2.5pt}}_{l}-c^{\scaleto{\text{ {\fontfamily{qcr}\selectfont P}}}{2.5pt}}_{l}+1 ; a^{\scaleto{\text{ {\fontfamily{qcr}\selectfont P}}}{2.5pt}}_{l}-b^{\scaleto{\text{ {\fontfamily{qcr}\selectfont P}}}{2.5pt}}_{l}+1 ; y^{\scaleto{\text{ {\fontfamily{qcr}\selectfont P}}}{2.5pt}}_{l}(B)^{-1}\right) - a^{\scaleto{\text{ {\fontfamily{qcr}\selectfont P}}}{2.5pt}}_{l} { }_{2} \tilde{F}_{1}\left(b^{\scaleto{\text{ {\fontfamily{qcr}\selectfont P}}}{2.5pt}}_{u}, b^{\scaleto{\text{ {\fontfamily{qcr}\selectfont P}}}{2.5pt}}_{u}-c^{\scaleto{\text{ {\fontfamily{qcr}\selectfont P}}}{2.5pt}}_{u}+1 ; b^{\scaleto{\text{ {\fontfamily{qcr}\selectfont P}}}{2.5pt}}_{u}-a^{\scaleto{\text{ {\fontfamily{qcr}\selectfont P}}}{2.5pt}}_{u}+1 ; y^{\scaleto{\text{ {\fontfamily{qcr}\selectfont P}}}{2.5pt}}_{u}(B)^{-1}\right) \\ \\
    & \left. { }_{2} \tilde{F}_{1}\left(a^{\scaleto{\text{ {\fontfamily{qcr}\selectfont P}}}{2.5pt}}_{l} + 1, a^{\scaleto{\text{ {\fontfamily{qcr}\selectfont P}}}{2.5pt}}_{l}-c^{\scaleto{\text{ {\fontfamily{qcr}\selectfont P}}}{2.5pt}}_{l}+1 ; a^{\scaleto{\text{ {\fontfamily{qcr}\selectfont P}}}{2.5pt}}_{l}-b^{\scaleto{\text{ {\fontfamily{qcr}\selectfont P}}}{2.5pt}}_{l}+1 ; y^{\scaleto{\text{ {\fontfamily{qcr}\selectfont P}}}{2.5pt}}_{l}(B)^{-1}\right)\right)\left(\left(\delta+\lambda\right)x^{*} { }_{2} \tilde{F}_{1}\left(b^{\scaleto{\text{ {\fontfamily{qcr}\selectfont P}}}{2.5pt}}_{l}, b^{\scaleto{\text{ {\fontfamily{qcr}\selectfont P}}}{2.5pt}}_{l}-c^{\scaleto{\text{ {\fontfamily{qcr}\selectfont P}}}{2.5pt}}_{l}+1 ; b^{\scaleto{\text{ {\fontfamily{qcr}\selectfont P}}}{2.5pt}}_{l}-a^{\scaleto{\text{ {\fontfamily{qcr}\selectfont P}}}{2.5pt}}_{l}+1 ; y^{\scaleto{\text{ {\fontfamily{qcr}\selectfont P}}}{2.5pt}}_{l}(x^{*})^{-1}\right) \right. \\ \\
    & \left. \left. \left. \left. \left. + c_{{\scaleto{T}{3pt}}} b^{\scaleto{\text{ {\fontfamily{qcr}\selectfont P}}}{2.5pt}}_{l} \left(B - x^{*}\right) { }_{2} \tilde{F}_{1}\left(b^{\scaleto{\text{ {\fontfamily{qcr}\selectfont P}}}{2.5pt}}_{l} + 1, b^{\scaleto{\text{ {\fontfamily{qcr}\selectfont P}}}{2.5pt}}_{l}-c^{\scaleto{\text{ {\fontfamily{qcr}\selectfont P}}}{2.5pt}}_{l}+1 ; b^{\scaleto{\text{ {\fontfamily{qcr}\selectfont P}}}{2.5pt}}_{l}-a^{\scaleto{\text{ {\fontfamily{qcr}\selectfont P}}}{2.5pt}}_{l}+1 ; y^{\scaleto{\text{ {\fontfamily{qcr}\selectfont P}}}{2.5pt}}_{l}(x^{*})^{-1}\right)\right)\right)\right)\right)\right]/\\ \\
    &\left[\left(\delta + \lambda\right) x^{*}   { }_{2} \tilde{F}_{1}\left(a^{\scaleto{\text{ {\fontfamily{qcr}\selectfont P}}}{2.5pt}}_{l}, a^{\scaleto{\text{ {\fontfamily{qcr}\selectfont P}}}{2.5pt}}_{l}-c^{\scaleto{\text{ {\fontfamily{qcr}\selectfont P}}}{2.5pt}}_{l}+1 ; a^{\scaleto{\text{ {\fontfamily{qcr}\selectfont P}}}{2.5pt}}_{l}-b^{\scaleto{\text{ {\fontfamily{qcr}\selectfont P}}}{2.5pt}}_{l}+1 ; y^{\scaleto{\text{ {\fontfamily{qcr}\selectfont P}}}{2.5pt}}_{l}(x^{*})^{-1}\right) \right. \\ \\
    & \left. + c_{{\scaleto{T}{3pt}}} a^{\scaleto{\text{ {\fontfamily{qcr}\selectfont P}}}{2.5pt}}_{l} \left(B - x^{*}\right) { }_{2} \tilde{F}_{1}\left(a^{\scaleto{\text{ {\fontfamily{qcr}\selectfont P}}}{2.5pt}}_{l} + 1, a^{\scaleto{\text{ {\fontfamily{qcr}\selectfont P}}}{2.5pt}}_{l}-c^{\scaleto{\text{ {\fontfamily{qcr}\selectfont P}}}{2.5pt}}_{l}+1 ; a^{\scaleto{\text{ {\fontfamily{qcr}\selectfont P}}}{2.5pt}}_{l}-b^{\scaleto{\text{ {\fontfamily{qcr}\selectfont P}}}{2.5pt}}_{l}+1 ; y^{\scaleto{\text{ {\fontfamily{qcr}\selectfont P}}}{2.5pt}}_{l}(x^{*})^{-1}\right)\right]
   \label{Appendix A: Mathematical Proofs-Equation16}
\end{align}
}

\normalsize

\vspace{0.3cm}

and

\footnotesize

\vspace{0.3cm}

{\allowdisplaybreaks
\begin{align}
    A^{\scaleto{\text{ {\fontfamily{qcr}\selectfont P}}}{2.5pt}}_{2,l}&= \left[ \lambda x^{*}  y^{\scaleto{\text{ {\fontfamily{qcr}\selectfont P}}}{2.5pt}}_{l}(B)^{-a^{\scaleto{\text{ {\fontfamily{qcr}\selectfont P}}}{2pt}}_{l}} y^{\scaleto{\text{ {\fontfamily{qcr}\selectfont P}}}{2.5pt}}_{l}(x^{*})^{a^{\scaleto{\text{ {\fontfamily{qcr}\selectfont P}}}{2pt}}_{l}} \Gamma\left(1 + a^{\scaleto{\text{ {\fontfamily{qcr}\selectfont P}}}{2.5pt}}_{l} - b^{\scaleto{\text{ {\fontfamily{qcr}\selectfont P}}}{2.5pt}}_{l}\right) \left(b^{\scaleto{\text{ {\fontfamily{qcr}\selectfont P}}}{2.5pt}}_{u} { }_{2} \tilde{F}_{1}\left(b^{\scaleto{\text{ {\fontfamily{qcr}\selectfont P}}}{2.5pt}}_{u} + 1, b^{\scaleto{\text{ {\fontfamily{qcr}\selectfont P}}}{2.5pt}}_{u}-c^{\scaleto{\text{ {\fontfamily{qcr}\selectfont P}}}{2.5pt}}_{u}+1 ; b^{\scaleto{\text{ {\fontfamily{qcr}\selectfont P}}}{2.5pt}}_{u}-a^{\scaleto{\text{ {\fontfamily{qcr}\selectfont P}}}{2.5pt}}_{u}+1 ; y^{\scaleto{\text{ {\fontfamily{qcr}\selectfont P}}}{2.5pt}}_{u}(B)^{-1}\right) \right. \right. \\ \\
    & \left. { }_{2} \tilde{F}_{1}\left(a^{\scaleto{\text{ {\fontfamily{qcr}\selectfont P}}}{2.5pt}}_{l}, a^{\scaleto{\text{ {\fontfamily{qcr}\selectfont P}}}{2.5pt}}_{l}-c^{\scaleto{\text{ {\fontfamily{qcr}\selectfont P}}}{2.5pt}}_{l}+1 ; a^{\scaleto{\text{ {\fontfamily{qcr}\selectfont P}}}{2.5pt}}_{l}-b^{\scaleto{\text{ {\fontfamily{qcr}\selectfont P}}}{2.5pt}}_{l}+1 ; y^{\scaleto{\text{ {\fontfamily{qcr}\selectfont P}}}{2.5pt}}_{l}(B)^{-1}\right) -a^{\scaleto{\text{ {\fontfamily{qcr}\selectfont P}}}{2.5pt}}_{l} { }_{2} \tilde{F}_{1}\left(b^{\scaleto{\text{ {\fontfamily{qcr}\selectfont P}}}{2.5pt}}_{u}, b^{\scaleto{\text{ {\fontfamily{qcr}\selectfont P}}}{2.5pt}}_{u}-c^{\scaleto{\text{ {\fontfamily{qcr}\selectfont P}}}{2.5pt}}_{u}+1 ; b^{\scaleto{\text{ {\fontfamily{qcr}\selectfont P}}}{2.5pt}}_{u}-a^{\scaleto{\text{ {\fontfamily{qcr}\selectfont P}}}{2.5pt}}_{u}+1 ; y^{\scaleto{\text{ {\fontfamily{qcr}\selectfont P}}}{2.5pt}}_{u}(B)^{-1}\right) \right. \\ \\
    & \left. \left. { }_{2} \tilde{F}_{1}\left(a^{\scaleto{\text{ {\fontfamily{qcr}\selectfont P}}}{2.5pt}}_{l} + 1, a^{\scaleto{\text{ {\fontfamily{qcr}\selectfont P}}}{2.5pt}}_{l}-c^{\scaleto{\text{ {\fontfamily{qcr}\selectfont P}}}{2.5pt}}_{l}+1 ; a^{\scaleto{\text{ {\fontfamily{qcr}\selectfont P}}}{2.5pt}}_{l}-b^{\scaleto{\text{ {\fontfamily{qcr}\selectfont P}}}{2.5pt}}_{l}+1 ; y^{\scaleto{\text{ {\fontfamily{qcr}\selectfont P}}}{2.5pt}}_{l}(B)^{-1}\right)\right) sin\left(\left(a^{\scaleto{\text{ {\fontfamily{qcr}\selectfont P}}}{2.5pt}}_{l}-b^{\scaleto{\text{ {\fontfamily{qcr}\selectfont P}}}{2.5pt}}_{l}\right) \pi \right) \right]/ \\ \\
    & \left[ \left(a^{\scaleto{\text{ {\fontfamily{qcr}\selectfont P}}}{2.5pt}}_{l} - b^{\scaleto{\text{ {\fontfamily{qcr}\selectfont P}}}{2.5pt}}_{l}\right) \pi \left(-y^{\scaleto{\text{ {\fontfamily{qcr}\selectfont P}}}{2.5pt}}_{l}(B)^{-b^{\scaleto{\text{ {\fontfamily{qcr}\selectfont P}}}{2pt}}_{l}} \left( \left(\delta + \lambda\right)x^{*} { }_{2} \tilde{F}_{1}\left(a^{\scaleto{\text{ {\fontfamily{qcr}\selectfont P}}}{2.5pt}}_{l}, a^{\scaleto{\text{ {\fontfamily{qcr}\selectfont P}}}{2.5pt}}_{l}-c^{\scaleto{\text{ {\fontfamily{qcr}\selectfont P}}}{2.5pt}}_{l}+1 ; a^{\scaleto{\text{ {\fontfamily{qcr}\selectfont P}}}{2.5pt}}_{l}-b^{\scaleto{\text{ {\fontfamily{qcr}\selectfont P}}}{2.5pt}}_{l}+1 ; y^{\scaleto{\text{ {\fontfamily{qcr}\selectfont P}}}{2.5pt}}_{l}(x^{*})^{-1}\right)  \right. \right. \right. \\ \\
    & \left. + c_{{\scaleto{T}{3pt}}} a^{\scaleto{\text{ {\fontfamily{qcr}\selectfont P}}}{2.5pt}}_{l} \left(B - x^{*}\right) { }_{2} \tilde{F}_{1}\left(a^{\scaleto{\text{ {\fontfamily{qcr}\selectfont P}}}{2.5pt}}_{l} + 1, a^{\scaleto{\text{ {\fontfamily{qcr}\selectfont P}}}{2.5pt}}_{l}-c^{\scaleto{\text{ {\fontfamily{qcr}\selectfont P}}}{2.5pt}}_{l}+1 ; a^{\scaleto{\text{ {\fontfamily{qcr}\selectfont P}}}{2.5pt}}_{l}-b^{\scaleto{\text{ {\fontfamily{qcr}\selectfont P}}}{2.5pt}}_{l}+1 ; y^{\scaleto{\text{ {\fontfamily{qcr}\selectfont P}}}{2.5pt}}_{l}(x^{*})^{-1}\right) \right)\\ \\
    & \left(b^{\scaleto{\text{ {\fontfamily{qcr}\selectfont P}}}{2.5pt}}_{u} { }_{2} \tilde{F}_{1}\left(b^{\scaleto{\text{ {\fontfamily{qcr}\selectfont P}}}{2.5pt}}_{u} + 1, b^{\scaleto{\text{ {\fontfamily{qcr}\selectfont P}}}{2.5pt}}_{u}-c^{\scaleto{\text{ {\fontfamily{qcr}\selectfont P}}}{2.5pt}}_{u}+1 ; b^{\scaleto{\text{ {\fontfamily{qcr}\selectfont P}}}{2.5pt}}_{u}-a^{\scaleto{\text{ {\fontfamily{qcr}\selectfont P}}}{2.5pt}}_{u}+1 ; y^{\scaleto{\text{ {\fontfamily{qcr}\selectfont P}}}{2.5pt}}_{u}(B)^{-1}\right) { }_{2} \tilde{F}_{1}\left(b^{\scaleto{\text{ {\fontfamily{qcr}\selectfont P}}}{2.5pt}}_{l}, b^{\scaleto{\text{ {\fontfamily{qcr}\selectfont P}}}{2.5pt}}_{l}-c^{\scaleto{\text{ {\fontfamily{qcr}\selectfont P}}}{2.5pt}}_{l}+1 ; b^{\scaleto{\text{ {\fontfamily{qcr}\selectfont P}}}{2.5pt}}_{l}-a^{\scaleto{\text{ {\fontfamily{qcr}\selectfont P}}}{2.5pt}}_{l}+1 ; y^{\scaleto{\text{ {\fontfamily{qcr}\selectfont P}}}{2.5pt}}_{l}(B)^{-1}\right) \right. \\ \\
    & \left. -b^{\scaleto{\text{ {\fontfamily{qcr}\selectfont P}}}{2.5pt}}_{l}{ }_{2} \tilde{F}_{1}\left(b^{\scaleto{\text{ {\fontfamily{qcr}\selectfont P}}}{2.5pt}}_{u}, b^{\scaleto{\text{ {\fontfamily{qcr}\selectfont P}}}{2.5pt}}_{u}-c^{\scaleto{\text{ {\fontfamily{qcr}\selectfont P}}}{2.5pt}}_{u}+1 ; b^{\scaleto{\text{ {\fontfamily{qcr}\selectfont P}}}{2.5pt}}_{u}-a^{\scaleto{\text{ {\fontfamily{qcr}\selectfont P}}}{2.5pt}}_{u}+1 ; y^{\scaleto{\text{ {\fontfamily{qcr}\selectfont P}}}{2.5pt}}_{u}(B)^{-1}\right){ }_{2} \tilde{F}_{1}\left(b^{\scaleto{\text{ {\fontfamily{qcr}\selectfont P}}}{2.5pt}}_{l} + 1, b^{\scaleto{\text{ {\fontfamily{qcr}\selectfont P}}}{2.5pt}}_{l}-c^{\scaleto{\text{ {\fontfamily{qcr}\selectfont P}}}{2.5pt}}_{l}+1 ; b^{\scaleto{\text{ {\fontfamily{qcr}\selectfont P}}}{2.5pt}}_{l}-a^{\scaleto{\text{ {\fontfamily{qcr}\selectfont P}}}{2.5pt}}_{l}+1 ; y^{\scaleto{\text{ {\fontfamily{qcr}\selectfont P}}}{2.5pt}}_{l}(B)^{-1}\right) \right)\\ \\
    & + y^{\scaleto{\text{ {\fontfamily{qcr}\selectfont P}}}{2.5pt}}_{l}(B)^{-a^{\scaleto{\text{ {\fontfamily{qcr}\selectfont P}}}{2pt}}_{l}} y^{\scaleto{\text{ {\fontfamily{qcr}\selectfont P}}}{2.5pt}}_{l}(x^{*})^{a^{\scaleto{\text{ {\fontfamily{qcr}\selectfont P}}}{2pt}}_{l} - b^{\scaleto{\text{ {\fontfamily{qcr}\selectfont P}}}{2pt}}_{l}} \left(b^{\scaleto{\text{ {\fontfamily{qcr}\selectfont P}}}{2.5pt}}_{u} { }_{2} \tilde{F}_{1}\left(b^{\scaleto{\text{ {\fontfamily{qcr}\selectfont P}}}{2.5pt}}_{u} + 1, b^{\scaleto{\text{ {\fontfamily{qcr}\selectfont P}}}{2.5pt}}_{u}-c^{\scaleto{\text{ {\fontfamily{qcr}\selectfont P}}}{2.5pt}}_{u}+1 ; b^{\scaleto{\text{ {\fontfamily{qcr}\selectfont P}}}{2.5pt}}_{u}-a^{\scaleto{\text{ {\fontfamily{qcr}\selectfont P}}}{2.5pt}}_{u}+1 ; y^{\scaleto{\text{ {\fontfamily{qcr}\selectfont P}}}{2.5pt}}_{u}(B)^{-1}\right) \right. \\ \\
    & { }_{2} \tilde{F}_{1}\left(a^{\scaleto{\text{ {\fontfamily{qcr}\selectfont P}}}{2.5pt}}_{l}, a^{\scaleto{\text{ {\fontfamily{qcr}\selectfont P}}}{2.5pt}}_{l}-c^{\scaleto{\text{ {\fontfamily{qcr}\selectfont P}}}{2.5pt}}_{l}+1 ; a^{\scaleto{\text{ {\fontfamily{qcr}\selectfont P}}}{2.5pt}}_{l}-b^{\scaleto{\text{ {\fontfamily{qcr}\selectfont P}}}{2.5pt}}_{l}+1 ; y^{\scaleto{\text{ {\fontfamily{qcr}\selectfont P}}}{2.5pt}}_{l}(B)^{-1}\right) - a^{\scaleto{\text{ {\fontfamily{qcr}\selectfont P}}}{2.5pt}}_{l} { }_{2} \tilde{F}_{1}\left(b^{\scaleto{\text{ {\fontfamily{qcr}\selectfont P}}}{2.5pt}}_{u}, b^{\scaleto{\text{ {\fontfamily{qcr}\selectfont P}}}{2.5pt}}_{u}-c^{\scaleto{\text{ {\fontfamily{qcr}\selectfont P}}}{2.5pt}}_{u}+1 ; b^{\scaleto{\text{ {\fontfamily{qcr}\selectfont P}}}{2.5pt}}_{u}-a^{\scaleto{\text{ {\fontfamily{qcr}\selectfont P}}}{2.5pt}}_{u}+1 ; y^{\scaleto{\text{ {\fontfamily{qcr}\selectfont P}}}{2.5pt}}_{u}(B)^{-1}\right)\\ \\
    & \left. { }_{2} \tilde{F}_{1}\left(a^{\scaleto{\text{ {\fontfamily{qcr}\selectfont P}}}{2.5pt}}_{l} + 1, a^{\scaleto{\text{ {\fontfamily{qcr}\selectfont P}}}{2.5pt}}_{l}-c^{\scaleto{\text{ {\fontfamily{qcr}\selectfont P}}}{2.5pt}}_{l}+1 ; a^{\scaleto{\text{ {\fontfamily{qcr}\selectfont P}}}{2.5pt}}_{l}-b^{\scaleto{\text{ {\fontfamily{qcr}\selectfont P}}}{2.5pt}}_{l}+1 ; y^{\scaleto{\text{ {\fontfamily{qcr}\selectfont P}}}{2.5pt}}_{l}(B)^{-1}\right)\right)\left(\left(\delta + \lambda\right)x^{*} { }_{2} \tilde{F}_{1}\left(b^{\scaleto{\text{ {\fontfamily{qcr}\selectfont P}}}{2.5pt}}_{l}, b^{\scaleto{\text{ {\fontfamily{qcr}\selectfont P}}}{2.5pt}}_{l}-c^{\scaleto{\text{ {\fontfamily{qcr}\selectfont P}}}{2.5pt}}_{l}+1 ; b^{\scaleto{\text{ {\fontfamily{qcr}\selectfont P}}}{2.5pt}}_{l}-a^{\scaleto{\text{ {\fontfamily{qcr}\selectfont P}}}{2.5pt}}_{l}+1 ; y^{\scaleto{\text{ {\fontfamily{qcr}\selectfont P}}}{2.5pt}}_{l}(x^{*})^{-1}\right) \right. \\ \\
    & \left. \left. \left. c_{{\scaleto{T}{3pt}}} b^{\scaleto{\text{ {\fontfamily{qcr}\selectfont P}}}{2.5pt}}_{l} \left(B - x^{*}\right) { }_{2} \tilde{F}_{1}\left(b^{\scaleto{\text{ {\fontfamily{qcr}\selectfont P}}}{2.5pt}}_{l} + 1, b^{\scaleto{\text{ {\fontfamily{qcr}\selectfont P}}}{2.5pt}}_{l}-c^{\scaleto{\text{ {\fontfamily{qcr}\selectfont P}}}{2.5pt}}_{l}+1 ; b^{\scaleto{\text{ {\fontfamily{qcr}\selectfont P}}}{2.5pt}}_{l}-a^{\scaleto{\text{ {\fontfamily{qcr}\selectfont P}}}{2.5pt}}_{l}+1 ; y^{\scaleto{\text{ {\fontfamily{qcr}\selectfont P}}}{2.5pt}}_{l}(x^{*})^{-1}\right) \right) \right) \right],
    \label{Appendix A: Mathematical Proofs-Equation17}
\end{align}
}

\normalsize

\vspace{0.3cm}

where ${ }_2 \tilde{F}_1(a, b ; c ; z)=\frac{_2 F_1(a, b ; c ; z)}{\Gamma(c)}$ denotes the Regularised Hypergeometric Function. Therefore, the Laplace transform of the trapping time is given by \eqref{TheTrappingTime-Subsection31-Equation3}.


\subsubsection{Proof of Theorem \ref{WhenandHowHouseholdsBecomeExtremelyPoor?-Section4-Theorem1}} \label{ProofofTheorem4.1}

Using similar arguments as those for Theorem \ref{WhenandHowHouseholdsBecomePoor?-Section3-Theorem1}, we know that for $x\geq B$, the capital immediately before the first capital loss is $h_{r}(t,x) = (x-x^{*})e^{rt}+x^{*}$ and the capital has three possibilities at time $t$, that it is more than $B$, that it is between $x^{*}$ and $B$, and that it is between $0$ and $x^{*}$. Thus, by conditioning on the time and the remaining proportion of the first capital loss and discounting the expected values to time $0$ at the force of interest $\delta$, when $x\geq B$ we obtain

\vspace{0.3cm}

{\allowdisplaybreaks
\begin{align}
	m^{\scaleto{\text{ {\fontfamily{qcr}\selectfont EP}}}{2.5pt}}_{\delta,u}(x)&=\int^{\infty}_{0}\lambda e^{-(\lambda + \delta)t}\left[\int^{x^{*}/h_{r}(t,x)}_{0} m^{\scaleto{\text{ {\fontfamily{qcr}\selectfont EP}}}{2.5pt}}_{\delta,l}( h_{r}(t,x) \cdot z) dG_{Z}(z) + \int^{B/h_{r}(t,x)}_{x^{*}/h_{r}(t,x)} m^{\scaleto{\text{ {\fontfamily{qcr}\selectfont EP}}}{2.5pt}}_{\delta,m}(h_{r}(t,x) \cdot z) dG_{Z}(z) \right. \\ \\
	& \left. + \int^{1}_{B/h_{r}(t,x)} m^{\scaleto{\text{ {\fontfamily{qcr}\selectfont EP}}}{2.5pt}}_{\delta, u} (h_{r}(t,x) \cdot z) dG_{Z}(z)\right] dt
	\label{Appendix A: Mathematical Proofs-Equation18}
\end{align}
}

\vspace{0.3cm}

Then, doing the change of variable $v_{u}=h_{r}(t,x)$ in the integrals with respect to $t$ from $0$ to $\infty$ in \eqref{Appendix A: Mathematical Proofs-Equation18}, we obtain \eqref{WhenandHowHouseholdsBecomeExtremelyPoor?-Section4-Equation2}.

For $x^{*}\leq x < B$, there are two possibilities. First, $t<\tau_{{\scaleto{B}{3pt}}}$ and the capital has not yet reached the capital barrier level $B$. In this case, we know the capital immediately before time $t$ is $h_{r-c_{{\scaleto{T}{2.5pt}}}}(t,x)=(x+x^{**})e^{(r-c_{{\scaleto{T}{2.5pt}}})t}-x^{**}$ and the capital has two possibilities at time $t$, that it is between $x^{*}$ and $B$, and that it is between $0$ and $x^{*}$. Second, for $t >\tau_{{\scaleto{B}{3pt}}}$, that is, no capital loss occurs before the capital exceeds the capital barrier $B$. In this case, we also know the capital immediately before time $t$ is $h_{r}(t- \tau_{{\scaleto{B}{3pt}}},B)=(B-x^{*})e^{r\left(t  - \tau_{{\scaleto{B}{2pt}}}\right)}+x^{*}$ and the accumulated capital has three possibilities at time $t$, that it is more than $B$, that it is between $x^{*}$ and $B$, and that it is between $0$ and $x^{*}$. Hence, by conditioning on the time and the remaining proportion of the first capital loss and discounting the expected values to time $0$ at the force of interest $\delta$, when $x^{*}\leq x < B$ we obtain

\vspace{0.3cm}

{\allowdisplaybreaks
\begin{align}
	&m^{\scaleto{\text{ {\fontfamily{qcr}\selectfont EP}}}{2.5pt}}_{\delta,m}(x)=\int^{\tau_{{\scaleto{B}{2pt}}}}_{0}\lambda e^{-(\lambda+\delta)t}\left[\int^{x^{*}/h_{\scaleto{r-c_{{\scaleto{T}{2.5pt}}}}{4pt}}(t,x)}_{0}m^{\scaleto{\text{ {\fontfamily{qcr}\selectfont EP}}}{2.5pt}}_{\delta,l}(h_{r-c_{{\scaleto{T}{2.5pt}}}}(t,x) \cdot z) dG_{Z}(z) \right.\\ \\
	& \left. + \int^{1}_{x^{*}/h_{\scaleto{r-c_{{\scaleto{T}{2.5pt}}}}{4pt}}(t,x)}m^{\scaleto{\text{ {\fontfamily{qcr}\selectfont EP}}}{2.5pt}}_{\delta,m}(h_{r-c_{{\scaleto{T}{2.5pt}}}}(t,x) \cdot z) dG_{Z}(z)\right]dt \\ \\ 
	& + \int^{\infty}_{\tau_{{\scaleto{B}{2pt}}}} \lambda e^{-(\lambda + \delta)t} \left[\int^{x^{*}/h_{r}(t- \tau_{{\scaleto{B}{2pt}}},B)}_{0} m^{\scaleto{\text{ {\fontfamily{qcr}\selectfont EP}}}{2.5pt}}_{\delta,l}(h_{r}(t- \tau_{{\scaleto{B}{3pt}}},B) \cdot z) dG_{Z}(z) 
	 \right. \\ \\ 
	 & \left. + \int^{B/h_{r}(t- \tau_{{\scaleto{B}{2pt}}},B)}_{x^{*}/h_{r}(t- \tau_{{\scaleto{B}{2pt}}},B)} m^{\scaleto{\text{ {\fontfamily{qcr}\selectfont EP}}}{2.5pt}}_{\delta,m}(h_{r}(t- \tau_{{\scaleto{B}{3pt}}},B) \cdot z) dG_{Z}(z)  +  \int^{1}_{B/h_{r}(t- \tau_{{\scaleto{B}{2pt}}},B)} m^{\scaleto{\text{ {\fontfamily{qcr}\selectfont EP}}}{2.5pt}}_{\delta,u}(h_{r}(t- \tau_{{\scaleto{B}{3pt}}},B) \cdot z) dG_{Z}(z) \right]dt \\ \\
	\label{Appendix A: Mathematical Proofs-Equation19}
\end{align}
}

\vspace{0.3cm}

Now, first changing variables $v_{m}=h_{r-c_{{\scaleto{T}{2.5pt}}}}(t,x)$ in the integrals with respect to $t$ from $0$ to $\tau_{{\scaleto{B}{3pt}}}$ in \eqref{Appendix A: Mathematical Proofs-Equation19}, and then changing variables $v_{u}=h_{r}(t- \tau_{{\scaleto{B}{3pt}}},B)$ in the integrals with respect to $t$ from $\tau_{{\scaleto{B}{3pt}}}$ to $\infty$ in \eqref{Appendix A: Mathematical Proofs-Equation19}, we obtain \eqref{WhenandHowHouseholdsBecomeExtremelyPoor?-Section4-Equation3}.

For $0 < x < x^{*}$, let $\tau_{x^{*}}:=\tau_{x^{*}}(x)$ be the solution to 

\vspace{0.3cm}

{\allowdisplaybreaks
\begin{align}
	h_{c_{{\scaleto{T}{2.5pt}}}}(t,x)=(x-B)e^{-c_{{\scaleto{T}{2.5pt}}}t}+B=x^{*}.
	\label{Appendix A: Mathematical Proofs-Equation20}
\end{align}
}

Namely, $\tau_{x^{*}}:=\tau_{x^{*}}(x)=-\frac{1}{c_{{\scaleto{T}{2.5pt}}}}ln\left(\frac{x^{*}-B}{x-B}\right)$, which is the time when the capital returns to the critical capital $x^{*}$ if no capital loss occurs prior to time $\tau_{x^{*}}$. Furthermore, $h_{c_{{\scaleto{T}{2.5pt}}}}(t,x)<x^{*}$ for $t<\tau_{x^{*}}$ and $h_{c_{{\scaleto{T}{2.5pt}}}}(\tau_{x^{*}},x)=x^{*}$. Moreover, $h_{c_{{\scaleto{T}{2.5pt}}}}(t,x)$ is the capital at time $t \leq \tau_{x^{*}}$ if no capital loss occurs prior to time $\tau_{x^{*}}$. 

Thus, for $0 < x < x^{*}$, there are three possibilities. First, $t<\tau_{x^{*}}$ and the capital up to time $t$ has not reached the critical capital $x^{*}$. In this case, the capital immediately before time $t$ is $h_{c_{{\scaleto{T}{2.5pt}}}}(t,x)=(x-B)e^{-c_{{\scaleto{T}{2.5pt}}}t}+B$. Second, $\tau_{x^{*}} \leq t< \tau_{x^{*}}+\tau_{{\scaleto{B}{3pt}}}\left(x^{*}\right)$ and the capital has not yet reached the capital barrier level $B$ and no capital loss occurs before the capital exceeds the critical capital $x^{*}$. In this case, the capital immediately before time $t$ is $h_{r-c_{{\scaleto{T}{2.5pt}}}}(t - \tau_{x^{*}},x^{*})=(x^{*}+x^{**})e^{(r-c_{{\scaleto{T}{2.5pt}}})\left(t-\tau_{x^{*}}\right)}-x^{**}$ and the capital up to time $t$ has two possibilities, that it is between $x^{*}$ and $B$, and that it is between $0$ and $x^{*}$. Third, $t\geq \tau_{x^{*}} + \tau_{{\scaleto{B}{3pt}}}\left(x^{*}\right)$, that is, no capital loss occurs before the capital up to time $t$ exceeds the capital barrier level $B$. In this case, the capital immediately before time $t$ is $h_{r}(t-\tau_{x^{*}} - \tau_{{\scaleto{B}{3pt}}}\left(x^{*}\right),B)=(B-x^{*})e^{r\left(t-\tau_{x^{*}} - \tau_{{\scaleto{B}{2pt}}}\left(x^{*}\right)\right)}+x^{*}$ and the capital up to time $t$ has three possibilities, that it is more than $B$, that it is between $x^{*}$ and $B$, and that it is between $0$ and $x^{*}$. Hence, by conditioning on the time and the remaining proportion of the first capital loss and discounting the expected values to time 0 at the force of interest $\delta$, when $0 < x < x^{*} $ we obtain

\vspace{0.3cm}

{\allowdisplaybreaks
\begin{align}
	&m^{\scaleto{\text{ {\fontfamily{qcr}\selectfont EP}}}{2.5pt}}_{\delta,l}(x)=\int^{\tau_{x^{*}}}_{0} e^{-(\lambda+\delta)t}e^{-\int^{t}_{0} \omega \left(h_{c_{{\scaleto{T}{2.5pt}}}}\left(y,x\right)\right) dy} \omega\left(h_{c_{{\scaleto{T}{2.5pt}}}}(t,x) \right) w^{\scaleto{\text{ {\fontfamily{qcr}\selectfont EP}}}{2.5pt}}\left( h_{c_{{\scaleto{T}{2.5pt}}}}(t,x), x^{*} - h_{c_{{\scaleto{T}{2.5pt}}}}(t,x)\right)dt\\ \\
	&+ \int^{\tau_{x^{*}}}_{0} \lambda e^{-(\lambda+\delta)t}e^{-\int^{t}_{0} \omega \left(h_{c_{{\scaleto{T}{2.5pt}}}}\left(y,x\right)\right) dy} \int^{1}_{0}m^{\scaleto{\text{ {\fontfamily{qcr}\selectfont EP}}}{2.5pt}}_{\delta,l}(h_{c_{{\scaleto{T}{2.5pt}}}}(t,x) \cdot z) dG_{Z}(z)dt\\ \\
	&+ \int^{\tau_{x^{*}} + \tau_{{\scaleto{B}{2pt}}}\left(x^{*}\right)}_{\tau_{x^{*}}} \lambda e^{-(\lambda+\delta)t}e^{-\int^{\tau_{x^{*}} }_{0} \omega \left(h_{c_{{\scaleto{T}{2.5pt}}}}\left(y,x\right)\right) dy} \left[ \int^{x^{*}/h_{r-c_{{\scaleto{T}{2.5pt}}}}(t - \tau_{x^{*}},x^{*})}_{0} m^{\scaleto{\text{ {\fontfamily{qcr}\selectfont EP}}}{2.5pt}}_{\delta,l}(h_{r-c_{{\scaleto{T}{2.5pt}}}}(t - \tau_{x^{*}},x^{*})\cdot z) dG_{Z}(z)\right.\\ \\
	&\left. + \int^{1}_{x^{*}/h_{r-c_{{\scaleto{T}{2.5pt}}}}(t - \tau_{x^{*}},x^{*})} m^{\scaleto{\text{ {\fontfamily{qcr}\selectfont EP}}}{2.5pt}}_{\delta,m}(h_{r-c_{{\scaleto{T}{2.5pt}}}}(t - \tau_{x^{*}},x^{*}) \cdot z) dG_{Z}(z) \right]dt + \int^{\infty}_{\tau_{x^{*}} + \tau_{{\scaleto{B}{3pt}}}\left(x^{*}\right)} \lambda e^{-(\lambda+\delta)t}e^{-\int^{\tau_{x^{*}} }_{0} \omega \left(h_{c_{{\scaleto{T}{2.5pt}}}}\left(y,x\right)\right) dy} \\ \\
	& \left[\int^{x^{*}/h_{r}(t-\tau_{x^{*}} - \tau_{{\scaleto{B}{2pt}}}\left(x^{*}\right),B)}_{0} m^{\scaleto{\text{ {\fontfamily{qcr}\selectfont EP}}}{2.5pt}}_{\delta,l}(h_{r}(t-\tau_{x^{*}} - \tau_{{\scaleto{B}{3pt}}}\left(x^{*}\right),B) \cdot z) dG_{Z}(z) \right. \\ \\
	& + \int^{B/h_{r}(t-\tau_{x^{*}} - \tau_{{\scaleto{B}{2pt}}}\left(x^{*}\right),B)}_{x^{*}/h_{r}(t-\tau_{x^{*}} - \tau_{{\scaleto{B}{2pt}}}\left(x^{*}\right),B)} m^{\scaleto{\text{ {\fontfamily{qcr}\selectfont EP}}}{2.5pt}}_{\delta,m}(h_{r}(t-\tau_{x^{*}} - \tau_{{\scaleto{B}{3pt}}}\left(x^{*}\right),B) \cdot z) dG_{Z}(z) \\ \\
	& \left. + \int^{1}_{B/h_{r}(t-\tau_{x^{*}} - \tau_{{\scaleto{B}{2pt}}}\left(x^{*}\right),B)} m^{\scaleto{\text{ {\fontfamily{qcr}\selectfont EP}}}{2.5pt}}_{\delta,u}(h_{r}(t-\tau_{x^{*}} - \tau_{{\scaleto{B}{3pt}}}\left(x^{*}\right),B) \cdot z) dG_{Z}(z) \right]dt
	\label{Appendix A: Mathematical Proofs-Equation21}
\end{align}
}

\vspace{0.3cm}

Now, first changing variables $v_{l}=h_{c_{{\scaleto{T}{2.5pt}}}}(t,x)$ and $u_{l}=h_{c_{{\scaleto{T}{2.5pt}}}}(y,x)$ in the integrals with respect to $t$ from $0$ to $\tau_{x^{*}}$ in \eqref{Appendix A: Mathematical Proofs-Equation21}, then changing variables $v_{m}=h_{r-c_{{\scaleto{T}{2.5pt}}}}(t - \tau_{x^{*}},x^{*})$ in the integrals with respect to $t$ from $\tau_{x^{*}}$ to $\tau_{x^{*}}+\tau_{{\scaleto{B}{3pt}}}\left(x^{*}\right)$ in \eqref{Appendix A: Mathematical Proofs-Equation21} and lastly changing variables $v_{u}=h_{r}(t-\tau_{x^{*}} - \tau_{{\scaleto{B}{3pt}}}\left(x^{*}\right),B)$ in the integrals with respect to $t$ from $\tau_{x^{*}}+\tau_{{\scaleto{B}{3pt}}}\left(x^{*}\right)$ to $\infty$ in \eqref{Appendix A: Mathematical Proofs-Equation21} we obtain \eqref{WhenandHowHouseholdsBecomeExtremelyPoor?-Section4-Equation4}.


\subsubsection{Proof of Proposition \ref{ExamplesofExtremePovertyRateFunctions-Subsection411-Proposition1}} \label{ProofofProposition4.1} 

When $Z_{i}\sim Beta(\alpha, 1)$, i.e. $g_{Z}(z)=\alpha z^{\alpha - 1}\mathbbm{1}_{\{0 < z < 1\}}$ with $\alpha>0$, equation \eqref{WhenandHowHouseholdsBecomeExtremelyPoor?-Section4-Equation7}, \eqref{WhenandHowHouseholdsBecomeExtremelyPoor?-Section4-Equation8} and \eqref{TheTimeofExtremePoverty-Subsection41-Equation1} can be written such that when $x \geq B$,

\vspace{0.3cm}

{\allowdisplaybreaks
\begin{align}
        0&= r(x-x^{*})m^{\scaleto{\text{ {\fontfamily{qcr}\selectfont EP}}}{2.5pt}}_{\delta,u}(x)_{\delta,u}'(x)-(\lambda + \delta)m^{\scaleto{\text{ {\fontfamily{qcr}\selectfont EP}}}{2.5pt}}_{\delta,u}(x) + \lambda \left[\int^{x^{*}/x}_{0} m^{\scaleto{\text{ {\fontfamily{qcr}\selectfont EP}}}{2.5pt}}_{\delta,l}(x\cdot z)\alpha z^{\alpha - 1}dz \right. \\ \\ 
	& \left. + \int^{B/x}_{x^{*}/x} m^{\scaleto{\text{ {\fontfamily{qcr}\selectfont EP}}}{2.5pt}}_{\delta,m} (x\cdot z) \alpha z^{\alpha - 1}dz + \int^{1}_{B/x} m^{\scaleto{\text{ {\fontfamily{qcr}\selectfont EP}}}{2.5pt}}_{\delta, u} (x\cdot z) \alpha z^{\alpha - 1} dz\right],  \label{Appendix A: Mathematical Proofs-Equation22}
\end{align}
}

\vspace{0.3cm}

when $x^{*} \leq x <  B$,

{\allowdisplaybreaks
\begin{align}
        0&=\left(r - c_{{\scaleto{T}{4pt}}}\right)\left(x + x^{**}\right)m'^{\scaleto{\text{ {\fontfamily{qcr}\selectfont EP}}}{2.5pt}}_{\delta,m}(x)-(\lambda + \delta)m^{\scaleto{\text{ {\fontfamily{qcr}\selectfont EP}}}{2.5pt}}_{\delta,m}(x) + \lambda \left[\int^{x^{*}/x}_{0}m^{\scaleto{\text{ {\fontfamily{qcr}\selectfont EP}}}{2.5pt}}_{\delta,l}(x\cdot z) \alpha z^{\alpha - 1} dz \right. \\ \\ 
	& \left. + \int^{1}_{x^{*}/x}m^{\scaleto{\text{ {\fontfamily{qcr}\selectfont EP}}}{2.5pt}}_{\delta,m}(x\cdot z)\alpha z^{\alpha - 1} dz\right],
        \label{Appendix A: Mathematical Proofs-Equation23}
\end{align}
}

\vspace{0.3cm}

and when $0 < x <  x^{*}$,

{\allowdisplaybreaks
\begin{align}
0&=c_{{\scaleto{T}{4pt}}}(x-B)m'^{\scaleto{\text{ {\fontfamily{qcr}\selectfont EP}}}{2.5pt}}_{\delta,l}(x)+[\lambda + \delta + \omega_{c}]m^{\scaleto{\text{ {\fontfamily{qcr}\selectfont EP}}}{2.5pt}}_{\delta,l}(x)-\omega_{c}-\lambda\int^{1}_{0}m^{\scaleto{\text{ {\fontfamily{qcr}\selectfont EP}}}{2.5pt}}_{\delta,l}(x\cdot z)\alpha z^{\alpha - 1}dz.
        \label{Appendix A: Mathematical Proofs-Equation24}
\end{align}
}

\vspace{0.3cm}

Applying the operator $\frac{d}{dx}$ to both sides of \eqref{Appendix A: Mathematical Proofs-Equation22}, \eqref{Appendix A: Mathematical Proofs-Equation23} and \eqref{Appendix A: Mathematical Proofs-Equation24}, together with a number of algebraic manipulations, yields to the following second order ODEs,

\vspace{0.3cm}

{\allowdisplaybreaks
\begin{align}
        x \geq B: 0 &= r(x^{2}-xx^{*})m''^{\scaleto{\text{ {\fontfamily{qcr}\selectfont EP}}}{2.5pt}}_{\delta,u}(x)+\left[(r(1+\alpha)-\delta-\lambda)x-r\alpha x^{*}\right]m'^{\scaleto{\text{ {\fontfamily{qcr}\selectfont EP}}}{2.5pt}}_{\delta,u}(x)-\alpha \delta m^{\scaleto{\text{ {\fontfamily{qcr}\selectfont EP}}}{2.5pt}}_{\delta,u}(x),  \quad \label{Appendix A: Mathematical Proofs-Equation25}
\end{align}
}

\vspace{0.3cm}

{\allowdisplaybreaks
\begin{align}
        x^{*} \leq x <  B: 0&= (r - c_{{\scaleto{T}{4pt}}})(x^{2}+xx^{**}) m''^{\scaleto{\text{ {\fontfamily{qcr}\selectfont EP}}}{2.5pt}}_{\delta,m}(x) \\ \\ 
        &+ \left[ \left(\left(r - c_{{\scaleto{T}{4pt}}}\right) \left( 1 + \alpha\right) - \delta - \lambda \right) x + \alpha \left(r - c_{{\scaleto{T}{4pt}}}\right)x^{**} \right] m'^{\scaleto{\text{ {\fontfamily{qcr}\selectfont EP}}}{2.5pt}}_{\delta,m}(x) - \alpha\delta m^{\scaleto{\text{ {\fontfamily{qcr}\selectfont EP}}}{2.5pt}}_{\delta,m}(x) \quad
        \label{Appendix A: Mathematical Proofs-Equation26} 
\end{align}
}

\vspace{0.3cm}        

and

\vspace{0.3cm}     

{\allowdisplaybreaks
\begin{align}        
        0 < x <  x^{*} : 0&= c_{{\scaleto{T}{4pt}}}(x^{2}-Bx) m''^{\scaleto{\text{ {\fontfamily{qcr}\selectfont EP}}}{2.5pt}}_{\delta,l}(x) \\ \\ 
        &+ \left[ \left(c_{{\scaleto{T}{4pt}}} \left( 1 + \alpha\right) + \delta + \lambda + \omega_{c} \right) x - \alpha c_{{\scaleto{T}{4pt}}}B \right] m'^{\scaleto{\text{ {\fontfamily{qcr}\selectfont EP}}}{2.5pt}}_{\delta,l}(x) + \alpha \left(\delta + \omega_{c}\right) m^{\scaleto{\text{ {\fontfamily{qcr}\selectfont EP}}}{2.5pt}}_{\delta,l}(x) - \alpha w_{c}. \\
        \label{Appendix A: Mathematical Proofs-Equation27}
\end{align}
}

\vspace{0.3cm}

Hence, for $0<x<x^{*}$, $m^{\scaleto{\text{ {\fontfamily{qcr}\selectfont EP}}}{2.5pt}}_{\delta,l}(x)$ satisfies the nonhomogeneous differential equation \eqref{Appendix A: Mathematical Proofs-Equation27}, when the extreme poverty rate function $\omega_{1}(x)=\omega_{c}$ (constant value) and the penalty function $w^{\scaleto{\text{ {\fontfamily{qcr}\selectfont EP}}}{2.5pt}}(x_{1}, x_{2})=1$. The particular solution of $m^{\scaleto{\text{ {\fontfamily{qcr}\selectfont EP}}}{2.5pt}}_{\delta,l}(x)$ is

\vspace{0.3cm}     

{\allowdisplaybreaks
\begin{align}        
        m^{* \scaleto{\text{ {\fontfamily{qcr}\selectfont EP}}}{2.5pt}}_{\delta,l}(x) = \frac{\omega_{c}}{\delta + \omega_{c}}.
        \label{Appendix A: Mathematical Proofs-Equation28}
\end{align}
}

\vspace{0.3cm}

Therefore, the general solution of $m^{\scaleto{\text{ {\fontfamily{qcr}\selectfont EP}}}{2.5pt}}_{\delta,l}(x)$ is given by 

{\allowdisplaybreaks
\begin{align}        
        m^{\scaleto{\text{ {\fontfamily{qcr}\selectfont EP}}}{2.5pt}}_{\delta,l}(x) =  h^{\scaleto{\text{ {\fontfamily{qcr}\selectfont EP}}}{2.5pt}}_{l}(x) + \frac{\omega_{c}}{\delta + \omega_{c}},
        \label{Appendix A: Mathematical Proofs-Equation29}
\end{align}
}

where $h^{\scaleto{\text{ {\fontfamily{qcr}\selectfont EP}}}{2.5pt}}_{l}(x)$ is the homogeneous solution of \eqref{Appendix A: Mathematical Proofs-Equation27}. Then, following a similar procedure to that of Proposition \ref{TheTrappingTime-Subsection21-Proposition1}, letting $f^{\scaleto{\text{ {\fontfamily{qcr}\selectfont EP}}}{2.5pt}}_{l}(y^{\scaleto{\text{ {\fontfamily{qcr}\selectfont EP}}}{2.5pt}}_{l}):= h^{\scaleto{\text{ {\fontfamily{qcr}\selectfont EP}}}{2.5pt}}_{l}(x)$, such that $y^{\scaleto{\text{ {\fontfamily{qcr}\selectfont EP}}}{2.5pt}}_{l}$ is associated with the change of variable $y^{\scaleto{\text{ {\fontfamily{qcr}\selectfont EP}}}{2.5pt}}_{l}:=y^{\scaleto{\text{ {\fontfamily{qcr}\selectfont EP}}}{2.5pt}}_{l}(x)=\frac{x}{B}$, the homogeneous part of equation \eqref{Appendix A: Mathematical Proofs-Equation27} reduces to equation \eqref{Appendix A: Mathematical Proofs-Equation8} for $c^{\scaleto{\text{ {\fontfamily{qcr}\selectfont EP}}}{2.5pt}}_{l}=\alpha$, $a^{\scaleto{\text{ {\fontfamily{qcr}\selectfont EP}}}{2.5pt}}_{l}=\frac{\alpha c_{{\scaleto{T}{2.5pt}}} + \lambda + \delta + \omega_{c} - \sqrt{(\alpha c_{{\scaleto{T}{2.5pt}}} + \lambda +\delta + \omega_{c})^{2}-4 \alpha c_{{\scaleto{T}{2.5pt}}}\left(\delta + \omega_{c}\right)}}{2 c_{{\scaleto{T}{2.5pt}}}}$ and $b^{\scaleto{\text{ {\fontfamily{qcr}\selectfont EP}}}{2.5pt}}_{l}=\frac{\alpha c_{{\scaleto{T}{2.5pt}}} + \lambda + \delta + \omega_{c} + \sqrt{(\alpha c_{{\scaleto{T}{2.5pt}}} + \lambda +\delta + \omega_{c})^{2}-4 \alpha c_{{\scaleto{T}{2.5pt}}}\left(\delta + \omega_{c}\right)}}{2 c_{{\scaleto{T}{2.5pt}}}}$, with regular singular points at $y^{\scaleto{\text{ {\fontfamily{qcr}\selectfont EP}}}{2.5pt}}_{l}=0, 1, \infty$ (corresponding to $x=0,B$ and $\infty$). A general solution of \eqref{Appendix A: Mathematical Proofs-Equation8} in the neighborhood of the singular point $y^{\scaleto{\text{ {\fontfamily{qcr}\selectfont EP}}}{2.5pt}}_{l}=0$ is given by 

\vspace{0.3cm}

\begin{align}
    f^{\scaleto{\text{ {\fontfamily{qcr}\selectfont EP}}}{2.5pt}}_{l}(y^{\scaleto{\text{ {\fontfamily{qcr}\selectfont EP}}}{2.5pt}}_{l}):=& h^{\scaleto{\text{ {\fontfamily{qcr}\selectfont EP}}}{2.5pt}}_{l}(x) = A^{\scaleto{\text{ {\fontfamily{qcr}\selectfont EP}}}{2.5pt}}_{1,l} { }_{2} F_{1}\left(a^{\scaleto{\text{ {\fontfamily{qcr}\selectfont EP}}}{2.5pt}}_{l}, b^{\scaleto{\text{ {\fontfamily{qcr}\selectfont EP}}}{2.5pt}}_{l} ; c^{\scaleto{\text{ {\fontfamily{qcr}\selectfont EP}}}{2.5pt}}_{l} ; y^{\scaleto{\text{ {\fontfamily{qcr}\selectfont EP}}}{2.5pt}}_{l}(x)\right)\\ \\
    &+A^{\scaleto{\text{ {\fontfamily{qcr}\selectfont EP}}}{2.5pt}}_{2,l}{y^{\scaleto{\text{ {\fontfamily{qcr}\selectfont EP}}}{2.5pt}}_{l}(x)}^{1-c^{\scaleto{\text{ {\fontfamily{qcr}\selectfont EP}}}{2pt}}_{l}} { }_{2} F_{1}\left(a^{\scaleto{\text{ {\fontfamily{qcr}\selectfont EP}}}{2.5pt}}_{l}-c^{\scaleto{\text{ {\fontfamily{qcr}\selectfont EP}}}{2.5pt}}_{l}+1, b^{\scaleto{\text{ {\fontfamily{qcr}\selectfont EP}}}{2.5pt}}_{l}-c^{\scaleto{\text{ {\fontfamily{qcr}\selectfont EP}}}{2.5pt}}_{l}+1 ; 2-c^{\scaleto{\text{ {\fontfamily{qcr}\selectfont EP}}}{2.5pt}}_{l} ; y^{\scaleto{\text{ {\fontfamily{qcr}\selectfont EP}}}{2.5pt}}_{l}(x)\right),
    \label{Appendix A: Mathematical Proofs-Equation30}
\end{align}

\vspace{0.3cm}

for arbitrary constants $A^{\scaleto{\text{ {\fontfamily{qcr}\selectfont EP}}}{2.5pt}}_{1,l},A^{\scaleto{\text{ {\fontfamily{qcr}\selectfont EP}}}{2.5pt}}_{2,l} \in \mathbb {R}$ (see for example, equations (15.5.3) and (15.5.4) of \cite{Book:Abramowitz1964}). Due to the fact that $m^{\scaleto{\text{ {\fontfamily{qcr}\selectfont EP}}}{2.5pt}}_{\delta,l}(x)$ is finite, we can then conclude that $A^{\scaleto{\text{ {\fontfamily{qcr}\selectfont EP}}}{2.5pt}}_{2,l}=0$, as the second term of \eqref{Appendix A: Mathematical Proofs-Equation30} is unbounded when $x \rightarrow 0^{+}$ for $\alpha > 0$. Thus, the solution of $m^{\scaleto{\text{ {\fontfamily{qcr}\selectfont EP}}}{2.5pt}}_{\delta,l}(x)$ is given by

\vspace{0.3cm}

\begin{align}
m^{\scaleto{\text{ {\fontfamily{qcr}\selectfont EP}}}{2.5pt}}_{\delta,l}(x) =  A^{\scaleto{\text{ {\fontfamily{qcr}\selectfont EP}}}{2.5pt}}_{1,l} { }_{2} F_{1}\left(a^{\scaleto{\text{ {\fontfamily{qcr}\selectfont EP}}}{2.5pt}}_{l}, b^{\scaleto{\text{ {\fontfamily{qcr}\selectfont EP}}}{2.5pt}}_{l} ; c^{\scaleto{\text{ {\fontfamily{qcr}\selectfont EP}}}{2.5pt}}_{l} ; y^{\scaleto{\text{ {\fontfamily{qcr}\selectfont EP}}}{2.5pt}}_{l}(x)\right) + \frac{\omega_{c}}{\delta + \omega_{c}}.
    \label{Appendix A: Mathematical Proofs-Equation31}
\end{align}

\vspace{0.3cm}

Then, following the proof of Proposition \ref{TheTrappingTime-Subsection21-Proposition1}, one can easily obtain the solutions for $m^{\scaleto{\text{ {\fontfamily{qcr}\selectfont EP}}}{2.5pt}}_{\delta,u}(x)$ and $m^{\scaleto{\text{ {\fontfamily{qcr}\selectfont EP}}}{2.5pt}}_{\delta,m}(x)$, when $x\geq B$ and  $x^{*}\leq x < B$, respectively.

Considering the continuity of $m^{\scaleto{\text{ {\fontfamily{qcr}\selectfont EP}}}{2.5pt}}_{\delta}(x)$ and $m'^{\scaleto{\text{ {\fontfamily{qcr}\selectfont EP}}}{2.5pt}}_{\delta}(x)$ at the critical capital $x^{*}$ and the capital barrier level $B$, that is, using \eqref{WhenandHowHouseholdsBecomeExtremelyPoor?-Section4-Equation5}, \eqref{WhenandHowHouseholdsBecomeExtremelyPoor?-Section4-Equation6}, \eqref{WhenandHowHouseholdsBecomeExtremelyPoor?-Section4-Equation10} and \eqref{WhenandHowHouseholdsBecomeExtremelyPoor?-Section4-Equation11}, one can derive a system of equations from which the unknown coefficients $A^{\scaleto{\text{ {\fontfamily{qcr}\selectfont EP}}}{2.5pt}}_{2,u}$, $A^{\scaleto{\text{ {\fontfamily{qcr}\selectfont EP}}}{2.5pt}}_{1,m}$, $A^{\scaleto{\text{ {\fontfamily{qcr}\selectfont EP}}}{2.5pt}}_{2,m}$ and $A^{\scaleto{\text{ {\fontfamily{qcr}\selectfont EP}}}{2.5pt}}_{1,l}$, can be determined to obtain an explicit solution for $m^{\scaleto{\text{ {\fontfamily{qcr}\selectfont EP}}}{2.5pt}}_{\delta}(x)$.


\subsubsection{Proof of Proposition \ref{ExamplesofExtremePovertyRateFunctions-Subsection411-Proposition2}} \label{ProofofProposition4.2} 

Following a similar procedure to that in Appendix \ref{ProofofProposition4.1}, for $0<x<x^{*}$, one can derive from \eqref{TheTimeofExtremePoverty-Subsection41-Equation1} the following nonhomogeneous second order ODE for $\psi^{\scaleto{\text{ {\fontfamily{qcr}\selectfont EP}}}{2.5pt}}_{l}(x)$, when the extreme poverty rate function $\omega_{2}(x)=\frac{\beta}{x}$ (exponential extreme poverty rate), the penalty function $w^{\scaleto{\text{ {\fontfamily{qcr}\selectfont EP}}}{2.5pt}}(x_{1}, x_{2})=1$ and the force of interest $\delta = 0$,

\vspace{0.3cm}

\begin{align}
x^{2}\left(x-B\right)\psi''^{\scaleto{\text{ {\fontfamily{qcr}\selectfont EP}}}{2.5pt}}_{l}(x)+x\left[\frac{c_{{\scaleto{T}{4pt}}}\left(1+\alpha\right) + \lambda}{c_{{\scaleto{T}{4pt}}}}x + \frac{\beta-\alpha B c_{{\scaleto{T}{4pt}}}}{c_{{\scaleto{T}{4pt}}}}\right]\psi'^{\scaleto{\text{ {\fontfamily{qcr}\selectfont EP}}}{2.5pt}}_{l}(x)+\frac{\beta\left(\alpha -1\right)}{c_{{\scaleto{T}{4pt}}}}\psi^{\scaleto{\text{ {\fontfamily{qcr}\selectfont EP}}}{2.5pt}}_{l}(x)- \frac{\beta\left(\alpha -1\right)}{c_{{\scaleto{T}{4pt}}}}=0.
    \label{Appendix A: Mathematical Proofs-Equation32}
\end{align}

\vspace{0.3cm}

Clearly, $\psi^{*\scaleto{\text{ {\fontfamily{qcr}\selectfont EP}}}{2.5pt}}_{l}(x) = 1$ is always a particular solution of equation \eqref{Appendix A: Mathematical Proofs-Equation32}, so that one can write

\vspace{0.3cm}

{\allowdisplaybreaks
\begin{align}        
        \psi^{\scaleto{\text{ {\fontfamily{qcr}\selectfont EP}}}{2.5pt}}_{l}(x) =  h^{\scaleto{\text{ {\fontfamily{qcr}\selectfont EP}}}{2.5pt}}_{l}(x) + 1,
        \label{Appendix A: Mathematical Proofs-Equation33}
\end{align}
}

where $h^{\scaleto{\text{ {\fontfamily{qcr}\selectfont EP}}}{2.5pt}}_{l}(x)$ is the homogeneous solution of \eqref{Appendix A: Mathematical Proofs-Equation32}. Now, making the substitution $h^{\scaleto{\text{ {\fontfamily{qcr}\selectfont EP}}}{2.5pt}}_{l}(x) = x^{1-\alpha}g^{\scaleto{\text{ {\fontfamily{qcr}\selectfont EP}}}{2.5pt}}_{l}(x)$, equation \eqref{Appendix A: Mathematical Proofs-Equation32} yields to the following second order ODE

\vspace{0.3cm}

\begin{align}
x\left(x-B\right)g''^{\scaleto{\text{ {\fontfamily{qcr}\selectfont EP}}}{2.5pt}}_{l}(x)+\left[\frac{c_{{\scaleto{T}{4pt}}}\left(3-\alpha\right) + \lambda}{c_{{\scaleto{T}{4pt}}}}x + \frac{B c_{{\scaleto{T}{4pt}}} \left(\alpha - 2\right) + \beta}{c_{{\scaleto{T}{4pt}}}}\right]g'^{\scaleto{\text{ {\fontfamily{qcr}\selectfont EP}}}{2.5pt}}_{l}(x)+ \frac{\left(1-\alpha\right)\left(c_{{\scaleto{T}{4pt}}}+\lambda\right)}{c_{{\scaleto{T}{4pt}}}}g^{\scaleto{\text{ {\fontfamily{qcr}\selectfont EP}}}{2.5pt}}_{l}(x)=0.
    \label{Appendix A: Mathematical Proofs-Equation34}
\end{align}

\vspace{0.3cm}

A second substitution, $y^{\scaleto{\text{ {\fontfamily{qcr}\selectfont EP}}}{2.5pt}}_{l}:=y^{\scaleto{\text{ {\fontfamily{qcr}\selectfont EP}}}{2.5pt}}_{l}(x)=\frac{x}{B}$, such that $f^{\scaleto{\text{ {\fontfamily{qcr}\selectfont EP}}}{2.5pt}}_{l}\left(y^{\scaleto{\text{ {\fontfamily{qcr}\selectfont EP}}}{2.5pt}}_{l}(x)\right) = g^{\scaleto{\text{ {\fontfamily{qcr}\selectfont EP}}}{2.5pt}}_{l}(x)$, produces equation \eqref{Appendix A: Mathematical Proofs-Equation8} for $a^{\scaleto{\text{ {\fontfamily{qcr}\selectfont EP}}}{2.5pt}}_{l}=1-\alpha$, $b^{\scaleto{\text{ {\fontfamily{qcr}\selectfont EP}}}{2.5pt}}_{l}=\frac{c_{{\scaleto{T}{2.5pt}}}+\lambda}{c_{{\scaleto{T}{2.5pt}}}}$ and $c^{\scaleto{\text{ {\fontfamily{qcr}\selectfont EP}}}{2.5pt}}_{l}=-\frac{B c_{{\scaleto{T}{2.5pt}}} \left(\alpha - 2\right) + \beta}{Bc_{{\scaleto{T}{2.5pt}}}}$, with regular singular points at $y^{\scaleto{\text{ {\fontfamily{qcr}\selectfont EP}}}{2.5pt}}_{l}=0,1,\infty$ (corresponding to $x=0,B$ and $\infty$). Thus, knowing that a general solution of \eqref{Appendix A: Mathematical Proofs-Equation8} in the neighborhood of the singular point $y^{\scaleto{\text{ {\fontfamily{qcr}\selectfont EP}}}{2.5pt}}_{l}=0$ is of the form \eqref{Appendix A: Mathematical Proofs-Equation30} and that $h^{\scaleto{\text{ {\fontfamily{qcr}\selectfont EP}}}{2.5pt}}_{l}(x) = x^{1-\alpha}g^{\scaleto{\text{ {\fontfamily{qcr}\selectfont EP}}}{2.5pt}}_{l}(x)$ one obtains the homogenous solution

\vspace{0.3cm}

\begin{align}
    h^{\scaleto{\text{ {\fontfamily{qcr}\selectfont EP}}}{2.5pt}}_{l}(x) & = A^{\scaleto{\text{ {\fontfamily{qcr}\selectfont EP}}}{2.5pt}}_{1,l}{y^{\scaleto{\text{ {\fontfamily{qcr}\selectfont EP}}}{2.5pt}}_{l}(x)}^{1-\alpha}  { }_{2} F_{1}\left(a^{\scaleto{\text{ {\fontfamily{qcr}\selectfont EP}}}{2.5pt}}_{l}, b^{\scaleto{\text{ {\fontfamily{qcr}\selectfont EP}}}{2.5pt}}_{l} ; c^{\scaleto{\text{ {\fontfamily{qcr}\selectfont EP}}}{2.5pt}}_{l} ; y^{\scaleto{\text{ {\fontfamily{qcr}\selectfont EP}}}{2.5pt}}_{l}(x)\right)\\ \\
    &+A^{\scaleto{\text{ {\fontfamily{qcr}\selectfont EP}}}{2.5pt}}_{2,l}{y^{\scaleto{\text{ {\fontfamily{qcr}\selectfont EP}}}{2.5pt}}_{l}(x)}^{2-c^{\scaleto{\text{ {\fontfamily{qcr}\selectfont EP}}}{2pt}}_{l} - \alpha} { }_{2} F_{1}\left(a^{\scaleto{\text{ {\fontfamily{qcr}\selectfont EP}}}{2.5pt}}_{l}-c^{\scaleto{\text{ {\fontfamily{qcr}\selectfont EP}}}{2.5pt}}_{l}+1, b^{\scaleto{\text{ {\fontfamily{qcr}\selectfont EP}}}{2.5pt}}_{l}-c^{\scaleto{\text{ {\fontfamily{qcr}\selectfont EP}}}{2.5pt}}_{l}+1 ; 2-c^{\scaleto{\text{ {\fontfamily{qcr}\selectfont EP}}}{2.5pt}}_{l} ; y^{\scaleto{\text{ {\fontfamily{qcr}\selectfont EP}}}{2.5pt}}_{l}(x)\right),
    \label{Appendix A: Mathematical Proofs-Equation35}
\end{align}

\vspace{0.3cm}

for arbitrary constants $A^{\scaleto{\text{ {\fontfamily{qcr}\selectfont EP}}}{2.5pt}}_{1,l},A^{\scaleto{\text{ {\fontfamily{qcr}\selectfont EP}}}{2.5pt}}_{2,l} \in \mathbb {R}$.  Due to the fact that $\psi^{\scaleto{\text{ {\fontfamily{qcr}\selectfont EP}}}{2.5pt}}_{l}(x)$ is finite, we can then conclude that $A^{\scaleto{\text{ {\fontfamily{qcr}\selectfont EP}}}{2.5pt}}_{1,l}=0$, as the first term of \eqref{Appendix A: Mathematical Proofs-Equation35} is unbounded when $x \rightarrow 0^{+}$ for $\alpha > 0$. Hence, the solution of $\psi^{\scaleto{\text{ {\fontfamily{qcr}\selectfont EP}}}{2.5pt}}_{l}(x)$ is given by

\vspace{0.3cm}

\begin{align}
\psi^{\scaleto{\text{ {\fontfamily{qcr}\selectfont EP}}}{2.5pt}}_{l}(x) =  A^{\scaleto{\text{ {\fontfamily{qcr}\selectfont EP}}}{2.5pt}}_{2,l}{y^{\scaleto{\text{ {\fontfamily{qcr}\selectfont EP}}}{2.5pt}}_{l}(x)}^{2-c^{\scaleto{\text{ {\fontfamily{qcr}\selectfont EP}}}{2pt}}_{l} - \alpha} { }_{2} F_{1}\left(a^{\scaleto{\text{ {\fontfamily{qcr}\selectfont EP}}}{2.5pt}}_{l}-c^{\scaleto{\text{ {\fontfamily{qcr}\selectfont EP}}}{2.5pt}}_{l}+1, b^{\scaleto{\text{ {\fontfamily{qcr}\selectfont EP}}}{2.5pt}}_{l}-c^{\scaleto{\text{ {\fontfamily{qcr}\selectfont EP}}}{2.5pt}}_{l}+1 ; 2-c^{\scaleto{\text{ {\fontfamily{qcr}\selectfont EP}}}{2.5pt}}_{l} ; y^{\scaleto{\text{ {\fontfamily{qcr}\selectfont EP}}}{2.5pt}}_{l}(x)\right) + 1.
    \label{Appendix A: Mathematical Proofs-Equation31}
\end{align}

\vspace{0.3cm}

As in Appendix \ref{ProofofProposition4.1}, following the proof of Proposition \ref{TheTrappingTime-Subsection21-Proposition1} for $\delta=0$, one can easily obtain the solutions for $\psi^{\scaleto{\text{ {\fontfamily{qcr}\selectfont EP}}}{2.5pt}}_{u}(x)$ and $\psi^{\scaleto{\text{ {\fontfamily{qcr}\selectfont EP}}}{2.5pt}}_{m}(x)$, when $x\geq B$ and $x^{*}\leq x < B$, respectively.

Finally, due to the continuity of $\psi^{\scaleto{\text{ {\fontfamily{qcr}\selectfont EP}}}{2.5pt}}(x)$ and $\psi'^{\scaleto{\text{ {\fontfamily{qcr}\selectfont EP}}}{2.5pt}}(x)$ at the critical capital $x^{*}$ and the capital barrier level $B$, that is, using \eqref{WhenandHowHouseholdsBecomeExtremelyPoor?-Section4-Equation5}, \eqref{WhenandHowHouseholdsBecomeExtremelyPoor?-Section4-Equation6}, \eqref{WhenandHowHouseholdsBecomeExtremelyPoor?-Section4-Equation10} and \eqref{WhenandHowHouseholdsBecomeExtremelyPoor?-Section4-Equation11} for $\delta=0$ and $w^{\scaleto{\text{ {\fontfamily{qcr}\selectfont EP}}}{2.5pt}}\left(x_{1}, x_{2}\right)=1$, one can derive a system of equations from which the unknown coefficients $A^{\scaleto{\text{ {\fontfamily{qcr}\selectfont EP}}}{2.5pt}}_{2,u}$, $A^{\scaleto{\text{ {\fontfamily{qcr}\selectfont EP}}}{2.5pt}}_{1,m}$, $A^{\scaleto{\text{ {\fontfamily{qcr}\selectfont EP}}}{2.5pt}}_{2,m}$ and $A^{\scaleto{\text{ {\fontfamily{qcr}\selectfont EP}}}{2.5pt}}_{2,l}$, can be determined to derive a closed-form expression for $\psi^{\scaleto{\text{ {\fontfamily{qcr}\selectfont EP}}}{2.5pt}}(x)$.

\newpage

\begin{textblock*}{15cm}(3cm,0.5cm) 
   \subsection{Effects of Underlying Factors on the Trapping Probability} \label{Appendix B: Effects of Underlying Factors on the Trapping Probability}

We consider the influence of the parameters on the trapping probability by varying them in a reasonable range, keeping all other parameters constant. The reference setup is given below.

\textbf{Reference setup:} $a = 0.1$, $b = 4$, $c_{{\scaleto{S}{4pt}}} = 0.4$, $Z_{i} \sim Beta(0.8, 1)$, $\lambda = 1$, $x^{*} = 1$, $B=2$ and $c_{{\scaleto{T}{4pt}}} = 1$.
\end{textblock*}

\newgeometry{hmargin=2.25cm,vmargin=4.7cm}

\begin{figure}[p]
	\vspace{-0.55cm}
  	\centering
  	\includegraphics[width=5.5cm, height=5.5cm]{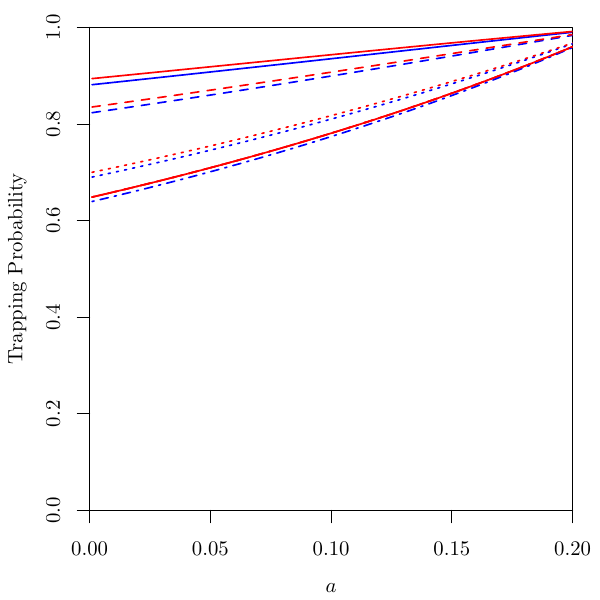}
 	 \hspace{0.01cm}
 	 \includegraphics[width=5.5cm, height=5.5cm]{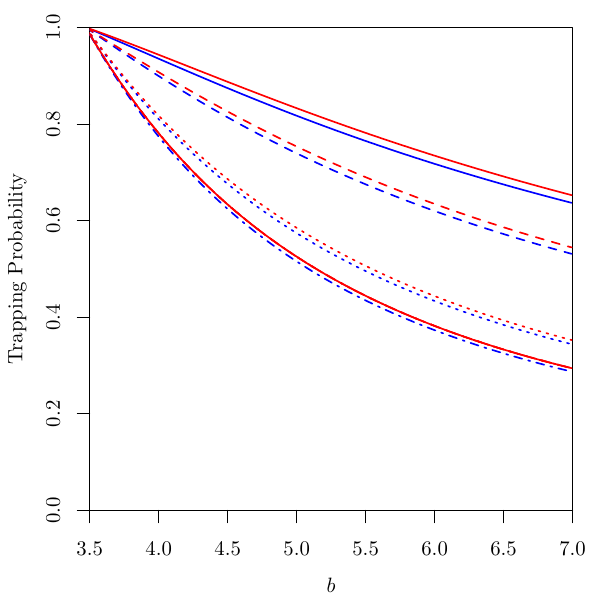}
  	 \hspace{0.01cm}
 	 \includegraphics[width=5.5cm, height=5.5cm]{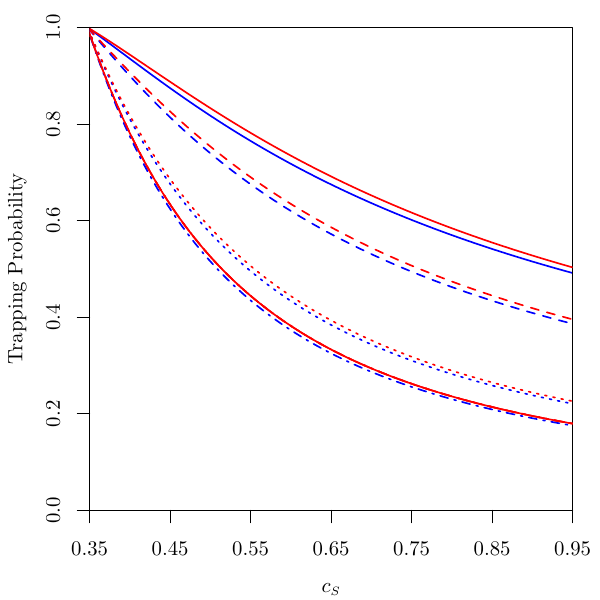}
	 \\ \vspace{0.1cm}
 	 \includegraphics[width=5.5cm, height=5.5cm]{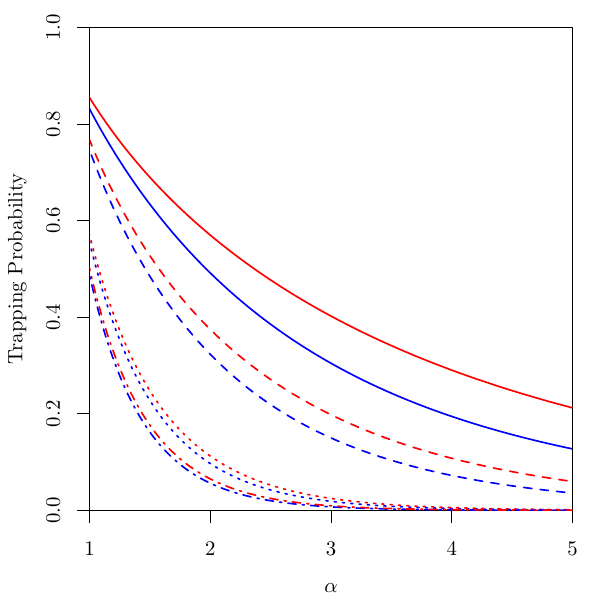}
 	 \hspace{0.01cm}
 	 \includegraphics[width=5.5cm, height=5.5cm]{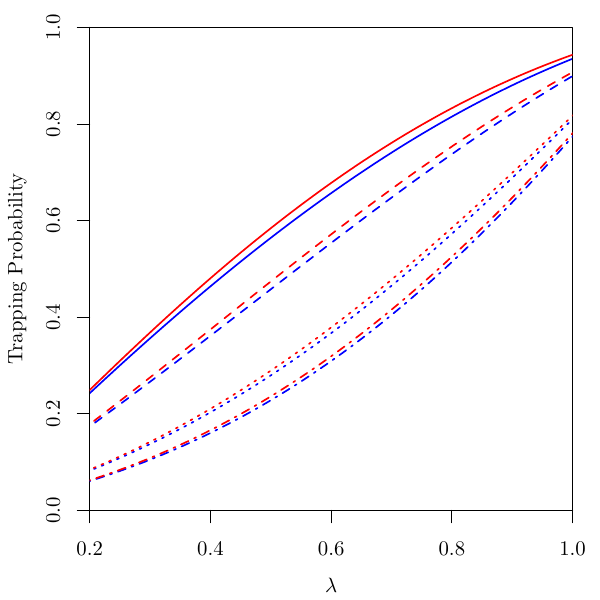}
  	\hspace{0.01cm}
 	 \includegraphics[width=5.5cm, height=5.5cm]{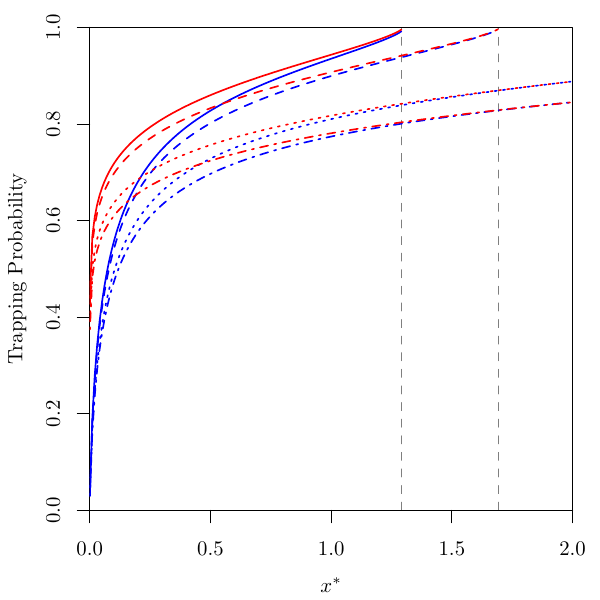}
	 \\ \vspace{0.1cm}
  	\hspace{-5.85cm} \includegraphics[width=5.5cm, height=5.5cm]{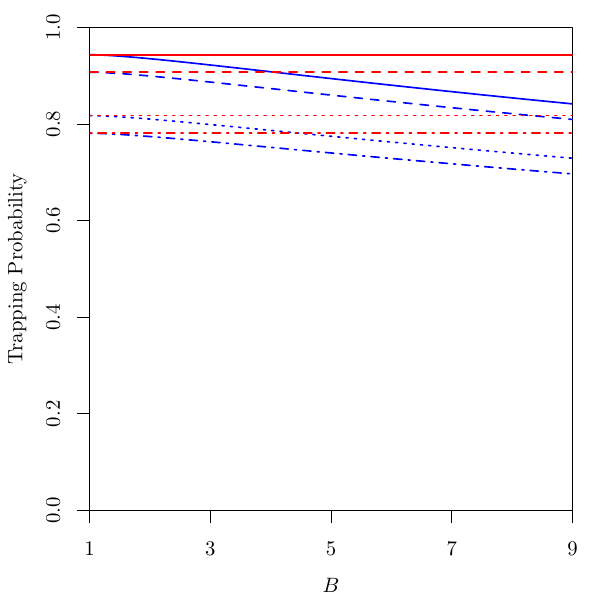}
  	\vspace{0.1cm}
  	\hspace{0cm} \includegraphics[width=5.5cm, height=5.5cm]{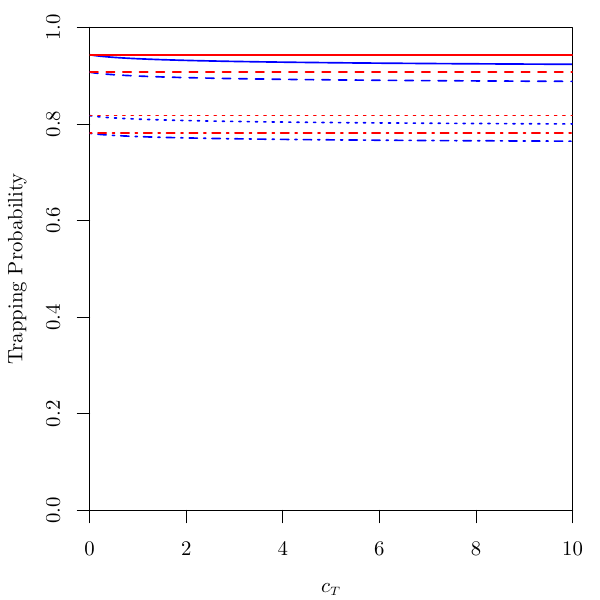} 
  \caption{Effects of the rate of consumption $(0<a<1)$, income generation $(b>0)$, investment or savings $(0<c_{{\scaleto{S}{4pt}}}<1)$, the parameter of the Beta distribution $(\alpha > 0)$ (i.e., expected remaining proportion of capital), the expected capital loss frequency $(\lambda>0)$, the critical capital $(x \geq x^{*})$, the capital barrier level $(B > x^{*})$ and the capital transfer rate $(c_{{\scaleto{T}{4pt}}} > 0)$ on the trapping probability of the original model obtained in \cite{Article:Henshaw2023} (in red) and on the trapping probability of the model with capital cash transfers (in blue) for initial capital $x = 1.3 \text{ (solid)}, 1.7 \text{ (dashed)}, 4.0 \text{ (dotted)}, 6.0 \text{ (dashed-dotted)}$.}
  \label{Appendix B: Effects of Underlying Factors on the Trapping Probability-Figure1}
\end{figure}


\restoregeometry

\newpage


\begin{textblock*}{15cm}(3cm,0.5cm) 
\subsection{Effects of Underlying Factors on the Probability of Extreme Poverty} \label{Appendix C: Effects of Underlying Factors on the Probability of Extreme Poverty}

We consider the influence of the parameters on the probability of extreme poverty by varying them in a reasonable range, keeping all other parameters constant. The reference setup is given below.

\textbf{Reference setup:} $a = 0.1$, $b = 4$, $c_{{\scaleto{S}{4pt}}} = 0.4$, $Z_{i} \sim Beta(0.8, 1)$, $\lambda = 1$, $x^{*} = 1$, $B=2$, $c_{{\scaleto{T}{4pt}}} = 1$, $\omega_{1} (x)= 0.05$ and $\omega_{2} (x)= \frac{0.05}{x}$.
\end{textblock*}

\newgeometry{hmargin=2.25cm,vmargin=4.7cm}
\begin{figure}[p]
\vspace{0.25cm}
	\vspace{-0.55cm}
  	\centering
  	\includegraphics[width=5.5cm, height=5.5cm]{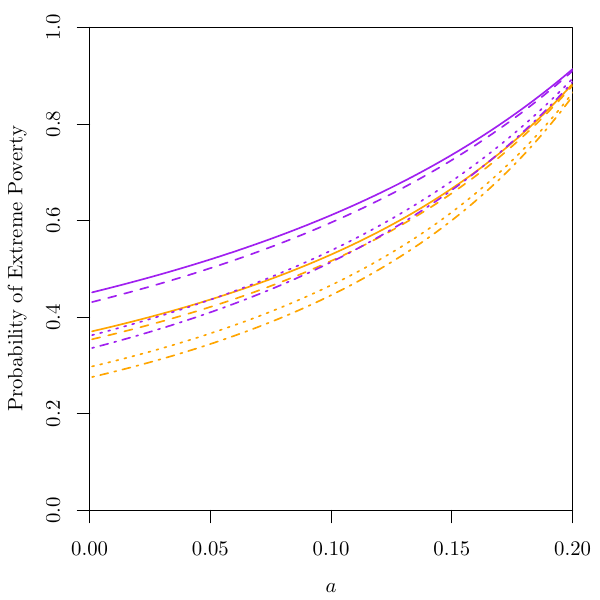}
 	 \hspace{0.01cm}
 	 \includegraphics[width=5.5cm, height=5.5cm]{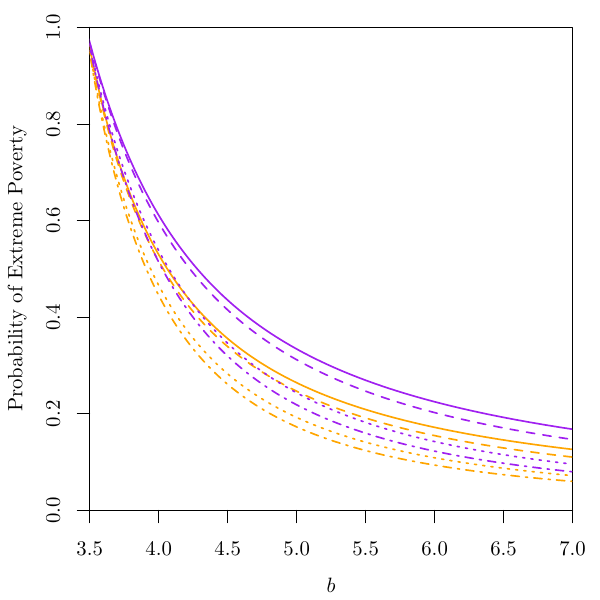}
  	 \hspace{0.01cm}
 	 \includegraphics[width=5.5cm, height=5.5cm]{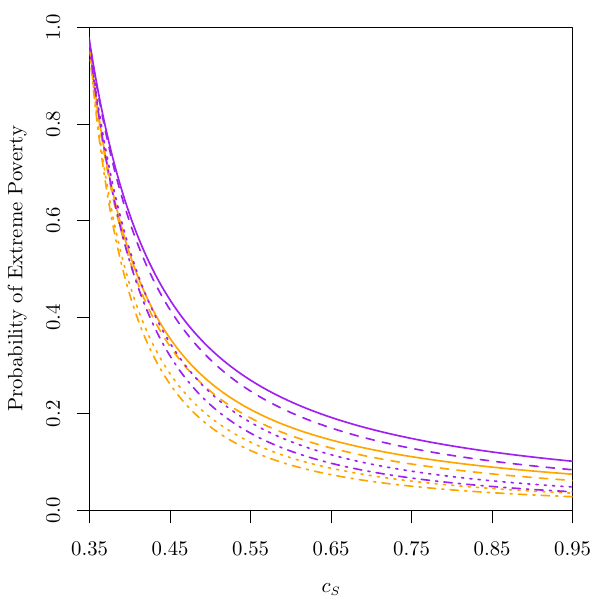}
	 \\ \vspace{0.1cm}
 	 \includegraphics[width=5.5cm, height=5.5cm]{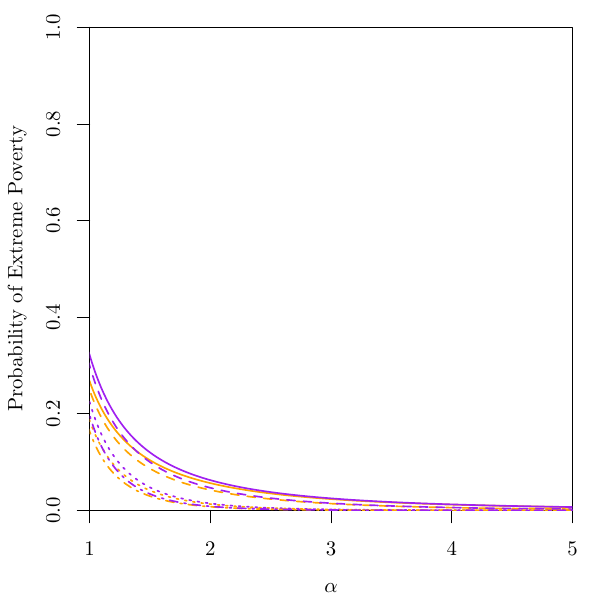}
 	 \hspace{0.01cm}
 	 \includegraphics[width=5.5cm, height=5.5cm]{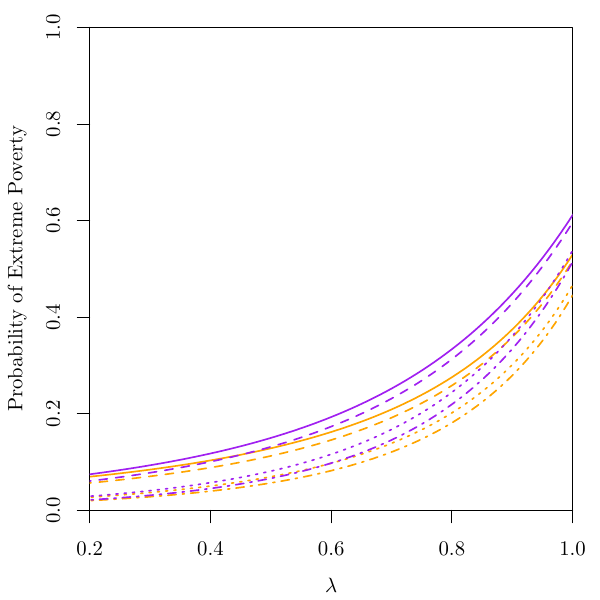}
  	\hspace{0.01cm}
 	 \includegraphics[width=5.5cm, height=5.5cm]{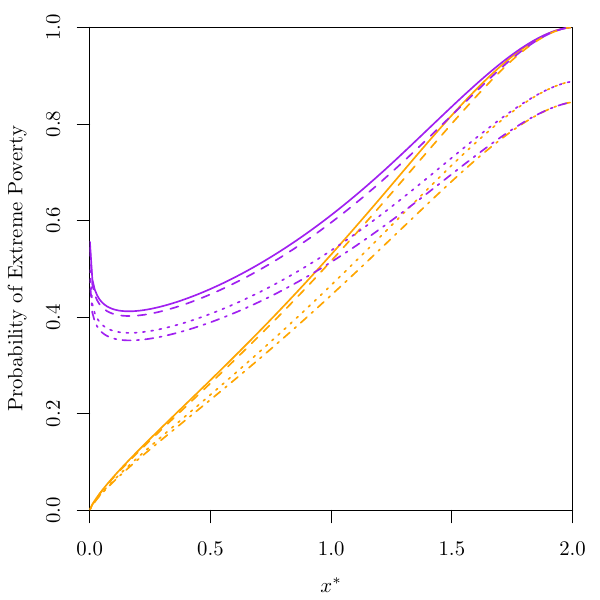}
	 \\ \vspace{0.1cm}
  	\includegraphics[width=5.5cm, height=5.5cm]{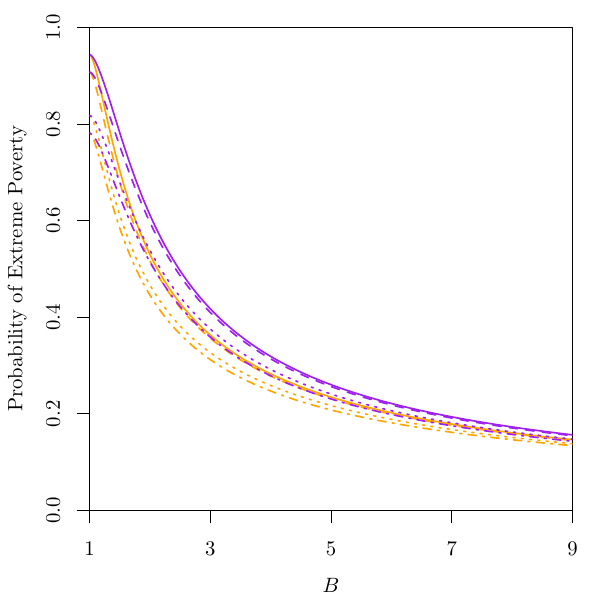}
  	\hspace{0.01cm}
   	 \includegraphics[width=5.5cm, height=5.5cm]{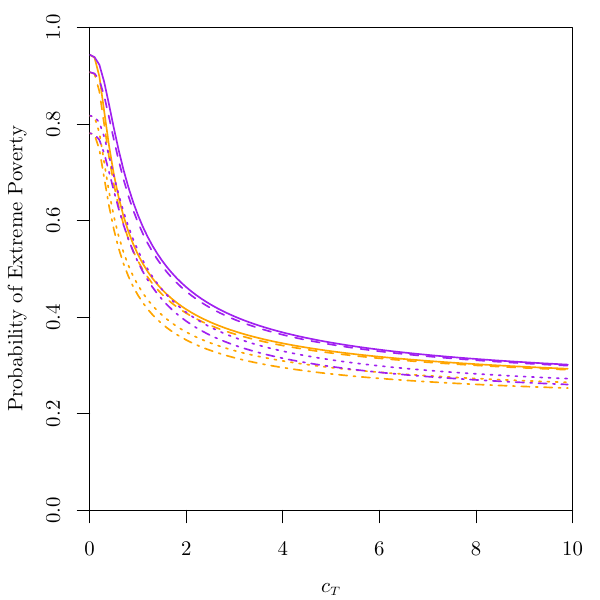}
  	\hspace{0.01cm}
  	\includegraphics[width=5.5cm, height=5.5cm]{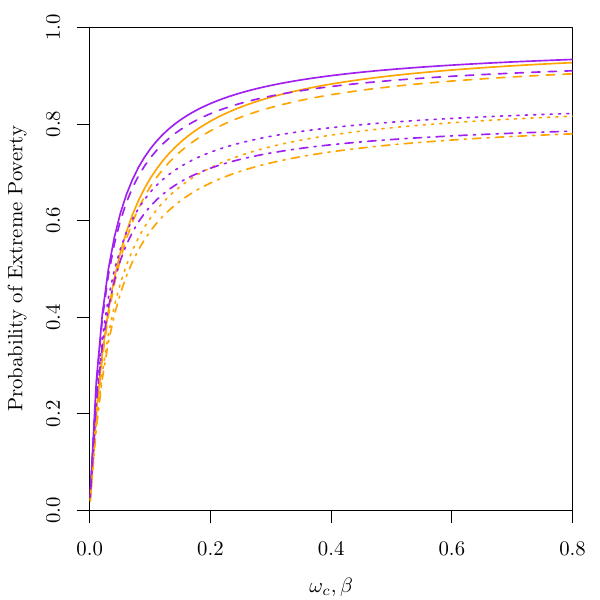} \\

  \caption{Effects of the rate of consumption $(0<a<1)$, income generation $(b>0)$, investment or savings $(0<c_{{\scaleto{S}{4pt}}}<1)$, the parameter of the Beta distribution $(\alpha > 0)$ (i.e., expected remaining proportion of capital), the expected capital loss frequency $(\lambda>0)$, the critical capital $(x \geq x^{*})$, the capital barrier level $(B > x^{*})$, the capital transfer rate $(c_{{\scaleto{T}{4pt}}} > 0)$ and the extreme poverty rate function on the probability of extreme poverty for a constant extreme poverty rate function (in orange) and an exponential extreme poverty rate function (in purple) for initial capital $x = 1.3 \text{ (solid)}, 1.7 \text{ (dashed)}, 4.0 \text{ (dotted)}, 6.0 \text{ (dashed-dotted)}$.}
  \label{Appendix C: Effects of Underlying Factors on the Probability of Extreme Poverty-Figure1}
\end{figure}


\restoregeometry

\typeout{get arXiv to do 4 passes: Label(s) may have changed. Rerun}

\end{document}